\documentclass[useAMS,usenatbib]{mn2e}
\usepackage{epsfig}
\usepackage{graphicx}
\usepackage{times}
\usepackage{color}

\def\cm3{cm$^{-3}$}
\def\kms{km~s$^{-1}$}
\def\msunyr{M$_{\odot}$\,yr$^{-1}$}
\def\lsun{L$_{\odot}$}
\def\rsun{R$_{\odot}$}
\def\mdot{$\dot{\rm M}$}

\def\msun{M$_{\odot}$}

\def\one{\ts {\,\sc i}}
\def\two{\ts {\,\sc ii}}
\def\three{\ts {\,\sc iii}}

\def\five{\ts {\sc v}}

\def\beq{\begin{equation}}
\def\eeq{\end{equation}}

\def\lesssim{\mathrel{\hbox{\rlap{\hbox{\lower4pt\hbox{$\sim$}}}\hbox{$<$}}}}
\def\gtrsim{\mathrel{\hbox{\rlap{\hbox{\lower4pt\hbox{$\sim$}}}\hbox{$>$}}}}

\def\lesssim{\mathrel{\hbox{\rlap{\hbox{\lower4pt\hbox{$\sim$}}}\hbox{$<$}}}}
\def\gtrsim{\mathrel{\hbox{\rlap{\hbox{\lower4pt\hbox{$\sim$}}}\hbox{$>$}}}}

\def\one{{\,\sc i}}
\def\two{{\,\sc ii}}
\def\three{{\,\sc iii}}

\def\five{{\sc v}}

\def\v1d{{\sc v1d}}
\def\mesa{{\sc mesa}}
\def\cmfgen{{\sc cmfgen}}
\def\heracles{{\sc heracles}}

\newcommand{\iso}[2]{\ensuremath{^{#1}\rm{#2}}}

\def\aj{AJ}

\def\apj{ApJ}
\def\apjs{ApJS}
\def\apjl{ApJL}
\def\aap{A\&A}

\def\mnras{MNRAS}

\def\jqsrt{JQSRT}

\title[Spectral diversity of SNe IIn]{Models of interacting supernovae and their spectral diversity}

\author[Dessart et al.]
{Luc Dessart,$^{1}$
D. John Hillier,$^{2}$
Edouard Audit,$^{3}$
Eli Livne,$^4$
and Roni Waldman$^4$
\\ \\
$^{1}$:
Laboratoire Lagrange, Universit\'e C\^{o}te d'Azur, Observatoire de la C\^{o}te d'Azur, CNRS,
Boulevard de l'Observatoire, CS 34229, 06304 Nice cedex 4, France. \\
$^2$: Department of Physics and Astronomy \& Pittsburgh Particle Physics,
Astrophysics, and Cosmology Center (PITT PACC),  University of Pittsburgh, \\
3941 O'Hara Street, Pittsburgh, PA 15260, USA. \\
$^{3}$: Maison de la Simulation, CEA, CNRS, Universit\'e Paris-Sud, UVSQ,
Universit\'e Paris-Saclay, 91191, Gif-sur-Yvette, France. \\
$^4$: Racah Institute of Physics, The Hebrew University, Jerusalem 91904, Israel. \\
}

\begin{document}

\date{Accepted . Received }

\pagerange{\pageref{firstpage}--\pageref{lastpage}} \pubyear{2015}

\maketitle
\label{firstpage}

\begin{abstract}
Using radiation-hydrodynamics and radiative-transfer simulations, we explore
the origin of the spectral diversity of interacting supernovae (SNe) of type IIn.
We revisit SN\,1994W and investigate the dynamical configurations that can give rise to
spectra with narrow lines at all times.
We find that a standard $\sim$\,10\,\msun\ 10$^{51}$\,erg SN ejecta
ramming into a 0.4\,\msun\ dense CSM is inadequate for SN\,1994W, as
it leads to the appearance of broad lines at late times. This structure, however, generates
spectra that exhibit the key morphological changes seen in SN\,1998S.
For SN\,1994W, we consider a completely different configuration, which involves
the interaction at a large radius of a low mass inner shell with a high mass outer shell.
Such a structure may arise in an 8-12\,\msun\ star from a nuclear flash (e.g., of Ne)
followed within a few years by core collapse.
Our simulations show that the large mass of the outer shell leads to the complete braking of the inner shell material,
the formation of a slow dense shell, and the powering of a luminous SN IIn, even for
a low inner shell energy.
Early on, our model line profiles are typical of SNe IIn, exhibiting narrow cores and broad
electron-scattering wings. As observed in SN1994W, they also remain narrow at late times.
Our SN\,1994W model invokes two low energy ejections, both atypical of observed massive stars,
and illustrates the diversity of configurations leading to SNe IIn.
These results also highlight the importance of spectra to constrain the dynamical properties
and understand the origin of SNe IIn.
\end{abstract}

\begin{keywords} radiative transfer -- radiation hydrodynamics --
supernovae: general -- supernovae: individual: 1994W, 1998S, 2011ht.
\end{keywords}

\section{Introduction}

Interacting SNe exhibit a wide range of radiative properties, covering extremes
that include both super-luminous events and ``impostors".
This diversity of bolometric displays reflects variations in the power source, and in
the context of interacting SNe, this is controlled by the outer shell mass and the inner shell
energy (we will loosely refer to the outer shell as circumstellar material; CSM).
More perplexing is the diversity of spectra associated with interacting SNe.
Multi-dimensional effects acting both on small scale (e.g., to produce clumping) and large
scale (e.g., an asymmetric CSM) may play a role.
However, a large variation of both inner and outer shell properties in
spherically-symmetric interactions can alone produce a wealth of
peculiar spectral signatures.
Multiple regions are likely to contribute simultaneously to the escaping radiation, including
the unshocked cool CSM, the ionised unshocked CSM, the interacting region between
the reverse and the forward shocks (which contains both shocked CSM and shocked ejecta material,
and bounds  the cold dense shell, CDS), as well as the inner ejecta in homologous expansion.
These distinct regions generally have different temperatures, density, velocity,
and they may also possess a different composition. Because of the evolving structure of the interaction,
the emitting regions at early times may be very distinct from those that contribute at late times. Consequently,
one expects significant spectral evolution from SNe IIn.

Interacting SNe are notoriously known for their spectral line profiles which exhibit
narrow cores and broad wings. This is the specific signature seen at early times
when the SN is discovered and the classification as IIn stems from that observation alone
\citep{dopita_etal_84, niemela_etal_85, schlegel_90}.
A few well known and recent examples are SN 1998S \citep{fassia_98S_00,leonard_98S_00},
1994W \citep{sollerman_etal_98}, 2006tf \citep{smith_06tf_08},
2009kn \citep{kankare_09kn}, 2010jl \citep{stoll_etal_11, zhang_etal_12},
or 2011ht \citep{roming_11ht_12,humphreys_etal_11ht, mauerhan_11ht_13}.
This special line profile morphology is controlled by the low velocity of the CSM
which gives rise to the narrow component, and by scattering with thermal electrons.
In SNe IIn the importance of electron scattering as a line broadening mechanism is enhanced
by the large electron scattering optical depth, the larger Sobolev lengths (since the velocities
are smaller), and the smaller bulk velocities which
can be smaller than the characteristic speed of thermal electrons
(of the order of 550\,$\sqrt{T/10^4 {\rm K}}$\,\kms).\footnote{Spectral line broadening
by thermal electrons is seen not just in supernovae, but also in partially-ionised dense slowly moving outflows
of stars (e.g., $\eta$ Car; \citealt{hillier_etal_01}).}
Modelling of this process in the SN context gives support to this interpretation
\citep{chugai_98S_01, dessart_etal_09}.

 \begin{table*}
 \caption{
 Summary of initial properties for the interaction models, separating those for the inner shell
 (which may be supernova ejecta or not) and for the outer shell (produced dynamically
 in an explosion or secularly through a long-lived super-wind phase).
For both shells, we give its kinematic age when reaching the initial interaction radius $R_{\rm t}$.
The quantity $V_{\rm m}$ represents the mean mass-weighted shell velocity.
Numbers in parenthesis correspond to powers of ten.
 \label{tab_mod}
 }
\begin{tabular}{l@{\hspace{12mm}}c@{\hspace{2mm}}c@{\hspace{2mm}}c@{\hspace{2mm}}c@{\hspace{2mm}}c@{\hspace{8mm}}c@{\hspace{8mm}}c@{\hspace{2mm}}
c@{\hspace{2mm}}c@{\hspace{2mm}}c@{\hspace{2mm}}c@{\hspace{2mm}}c@{\hspace{2mm}}}
\hline
 & & \multicolumn{3}{c}{Inner shell}   & &  & & \multicolumn{3}{c}{Outer shell}  \\
 \hline
    Model   &  Type &   Age   & $E_{\rm kin}$     & $M_{\rm tot}$  &  $V_{\rm m}$ & $R_{\rm t}$ & Type &  Age & $E_{\rm kin}$ & $M_{\rm tot}$  & \mdot  & $V_{\rm m}$  \\
                 &            &   [d]     & [erg]   & [\msun]            & [\kms]    &     [10$^{15}$\,cm] &    & [d] &  [erg]   & [\msun]  &  [\msunyr] & [\kms]  \\
\hline
      A   &      Ejecta  &  12.0 & 1.00(51)       &  9.96  &   3000 &  1.00   &  Wind   &  1157.4 &   3.49(46)   &  0.35   &  0.1   &  100  \\
\hline
      B1 & Ejecta &  11.5 &     4.95(49) &       0.98 &        2189    &   0.85 & Wind &   167.2 &    3.56(48)   &  1.0 &  1.0&        600   \\
      B2 & Ejecta &  22.9 &     9.57(49) &       9.20 &        970      &  1.29  & Wind &    253.3 &    3.56(47) &    0.1 &  0.1&        600  \\
      B3 & Ejecta &   6.9 &     9.56(50) &        9.50 &        3042    & 1.18  & Wind &    233.4 &    3.56(47) &     0.1 &  0.1 &        600 \\
\hline
      R1 & Ejecta &  36.2 &    4.70(49) &       1.0 &        2150 & 1.00 &Wind &       1971.1 &    3.39(46) &       1.0 &        1.0 &        60 \\
      R2 & Ejecta &   55.1&     8.90(49) &        9.54 &      937 & 0.97&Wind &       1889.0  &    3.42(45) &      0.1 &      0.1 &        60 \\
      R3 & Ejecta &   22.2 &     9.28(50) &    9.54 &        3027 & 1.33 &Wind &       2598.8 &     3.42(45) &      0.1 &      0.1 &        60 \\
\hline
      C   &      Ejecta  & 2.3   &   8.56(49)       &   0.31 &   4755 &  0.20 &    Ejecta  &   127.9    & 1.19(49)  &  6.3      & \dots &  396 \\
      D   &      Ejecta   & 23.4&   6.77(49)       &   0.29 &   4730  & 1.80    &   Ejecta   &  941.7  & 0.94(49)  &  6.1      & \dots &  377 \\
\hline
\end{tabular}
 \end{table*}

However, the relative spectral uniformity of SNe IIn (with line profiles showing narrow cores
and broad symmetric wings) holds only at early times.
Eventually their spectra start showing very diverse line profiles, in some cases unique.
So far, the modelling of SNe IIn spectra has been focused on the early epochs when the spectra
show the distinct ``IIn" signature.
But to build a consistent picture of interacting SNe it is crucial to understand the entire evolution,
from early to late times.
As we discuss here, the late-time line profiles offer key constraints, not available at early times
when the optical depth is large and frequency redistribution through electron scattering is strong.

Here, we focus on the behaviour of line-profile widths after maximum light and demonstrate
how line-profile morphology can be used to distinguish interacting models, thereby lifting
the degeneracy inherent to SN light curves.
Specifically, we want to understand what distinguishes events like SN\,1998S
or SN\,2010jl, which show broad emission lines at late times --- and in particular H$\alpha$, from events like
SNe 1994W, 2009kn, or 2011ht  which show narrow line profiles that become systematically narrower with time.

This issue emerged  in \citet{dessart_etal_09} when modelling SN\,1994W. Although our models were
not based on hydrodynamical simulations, we reproduced the spectral evolution of the event by invoking slow
emitting material at {\it all times} and we noted the remarkable absence of broad lines.
This suggested that while fast material may have been present early on, it must have been quickly decelerated
to a low velocity, or that this fast material was somehow dark (i.e., had a low emissivity), or that at all times
all emission from the fast-moving material was reprocessed by the slowly moving CSM (i.e., because of
an optical depth effect).

In the model proposed by \citet{chugai_etal_04}, a SN ejecta of 7-12\,\msun\ rams into a 0.4\,\msun\ extended
shell, producing a light curve similar to SN\,1994W. In this model, the energy/momentum
is stored primarily in the inner shell.
The moderate deceleration of the inner shell by the outer shell leads to a moderate luminosity boost
compared to standard non-interacting SNe.
It also leads to the formation of a CDS moving at $\sim$\,4000\,\kms, which implies
that a lot of fast moving material survives
the interaction with the CSM.
In contrast, the model of \citet{chugai_etal_04} predicts an H$\alpha$ line width of the order of 1000\,\kms\
throughout the high-brightness
phase of SN\,1994W, including late times when the CSM optical depth in their model is below unity.
This result is surprising, because their hydrodynamical model predicts a fast moving CDS :
photons emitted from the CDS are Doppler shifted and this should eventually become noticeable
in the emerging radiation as Doppler-broadened line profiles.
Doppler-broadened profiles are seen (and explained) at late times in some SNe IIn, for example 2010jl \citep{d15_10jl},
so this mechanism is not unexpected in interacting SNe.

Since SN\,1994W was discovered, other SNe have been identified with similar properties, in particular
SN\,2009kn \citep{kankare_09kn} and SN\,2011ht \citep{roming_11ht_12,humphreys_etal_11ht, mauerhan_11ht_13}.
SN\,2011A seems to be analogous to these events too \citep{dejaeger_11A_15}.
A separate SN IIn-P classification has been proposed for these events \citep{mauerhan_11ht_13}. Here,
we wish to explore what hydrodynamical configurations can produce this distinct class of events.
Besides modelling light curves, which offer ambiguous constraints when used alone, we also study the spectral
evolution of our radiation-hydrodynamical models.
We use the same numerical approach as in \citet{d15_10jl} for the study of super-luminous interacting SNe
like SN\,2010jl.
Starting from various initial configurations, we perform multi-group radiation hydrodynamics simulations
of the interaction between an inner shell and an outer shell. We then post-process these radiation hydrodynamical
simulations (reading off the radius, velocity, temperature, and density of the simulated domain at a given epoch)
with the non-Local-Thermodynamic-Equilibrium (nLTE) code \cmfgen. As in \citet{d15_10jl}, the radiative
transfer works for an arbitrary velocity field and can therefore retain the complex dynamical properties of the
interaction region. Because the simulation with \cmfgen\ assumes steady-state, we repeat
the calculation for multiple epochs to capture the spectral evolution of the interacting SN model.

In the next section we provide a brief summary of the numerical procedure, in particular the small
improvements over the approach presented in \citet{d15_10jl}.
In Section~\ref{sect_obs}, we give the sources of observational data we use, as well as the
adopted distance, reddening, and explosion time.
In Section~\ref{sect_moda}, we start by revisiting the model of \citet{chugai_etal_04},
which involves an energetic massive inner shell ramming into a low mass extended CSM.
We show how that model is more suitable to explain events like SN\,1998S.
We then present, in Section~\ref{sect_grid}, a grid of simulations in which we vary
the inner and outer shell properties to identify a configuration compatible with the observations of SN\,1994W.
In Section~\ref{sect_modc}, we propose a scenario for SN\,1994W.
It involves an energetic low-mass inner shell that is much less massive than the slow moving massive
outer shell. The reversed mass balance between inner/outer shells gives rise to a number of interesting properties
that can help explain the light curve and spectra of events like SN\,1994W.
As discussed by \citet{WH15}, various aspects of this scenario may be encountered in the final
stages of evolution of 9-11\,\msun\ stars.

While we were finalizing this manuscript and preparing for submission,
a study of SN\,2011ht by \citet{chugai_15} came out. His approach is very different
from ours but our respective conclusions agree --- the simulations and results presented here
have been obtained over the past year and our conclusions have not been influenced by \citet{chugai_15}.

\section{Numerical approach}
 \label{sect_setup}

   The simulations presented in this paper are performed with a variety of numerical tools, including
   \heracles\ \citep{gonzalez_etal_07,vaytet_etal_11}, \mesa\ \citep{mesa1,mesa2,mesa3},
   \v1d\ \citep{livne_93,dlw10a, dlw10b}, and \cmfgen\ \citep{HM98, DH05a, DH08, DH10, HD12}.

   \heracles\ is a Eulerian multi-dimensional radiation-hydrodynamics code \citep{gonzalez_etal_07},
  with the possibility for multi-group radiation transport \citep{vaytet_etal_11}.
The hydrodynamics is treated using a standard second order Godunov scheme. For the radiation transfer,
the multi-group M1 moment model \citep{m1_model} is used.
In all \heracles\ simulations, we adopt a uniform H-rich composition suitable for the study of SNe IIn.
In practice, we use a H mass fraction of 0.633, He mass fraction of 0.36564, and an iron mass fraction of 0.00136.
Instead of assuming an ideal gas, we use a general equation of state that accounts for the contributions
from atoms, ions, and electrons, including the contribution from ionisation energy. We have done comparison
tests with simulations that adopt an ideal gas equation of state and we find small differences only,
probably because the shocked low-density plasmas studied here are radiation dominated --- the thermal
energy is a small fraction of the energies (radiation or kinetic) involved.
Thermodynamic quantities (energy, pressure, sound speed, heat capacity, temperature) are tabulated in
density-temperature and density-energy space  for convenient use in \heracles.
As in \citet{d15_10jl}, we supply the code with an opacity table for our adopted composition.
We compute our opacities as a function of density, temperature and energy group.
Energy groups are positioned at strategic locations to capture the strong variation
in absorptive opacity with wavelength. We use one group for the entire Lyman continuum
(including the X-ray range), two groups for the Balmer continuum, two for the Paschen continuum,
and three groups for the Brackett continuum and beyond.

    Our \heracles\ simulations are 1-D and use a uniform radial grid with 1200-2400 cells (more grid
cells are employed when the radial coverage is larger). Such a resolution is sufficient to model
adequately the overall dynamics and energetics of the interaction and study the spectral
properties with \cmfgen.\footnote{With a higher resolution,
we resolve better the velocity jumps, for example associated with the weak reverse shock.
The Zeldovich spike ahead of the forward shock also peaks at a higher temperature. But the global energetics (e.g., the
bolometric light curve, the conversion efficiency) and dynamics (shock propagation speed) are only
weakly altered. See \citet{d15_10jl} for discussion.}

    The initial configuration of the interaction is produced in various ways. In all cases, we consider interactions
    between two shells at large distances -- the junction between the inner and outer shells is far beyond
    the initial radius of the progenitor star, and typically in the range 0.2--1.8$\times$\,10$^{15}$\,cm.
    The parameter space is large so other configurations will be considered later.

    The internal energy budget at the onset of
    interaction will differ whether we consider eruptions/explosions
    from blue-supergiant (BSG) or red-supergiant (RSG) stars because of the different magnitude of expansion cooling.
    In \citet{d15_10jl}, we ignored such considerations because we focused on super-luminous SNe IIn, events in
    which the power released by the interaction completely overwhelms the internal energy of the inner and outer shells
    at the onset of interaction.
    This may no longer hold when considering SNe IIn with a luminosity closer to non-interacting SNe.
    As we ignore any radiation from the system emitted prior to the
    onset of interaction our results underestimate the true luminosity from the system at early times.
    This matters if one is concerned with the quantitative aspects of the problem; here we are more focused
    on understanding the fundamental qualitative differences between SNe IIn. One can recall for example
    the strong offset in bolometric light curve between SN1994W and SN2011ht, while the two events
    show a similar spectral evolution.

    In practice, we use two different approaches for the initial setup. The first one, also used in \citet{d15_10jl},
    is simplistic since we set analytically the values for the fluid quantities (our choices are guided by simulations
    of core collapse SNe, e.g., \citealt{DH11}). This approach is used in Section~\ref{sect_moda}.
    One drawback is that we have to guess the initial temperature.
    If the outer shell arises from a pre-SN wind located at large distances, its age is old, and its temperature
    is low --- how low is irrelevant and we just adopt our floor temperature of 2000\,K.
    If the outer shell arises instead from a recent ejection, the temperature may still be high, i.e.,
    well above the recombination temperature of hydrogen.
    Further, the structure of the outer shell may also be significantly affected by radiation
    arising from the inner shell emitted prior to the onset of interaction.
    For the inner shell, a proper account of the ejecta energy/temperature is relevant because
    the interaction may start when the ejecta is still hot and optically thick.

    In Sections~\ref{sect_grid}--\ref{sect_modc}, we use a more physical approach for the initial setup.
    We prepare initial conditions for the \heracles\ simulations
    using \mesa\ \citep{mesa1,mesa2,mesa3} and \v1d\ \citep{livne_93,dlw10a, dlw10b}.
    In practice, we evolve a massive star from the main sequence until the BSG phase,
    the RSG phase, or until core-collapse. We then trigger an explosion/eruption as desired, by depositing a
    prescribed energy at a prescribed radius/mass cut and on a very small time scale. We then let the
    resulting ejecta evolve until it reaches the desired radius for our interaction setup. The approach is still
    artificial in the sense that we do not model the explosion mechanism, but given that trigger, the model
    is energetically and dynamically consistent.

\begin{figure*}
\epsfig{file=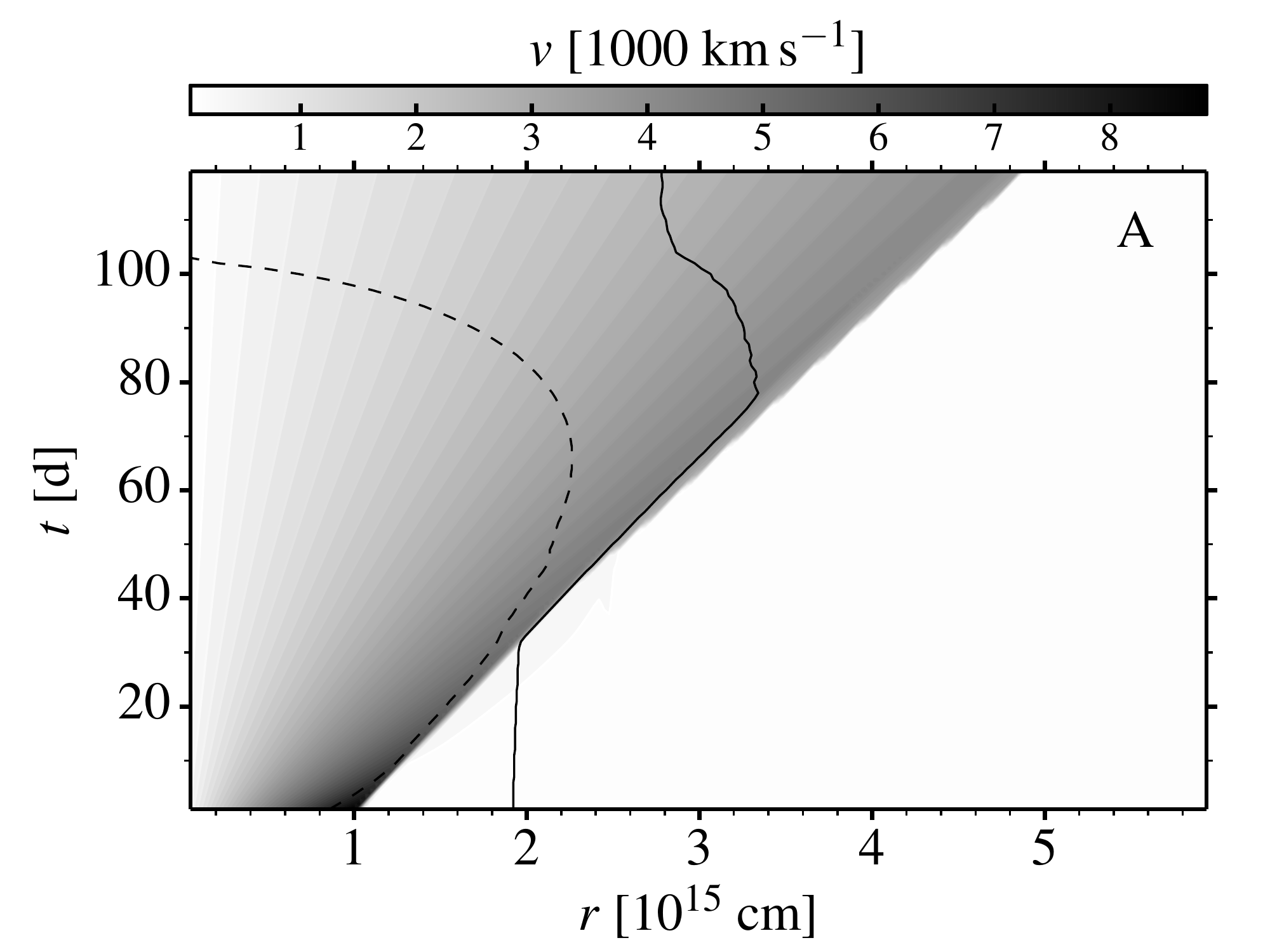,width=0.45\textwidth}
\epsfig{file=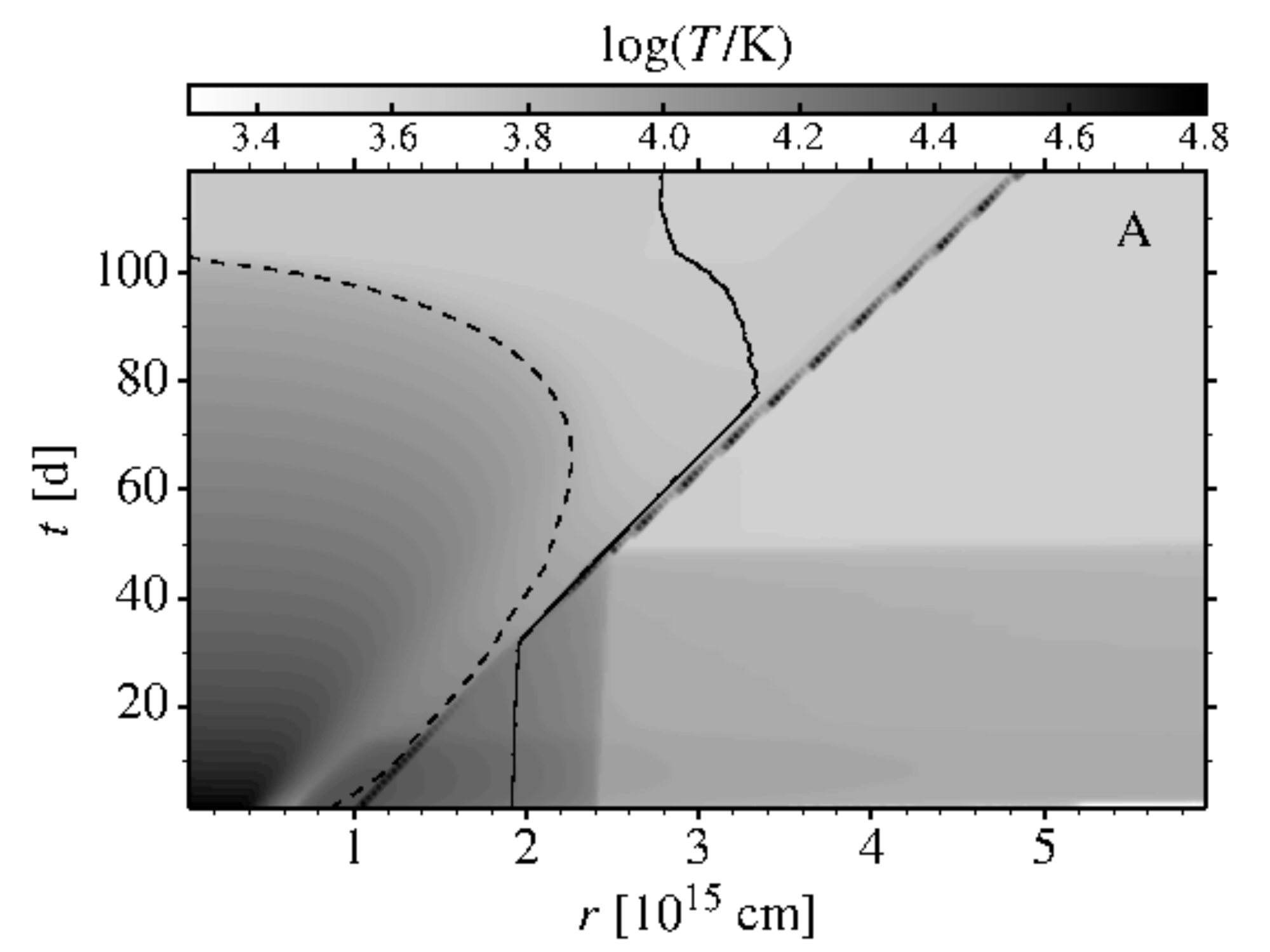,width=0.45\textwidth}
\epsfig{file=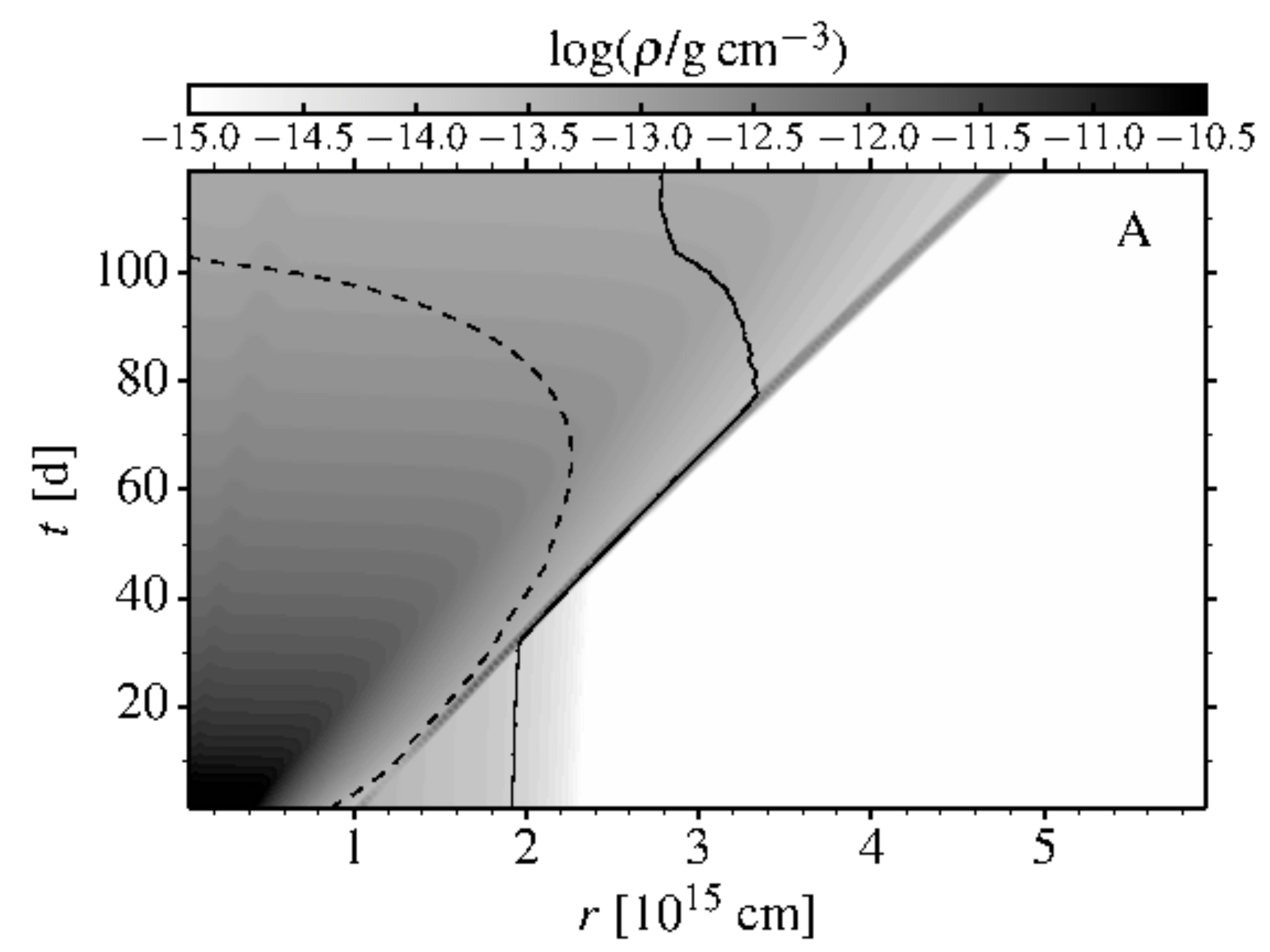,width=0.45\textwidth}
\epsfig{file=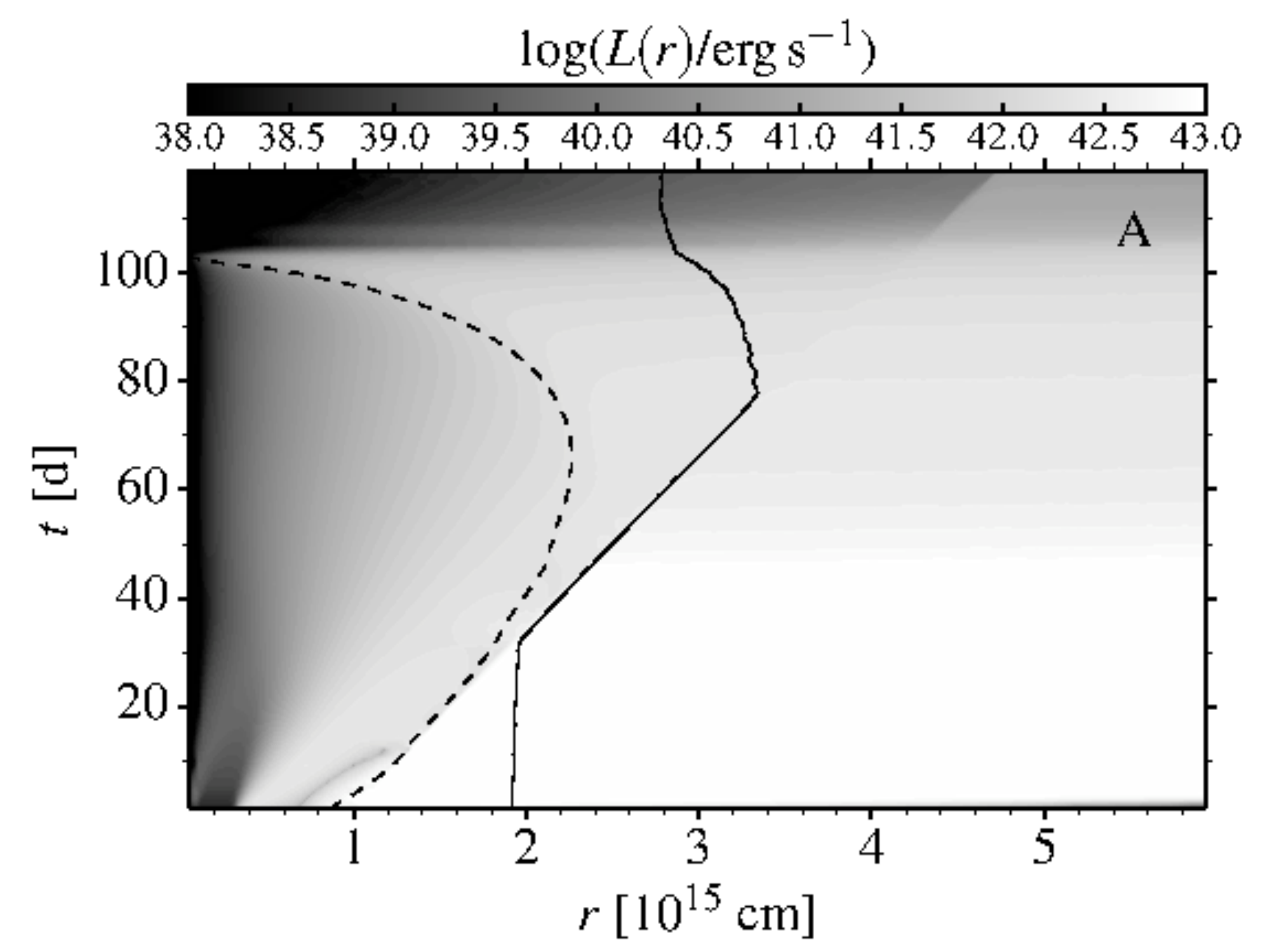,width=0.45\textwidth}
\caption{Greyscale image showing the velocity, temperature, density, and local luminosity versus radius and time,
as computed by \heracles\ for model A. The solid line traces the photosphere
(we refer here to the location where the electron-scattering optical depth integrated inwards from $r_{\rm max}$ is equal
to 2/3) and the dashed line the location where the optical depth is 10 (for both quantities, we use the opacity
from electron scattering only).
\label{fig_map_moda}
}
\end{figure*}

    In the present simulations, we ignore $^{56}$Ni and any contribution from radioactive decay, so that radiation originates
    from the release of internal (i.e., primarily radiation) energy originally present in the inner/outer shell
    and from direct (or re-processed) emission from the shock.

    To post-process the \heracles\ simulations, we perform steady-state nLTE radiative-transfer simulations
    with \cmfgen, with allowance for arbitrary velocity fields.  At multiple epochs, we remap
    the radius, velocity, temperature and density computed by \heracles\ for a wide range of interaction configurations
    and solve for the radiative transfer by holding the temperature fixed.
    The rate equations are solved using the Sobolev approximation. More quantitative calculations will require
    the inclusion of line blanketing, and the adoption a more realistic nLTE cooling function in the hydrodynamic
    simulations.
        The assumption of steady state in \cmfgen\ implies that we ignore the light-travel time to the outer boundary
        and any explicit delay associated with optical-depth effects. However, because \cmfgen\ uses the temperature
        structure from \heracles, time delays associated with the diffusion of radiation through the optically thick
        ejecta/CSM are taken into account.
        Because \heracles\ explicitly allows for time dependence, there is an inherent time offset, of the order of a few days
        around bolometric maximum, between the escaping radiation at a given time in \heracles\ and its computation
        with \cmfgen\ based on the \heracles\ snapshot at that time.

    In contrast to \citet{d15_10jl}, we employ
    a more complex model atom, including H\one, He\one, He\two,  C\one--C\three, N\one--\three,
    O\one, O\two, Na\one, Si\two, Ca\two, Sc\two, Ti\two, and Fe\two--\five. We use a modest number of levels for
    each atom/ion (comparable to the simulations reported in \citealt{DH11}) because our focus is on the
    spectral evolution --- we search for signatures of the interaction dynamics
    on the H$\alpha$ profile morphology so that the exact magnitude of line blanketing, for example,
    does not need to be known accurately for our purposes.

\section{Observational data}
\label{sect_obs}

We compare our models to SN\,1994W, SN\,1998S, and SN\,2001ht.

For SN\,1994W, the photometry is from \citet{sollerman_etal_98}, combined with spectra from
\citet{chugai_etal_04}. We use a distance of 25.4\,Mpc, a reddening $E(B-V)=0.15$\,mag, and adopt the same
time reference (i.e., 14th of July 1994) as in \citet{sollerman_etal_98}.

For SN\,2011ht, the photometry is from \citet{roming_11ht_12}.
We use a distance of 19.2\,Mpc, neglect reddening, and use JD\,2455833.0
for the time origin \citep{roming_11ht_12}.

For SN\,1998S, the photometry is from \citet{fassia_98S_00}, and the spectra are from \citet{fassia_98S_01} and
\citet{leonard_98S_00}.
We use a distance of 17.0\,Mpc, a reddening $E(B-V)=0.22$\,mag, and adopt the same
time reference (i.e., JD\,2450875.2) as in \citet{fassia_98S_00}.

\section{Model A:  a massive energetic ejecta ramming into a 0.4\,\msun\ extended CSM}
\label{sect_moda}

In this first section, we discuss our model A, which corresponds to an interaction configuration similar to that proposed
for SN\,1994W by \citet{chugai_etal_04}.

\begin{figure}
\epsfig{file=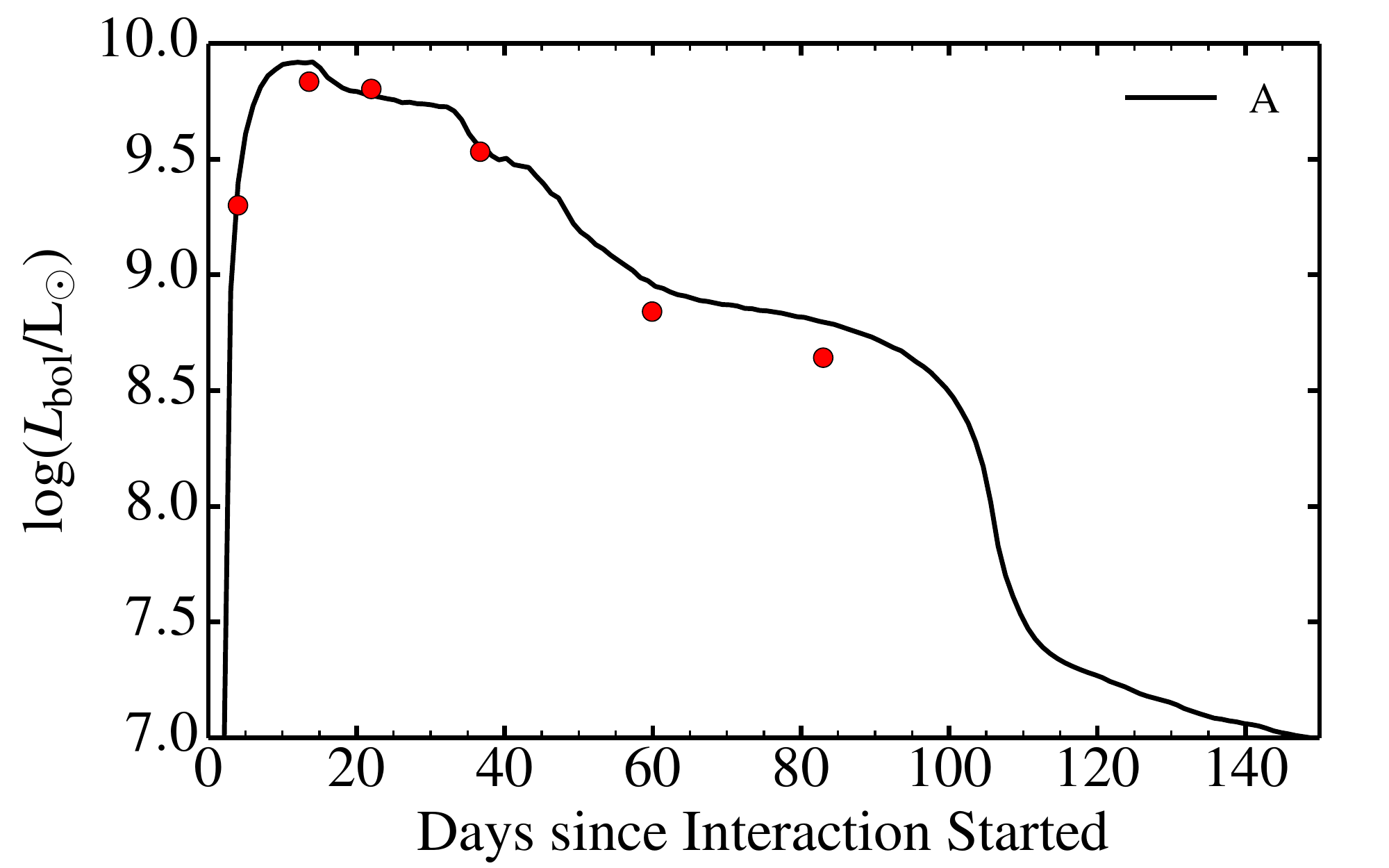,width=0.47\textwidth}
\caption{Bolometric light curve for model A. The dots
correspond to the luminosity computed at a few epochs with \cmfgen.
\label{fig_lbol_moda}
}
\end{figure}

\subsection{Interaction configuration}

The inner shell is massive ($\sim$\,10\,\msun) and energetic (10$^{51}$\,erg kinetic energy), quite typical of a standard
Type II SN ejecta.
The outer shell is assumed to have formed through a wind, with a mass loss rate of 0.1\,\msunyr\ which lasted 3.5\,yr
and abated $\sim$\,3\,yr before the inner shell exploded. With a speed of 100\,\kms, it stretches
from 1.0 to 2.1$\times$10$^{15}$\,cm. Beyond that (in the outer CSM region produced before the phase of intense mass loss),
the density drops by four orders of magnitude.
At the onset of interaction, the inner shell age is $\sim$\,12.0\,d.
A figure showing the initial configuration of this interaction in given in the appendix (Fig.~\ref{appendix_fig_moda_init}).

\subsection{Dynamical properties}

\begin{figure*}
\epsfig{file=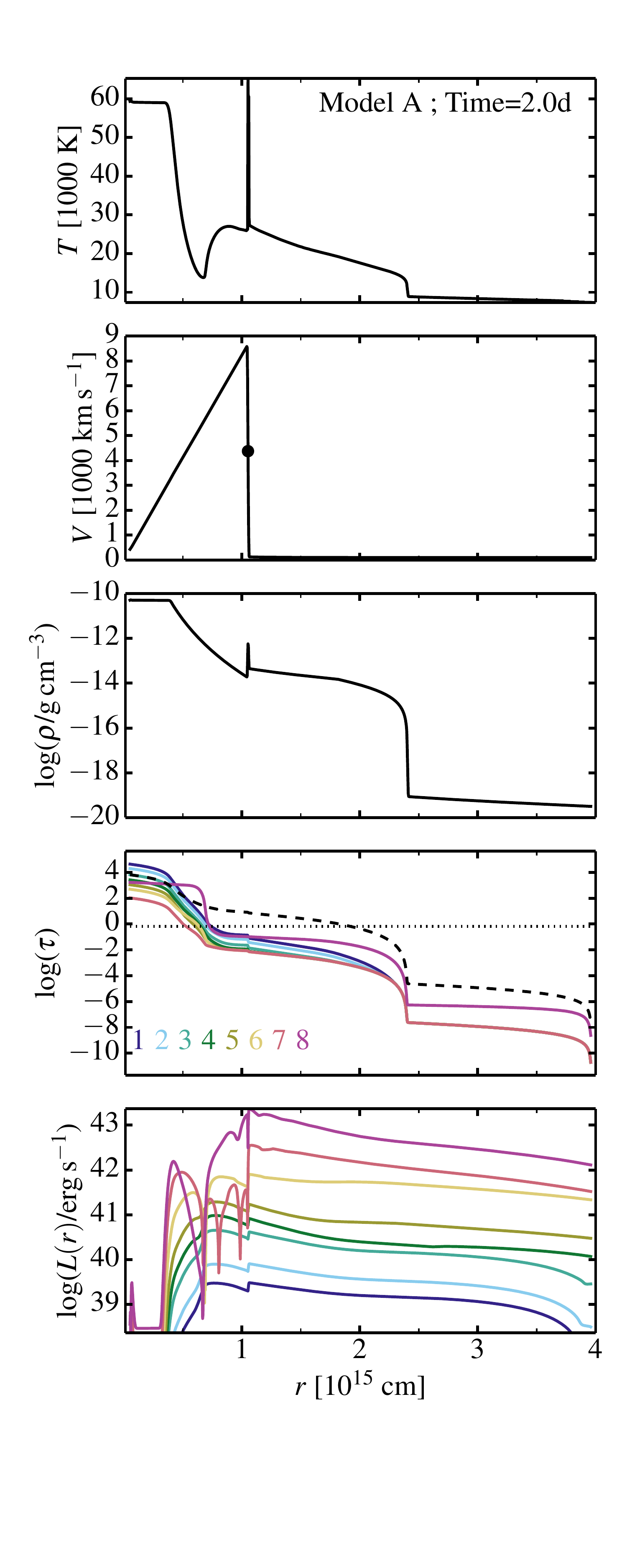,width=0.47\textwidth}
\epsfig{file=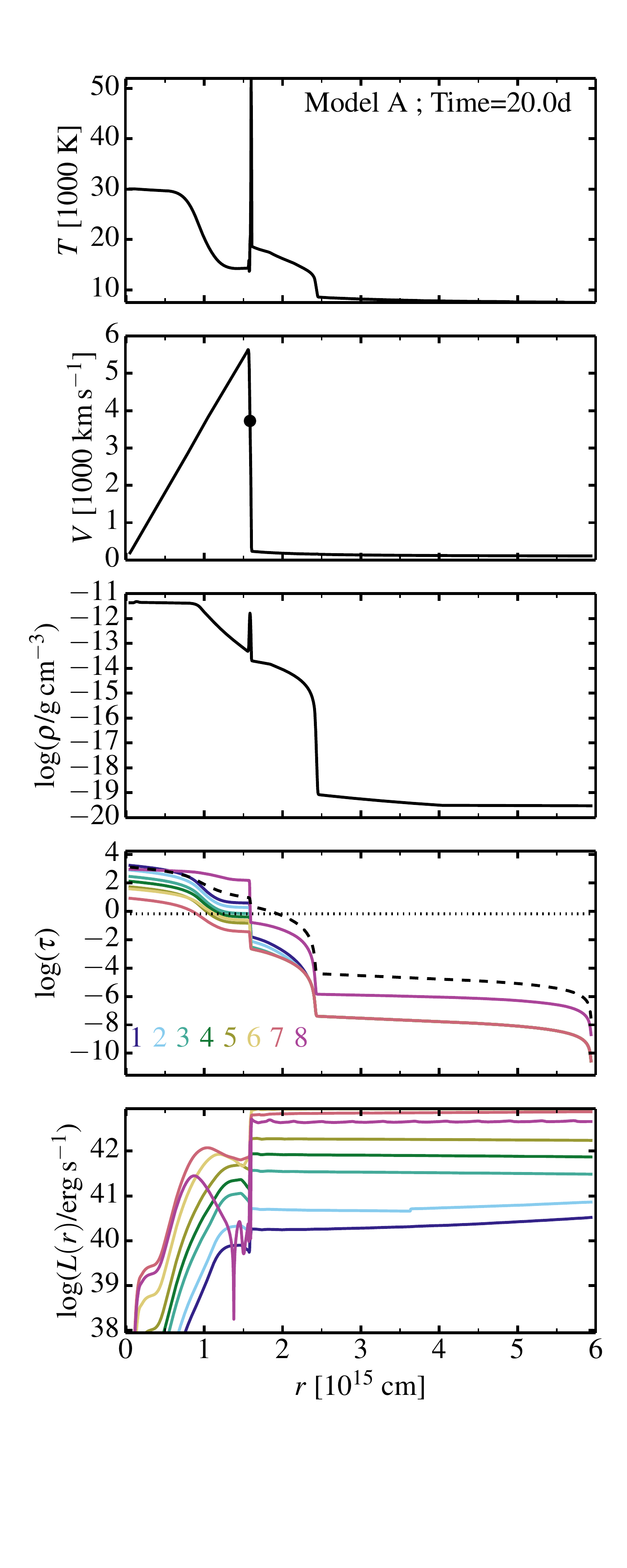,width=0.47\textwidth}
\vspace{-2cm}
\caption{
Left:
Properties of the interaction Model A computed with \heracles\ at 2.0\,d after the start of the interaction.
We show the gas temperature, the velocity (the dot corresponds to the velocity of the CDS),
the mass density, the optical depth, and the radiative luminosity versus radius.
For the optical-depth panel, we show that quantity for each energy group (coloured line;
the group energy increases with the group number; see Section~\ref{sect_setup} for details)
and for electron scattering (dashed line) --- the dotted line corresponds to an optical depth of 2/3.
At this time, the CSM is optically thick. The photosphere is located at $\sim$\,2.0$\times$10$^{15}$\,cm.
The bolometric luminosity is rising to maximum as radiation diffuses through the CSM.
Right: Same as left, but now at 20.0\,d after the onset of the interaction.
The CSM is still optically thick but the CDS is closer to the photosphere.
\label{fig_moda_1}
}
\end{figure*}

\begin{figure*}
\epsfig{file=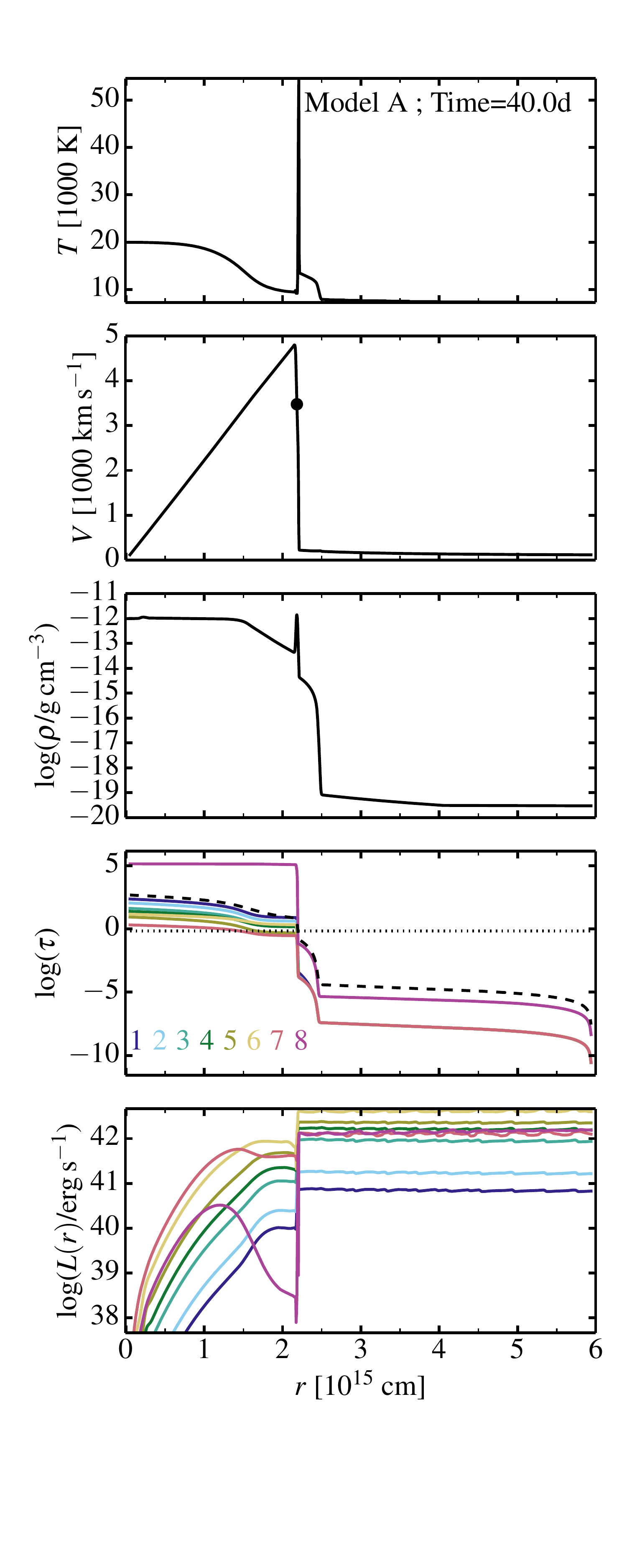,width=0.47\textwidth}
\epsfig{file=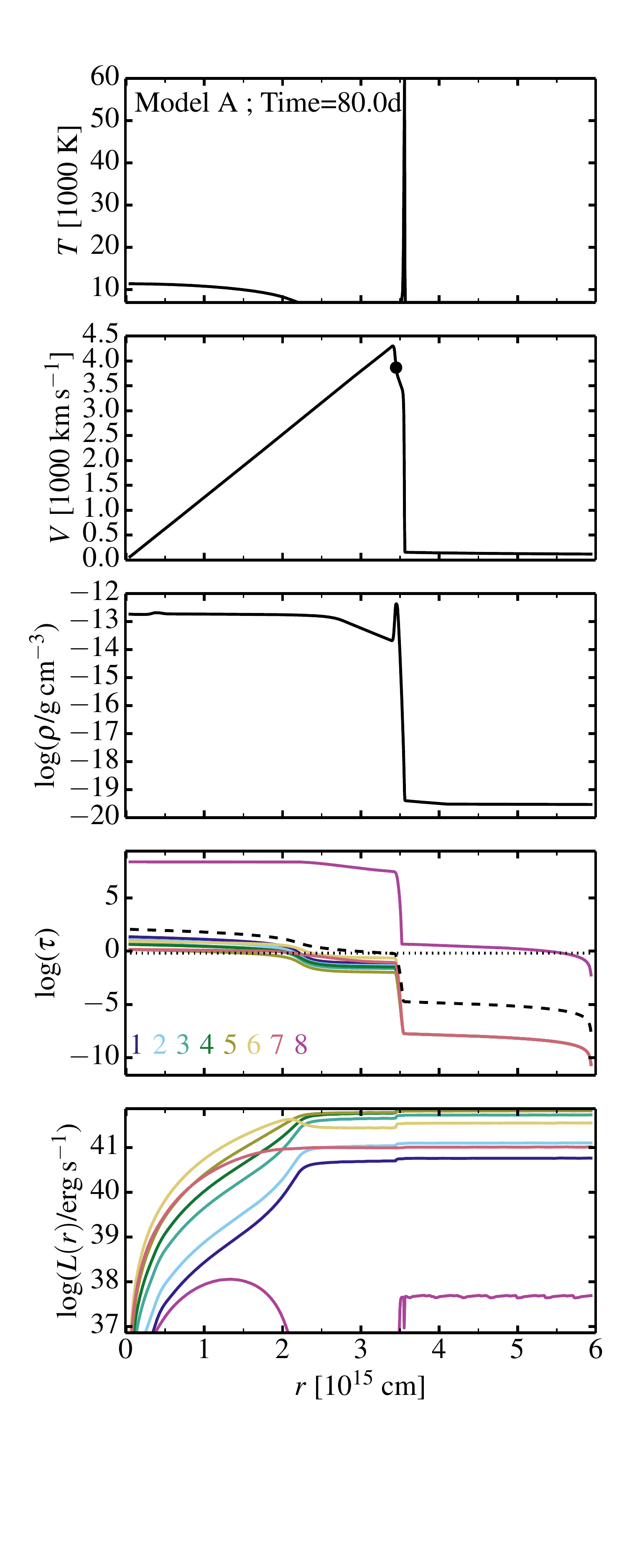,width=0.47\textwidth}
\vspace{-2cm}
\caption{Left: Same as for Fig.~\ref{fig_moda_1}, but now at 40.0\,d after the onset of the interaction.
The photosphere is now within the CDS --- the CSM above the CDS is optically-thin.
Right: Same as left, but now at 80.0\,d after the onset of the interaction.
The CDS is becoming optically thin and the bulk of the radiation stems from the optically-thick
ejecta located at smaller radii.
\label{fig_moda_2}
}
\end{figure*}

The evolution of the properties of the radiation-hydrodynamics simulations of model A with \heracles\ is shown in three
sets of illustrations.
Figure~\ref{fig_map_moda} shows greyscale images of the velocity, temperature, density, and radiative luminosity
versus radius and time.
Figure~\ref{fig_lbol_moda} shows the \heracles\ bolometric light curve for model A
(the dots correspond to the luminosity computed at various epochs with \cmfgen\ --- see Section~\ref{sect_moda_cmfgen}).
Figures~\ref{fig_moda_1}--\ref{fig_moda_2} show results from the \heracles\ simulation at selected epochs.
Each figure contains five panels illustrating the gas temperature, the velocity, the mass density, the optical depth,
and the radiative luminosity versus radius.

The early evolution is quite typical of interacting SNe
\citep{chugai_etal_04,whalen_2n_13,moriya_etal_13b,d15_10jl}. The initial shock luminosity gives rise to a burst of radiation
at the ejecta/CSM interface, which propagates through the cold and dense CSM on a free-flight time. Radiation
raises the ionisation of the CSM, enhancing its electron-scattering optical depth. By 2.0\,d after the onset of the
interaction, the CSM is optically thick out to $\lesssim$\,2.0$\times$10$^{15}$\,cm (this is
where the electron-scattering optical depth integrated inwards from $r_{\rm max}$ is equal to 2/3 ---
we define this location as the photosphere), and the total CSM electron-scattering optical depth is of the order of 10
(these different regions are clearly identified in Fig.~\ref{fig_map_moda}).
Subsequently, radiation injected at the shock has to diffuse through the CSM.
The emergent bolometric luminosity thus rises to maximum after about a diffusion time of 10\,d, and then drops.
A dense shell has formed and contains about 5\% of the swept-up CSM mass  (left column of Fig.~\ref{fig_moda_1}).

At 20.0\,d after the onset of interaction (right column of Fig.~\ref{fig_moda_1}), the ejecta/CSM interface has
reached $\sim$\,1.6$\times$10$^{15}$\,cm.
This shocked region is bounded by a reverse shock and a forward shock and is confined within a CDS of about 15000\,K,
significantly cooler than the temperature spike just exterior to the CDS (at 50000\,K).
This CDS is moving at $\sim$\,3500\,\kms.
In contrast, the photosphere has essentially not moved because it resides in the slowly moving CSM, but the CDS is now much
closer to that photosphere. Indeed, the CSM electron-scattering optical depth (between the photosphere and the CDS) is only of a few.
All the energy is emitted from the CDS region so that the luminosity is constant beyond it, in particular between the CDS and the photosphere.

At 40.0\,d after the onset of the interaction (left column of Fig.~\ref{fig_moda_2}), the CDS has reached
$\sim$\,2.1$\times$10$^{15}$\,cm and
has overtaken the former location of the photosphere within the CSM. The photosphere is at that time
within the CDS. Its temperature is about 10000\,K, and it is characterised by a very steep density profile. For an external
observer, the SN would look like a radiating sphere with no extension, i.e., with a very sharp edge.
Up until that time, the luminosity of all energy groups shows a pronounced jump across the CDS, implying that
the bulk of the radiation emerges from the CDS --- the emitting region is moving outwards with a typical
velocity of 3500\,\kms.

At 80.0\,d after the onset of the interaction (right column of Fig.~\ref{fig_moda_2}), the CDS has reached $\sim$\,3.5$\times$10$^{15}$\,cm and
is progressing unimpeded through the low density outer CSM. Its speed has remained essentially constant
throughout and is of the order of 3500\,\kms. The CDS is now cooler, with a temperature of $\sim$\,6000\,K,
and its electron-scattering optical depth is $\lesssim$\,1. The CDS is optically thick to Lyman continuum photons but is transparent
to Balmer continuum photons. From 50--60\,d onwards, the luminosity shows only a modest jump across
the CDS, implying that the bulk of the radiation now comes from the inner ejecta, which has not taken part
(and will not take part) in the interaction. These emitting layers are moving at a few 1000\,\kms.

The bolometric light curve therefore shows two distinct phases (Fig.~\ref{fig_lbol_moda}).
The first phase, up until about 40-50\,d, is dominated
by the interaction as the shock is making its way through the dense CSM. During this phase, the CDS is optically
thick. After that, the bolometric luminosity comes from the CDS as well as the inner ejecta, which is no longer obscured.
This second phase corresponds to the plateau light curve that this model would have had in the absence of the
interaction\footnote{As noted earlier we have ignored the energy contribution arising from
radioactive decay, and this may affect the light curve.} ---
the contribution from the CDS and the interaction during that second phase is modest
(it depends on the wind mass loss rate for the outer CSM, which is chosen here to be small).

\subsection{Spectral evolution}
\label{sect_moda_cmfgen}

The second part of our study is to post-process these \heracles\ simulations with \cmfgen, following the
procedure discussed in \citet{d15_10jl} and outlined in Section~\ref{sect_setup}.
We choose representative epochs. We compute \cmfgen\ models at 2.0\,d (CDS hardly formed, CSM optically thick,
source of radiation deeply embedded within this CSM), at 11.6\,d (around maximum light; CSM still optically thick),
at 20.0 and 34.7\,d (as the CDS gets close to the photosphere located at 2.1$\times$\,10$^{15}$\,cm),
and at 57.9 and 81.0\,d (when the bulk of  the radiation comes from ``unshocked" ejecta, below the CDS; CSM optically thin).
Great care is taken when remapping the \heracles\ model onto the \cmfgen\ grid to resolve the strong variations in density,
temperature, and velocity of the dynamical model.

\begin{figure*}
\epsfig{file=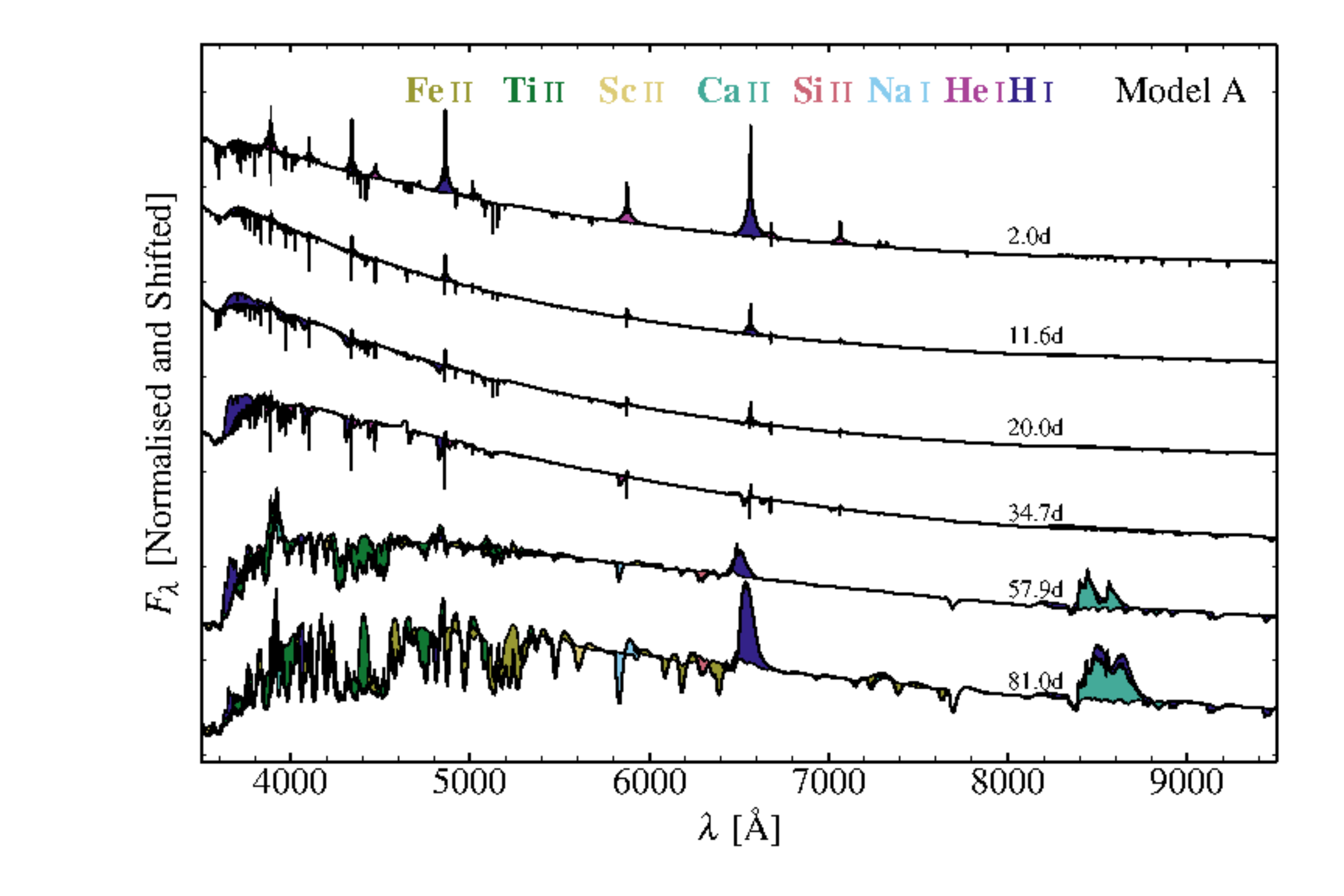,width=\textwidth,clip}
\hbox{}\vspace{-1.0cm}
\epsfig{file=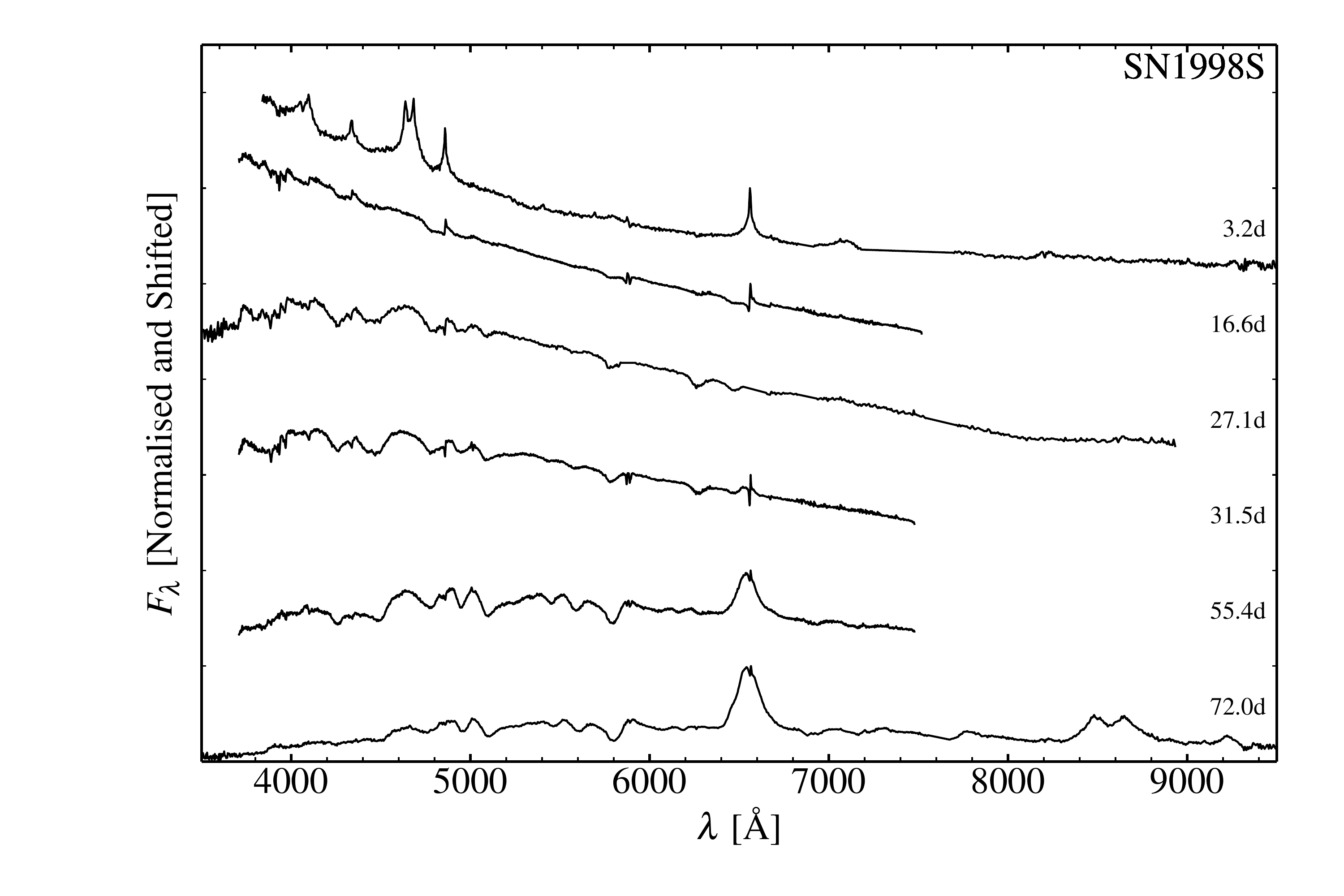,width=\textwidth,clip}
\vspace{-1.0cm}
\caption{Top: Multi-epoch spectra for model A during the high-brightness phase (time is since the onset of interaction).
Bottom: Multi-epoch spectra of SN\,1998S (corrected for redshift but not for extinction)
covering the early phase with narrow lines and broad wings (3.2\,d),
pure absorption spectrum (16.6, 27.1, and 31.5\,d), absorption spectrum with strong and broad emission lines (55.4 and 72.0\,d).
The axis range for the top and bottom panels is the same so one can compare the models and observations line by line.
\label{fig_spec_moda_98s}
}
\end{figure*}

The resulting spectral evolution is shown in the top panel of Fig~\ref{fig_spec_moda_98s}.
The first two epochs show the typical IIn spectral morphology. Line profiles are centred at the rest
wavelength. They show a strong and narrow central peak with extended wings. The line broadening mechanism
is dominated by electron scattering \citep{chugai_98S_01, dessart_etal_09}. This is well understood since
the bulk of the radiation is injected deep within the CSM. The slow expansion of the CSM in this model (100\,\kms)
is much smaller than the electron thermal speed of  $\sim$\,500\,\kms, and its optical depth of about ten makes
electron scattering an important process for frequency redistribution.
The spectra are quite blue at those epochs, with lines of He\one\ and H\one\ primarily.
There is also a large number of very narrow lines, primarily from Fe\two, which are formed in the cool part
of the unshocked CSM. This is a clear signature that the spectrum forms in two distinct regions, one
hot and partially ionised (associated with H\one\ and He\one\ lines in particular), and the other
cool and recombined (associated with very narrow lines of H\one\ or Fe\two).

However, the spectra do not retain this IIn morphology long.
At 20.0 and 34.7\,d, the IIn signature is gone and we observe
instead line profiles with blue-shifted absorption and weak emission. The bulk
of the radiation is now coming from the outer edge of the fast moving optically-thick CDS,
which has a steep density profile. The emission part of the line is therefore dwarfed
and the absorption is blue-shifted. The CDS temperature is still of the order of 10000\,K,
so He\one\ lines continue to be seen (primarily as blue-shifted absorption).
As for previous epochs, narrow H\one\ and Fe\two\ lines are predicted.

At the last two epochs shown in Fig.~\ref{fig_spec_moda_98s}, the morphology has drastically changed. The spectra
are now analogous to SNe II-P during the late plateau phase, but H$\alpha$ and the Ca\two\ triplet at  8500\,\AA\
show strong and broad emission with weak or no associated absorption.\footnote{At late times, a treatment
of the radiative transfer with time-dependent terms would probably yield a stronger H$\alpha$ line,
both in absorption and emission \citep{UC05,DH08}.}
We are now seeing a hybrid source of emission, with a dominant contribution from the inner ejecta and
a more modest contribution from the more optically-thin CDS. The dense and fast CDS contributes most notably
through strong and broad H$\alpha$ and Ca\two\ emission, with little associated absorption.
It is emission from the CDS that fills in the inner ejecta absorption associated with H$\alpha$ and Ca\two\
(Figs.~\ref{fig_dfr_w0s_57p9} \& \ref{fig_dfr_w0s_81p0}).

To facilitate the line identification we have calculated synthetic spectra  with  selected species excluded
from the calculation, and then coloured the offset between the resulting flux and the total flux (top panel
of Fig.~\ref{fig_spec_moda_98s}). This exercise reveals the dominance of H\one\ and He\one\ at early
times (up until epoch 34.7\,d) when the spectrum is relatively blue (e.g., there is a large flux short-ward of the Balmer edge).
As the color temperature drops and the
material recombines in the spectrum formation region, lines of Ti\two, Sc\two, and Fe\two\ appear
and strengthen. The drastic evolution in line profile shapes, especially for H$\alpha$, is evident.
We also note the presence of Si\two\,6347-6371\,\AA. This doublet is present at 34.7\,d as two
weak absorptions, displaced from rest wavelength by $\sim$\,2500\,\kms. At the next two epochs,
this Si\two\ doublet appears as one broad and strong absorption, with its maximum still around $-$2500\,\kms\
from rest wavelength.

\begin{figure*}
\epsfig{file=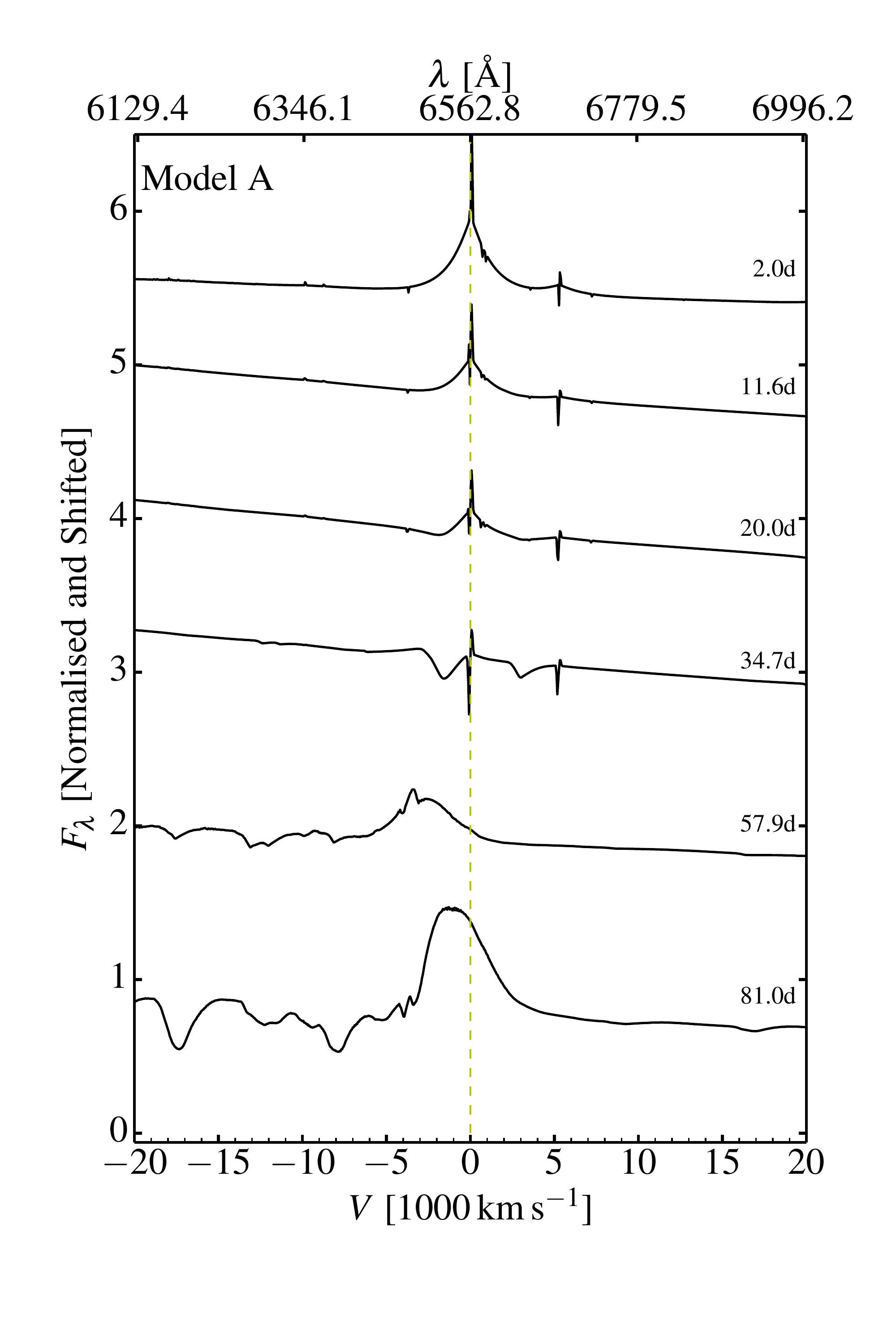,width=0.45\textwidth}
\epsfig{file=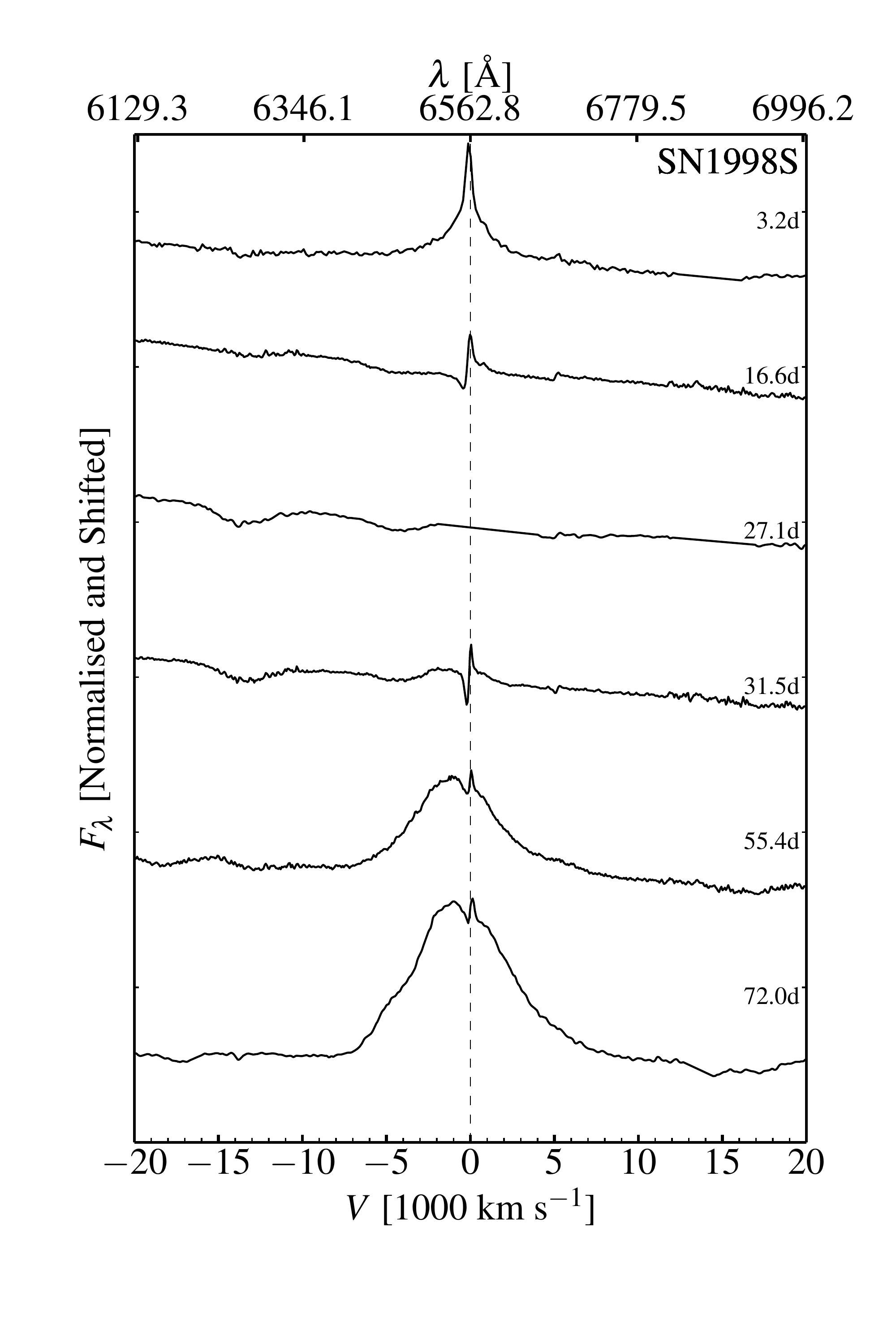,width=0.45\textwidth}
\vspace{-1.cm}
\caption{On the left we illustrate multi-epoch H$\alpha$ spectra for model A during the high-brightness
phase where the time is measured relative to the onset of interaction. On the right
we show data for SN1998S. Notice the drastic evolution in the profiles of H$\alpha$, and
the striking qualitative similarity in the temporal evolution for the model and SN1998S.
\label{fig_spec_moda_98s_ha}
}
\end{figure*}

The left panel of Fig.~\ref{fig_spec_moda_98s_ha} shows the H$\alpha$ region at multiple epochs spanning
the high-brightness phase of model A (same epochs as for Fig.~\ref{fig_spec_moda_98s}), in both
velocity (bottom $x$-axis) and wavelength (top $x$-axis) space.
It illustrates the narrow-core broad-wing symmetric line-profile shape at early times (2.0 and 11.6\,d);
the transition to a P-Cygni profile with weak emission and little sign of broadening from electron scattering (20.0 and 34.7\,d);
the appearance of a blue-shifted and strong emission line, dominated by Doppler broadening (57.9 and 81.0\,d).
The emission blueshift decreases as the CDS optical depth drops \citep{d15_10jl}.
The first four spectra show the clear presence of narrow components in H$\alpha$ and He\one\,6678\,\AA,
which arise from the cool unshocked slow ($\sim$\,100\,\kms) CSM --- these features are absent at later times because
the dense part of the CSM has been completely swept up into the fast-moving CDS (the only slow CSM left has
a very low density in our setup --- see Fig.~\ref{appendix_fig_moda_init}).

In the appendix, we provide additional figures describing the origin of the flux versus wavelength and radius,
and at multiple epochs (Figs.~\ref{fig_dfr_w0s_11p6}--\ref{fig_dfr_w0s_81p0}).
Viewed this way in two-dimensional space, one gets a better sense of the complicated
spectrum formation process, with the distinct contribution each region makes to the emergent light.

\subsection{Comparison to observations}

In Fig.~\ref{fig_lbol_moda}, we showed the bolometric
luminosity computed by \cmfgen\ at multiple epochs (dots).
There is a good agreement between the results from \heracles\ and \cmfgen\ despite the different numerics and physics of each code.
We can therefore use the \cmfgen\ models to compute the $V$-band magnitude and compare to observations.

\begin{figure}
\epsfig{file=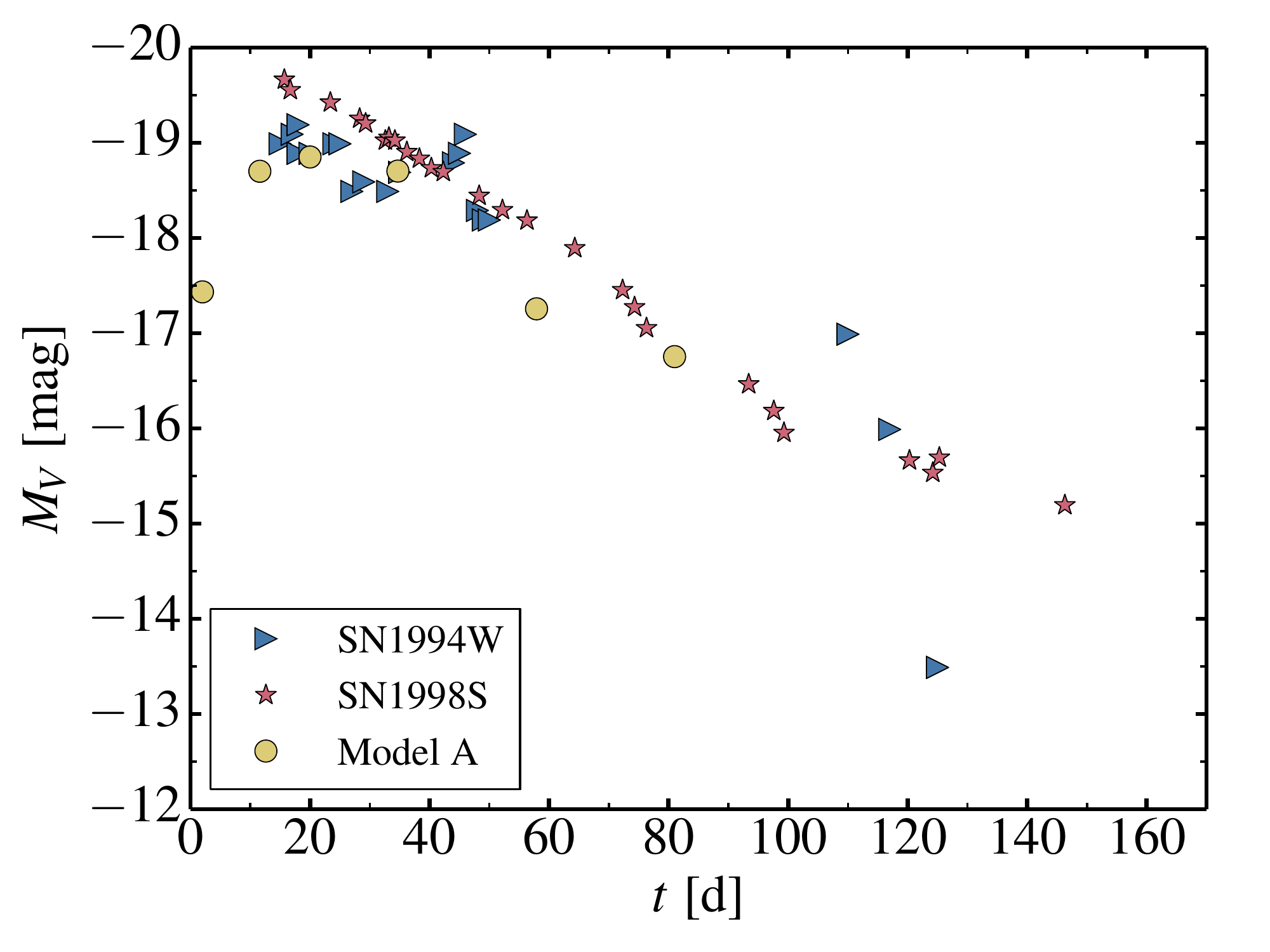,width=0.45\textwidth}
\caption{
Absolute $V$-band light curve for SN\,1994W and SN\,1998S compared to our model A.
See Section~\ref{sect_obs} for details on the source of observations, distances, reddening, and inferred
time of explosion.
For the model, the time is given with respect to the onset of interaction.
\label{fig_lc_moda_obs}
}
\end{figure}

Figure~\ref{fig_lc_moda_obs} shows the good agreement between model A and the observed $V$-band
light curve of SN\,1994W, as also found by \citet{chugai_etal_04}.
This is expected since the interaction configuration of our model A corresponds to the model of
\citet{chugai_etal_04}, and our predictions for the dynamics and energetics of that model are comparable
to theirs.

But spectroscopically, model A
is not compatible with the multi-epoch spectral observations of SN1994W.
While model A shows narrow lines at early times, when the IIn morphology prevails, it evolves to
a broad line spectrum at later epochs during the high brightness phase. In contrast,
SN\,1994W shows line profiles that are initially narrow and become increasingly more narrow
as time progresses \citep{chugai_etal_04}.

Interestingly, model A reproduces the evolution of SN\,1998S.
Model A matches roughly the $V$-band light curve of SN\,1998S (Fig.~\ref{fig_lc_moda_obs})  --- SN\,1998S and
SN\,1994W have comparable light curves.\footnote{The ambiguous information conveyed by light curves in
this case is striking because SN\,1998S and SN\,1994W have completely different spectral evolution.}
But more  importantly, Model A reproduces the overall spectral evolution of SN\,1998S
(compare the two panels in Fig.~\ref{fig_spec_moda_98s}), as well as the morphological changes
in the line profiles (e.g., for H$\alpha$; Fig.~\ref{fig_spec_moda_98s_ha}).

In our model A the absorptions blue-ward of H$\alpha$ are due to Fe\two\ and Si\two\
(see above). In the observations of SN\,1998S, a sizeable dip in the spectrum coincides with the location of
Si\two\ in our model (see also \citealt{leonard_98S_00,fassia_98S_01}) --- the feature is
not caused by high-velocity H$\alpha$ absorption.\footnote{The fundamental impact of interaction
is the deceleration of fast material. Any fast material from the SN ejecta
piles up in the CDS (and the fastest material is the first to disappear). The fastest material emitting/absorbing H$\alpha$
photons is tied to the CDS and it thus carries a maximum Doppler shift corresponding to the CDS velocity.}

Quantitatively, model A shows some offsets with the observations of SN\,1998S. The model light curve is
too faint initially, but this could be remedied by adjusting the location of the CSM (closer
to the progenitor star, as in \citealt{shivvers_98S_15}) as well as accounting for the ejecta radiation released since shock
break out (we only start accounting for radiative losses at the onset of the interaction).
The model light curve is also too faint at late times, but this could be remedied by adopting a higher mass loss rate
for the distant CSM (or invoking some power from radioactive decay).
So, both offsets could be cured by simple adjustments to the adopted interaction configuration.

Spectroscopically, there are also some discrepancies. The earliest spectrum of SN\,1998S shows
lines of more ionised species than predicted in our models.
This should be cured by moving the CSM closer to the exploding star, as in \citet{shivvers_98S_15}.
Indeed, given a radiation energy $\delta E_{\rm rad}$ stored in the optically-thick CSM volume $V$
the gas temperature (which equates roughly the radiation temperature) will scale as
$(\delta E_{\rm rad} / aV)^{1/4}$. So, the closer the interaction site is to the progenitor surface, the higher
the CSM temperature and ionisation, all else being the same. In the future, we will design a more suitable model
to match the observed properties of SN\,1998S, by adjusting the velocity and density structure of the
adopted CSM, as well as the properties of the underlying ejecta (in the absence of interaction, SN\,1998S
would have been a II-L, not a II-P).

These discrepancies are merely small quantitative offsets
and we conclude that the model of \citet{chugai_etal_04}, or our model A, is more suitable to explain events like SN\,1998S.
The fundamental conflict between this model and the observations of SN1994W is that it systematically
leads to the production of broad lines at late times, something that we already emphasised, without proof,
in \citet{dessart_etal_09}.

In their modelling of H$\alpha$ for SN\,1994W, \citet{chugai_etal_04} adopted free expansion for the CSM with a value
of 1100\,\kms\ at 5.4$\times$10$^{15}$\,cm (any emission occurring at smaller radii arises from
slower regions). They do not seem to consider emission from the fast moving CDS at $\sim$\,3500\,\kms\ (all emission
seems to arise from within the slow moving CSM) and the strongly non-monotonic velocity structure of this configuration.
This likely explains the origin of the relatively narrow H$\alpha$ profiles they produce, even at 89\,d
when the CSM optical depth is merely 0.54.
At such a low optical depth, the CSM material cannot efficiently absorb and re-emit the photons arising from the CDS.
In our simulations, we include the whole system (i.e., unshocked ejecta,
shocked ejecta, shocked CSM, unshocked CSM), and find that the contribution from the ejecta and/or CDS eventually dominates
in this configuration when the CSM optical depth drops below a few.
The associated signature is the production of broad lines at late times.
Such broad lines are seen at late times in SN\,1998S. For the same physical reason, they are seen (and explained)
in SN\,2010jl \citep{d15_10jl}.

\begin{figure*}
\epsfig{file=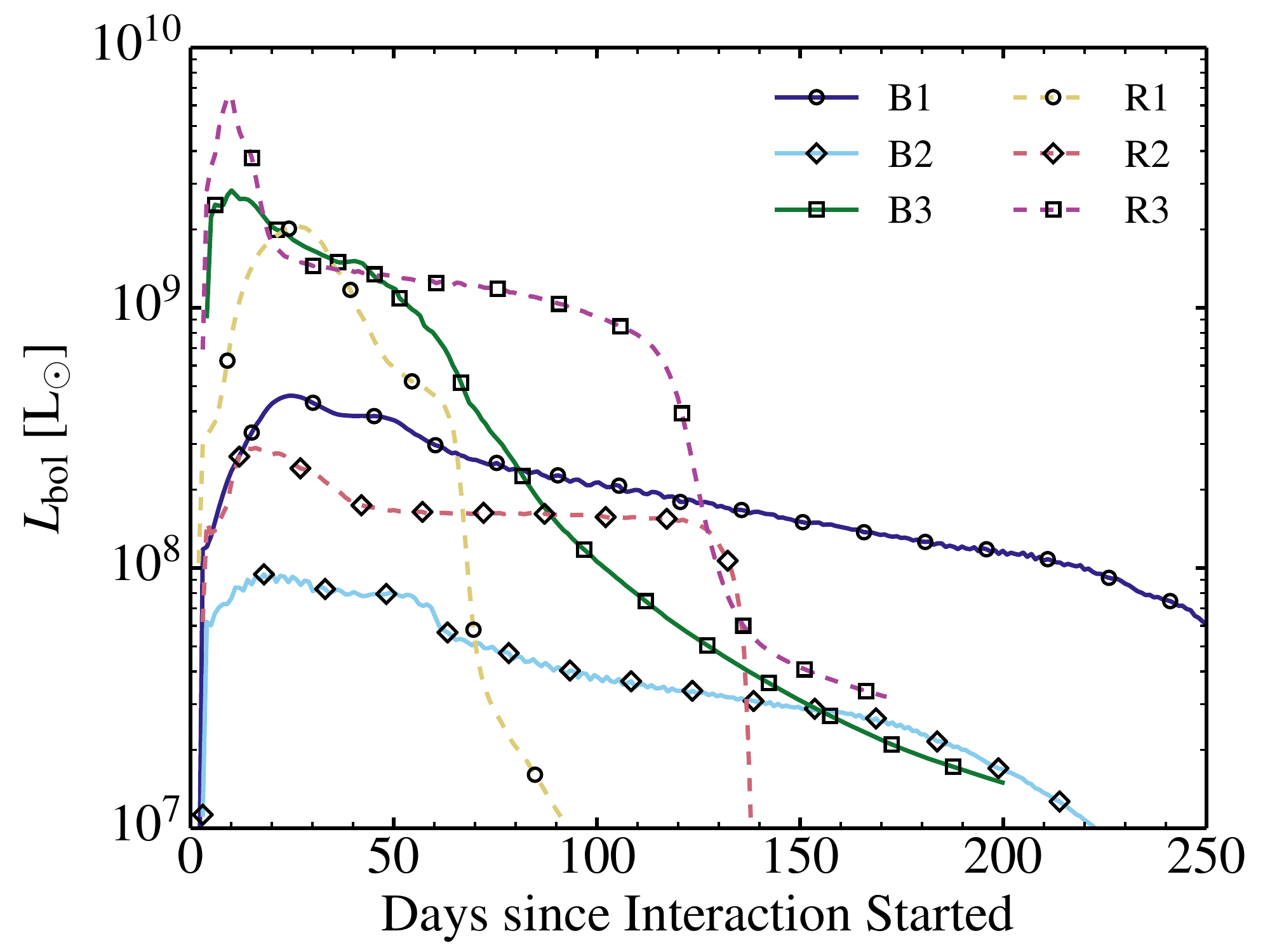, width=0.45\textwidth}
\epsfig{file=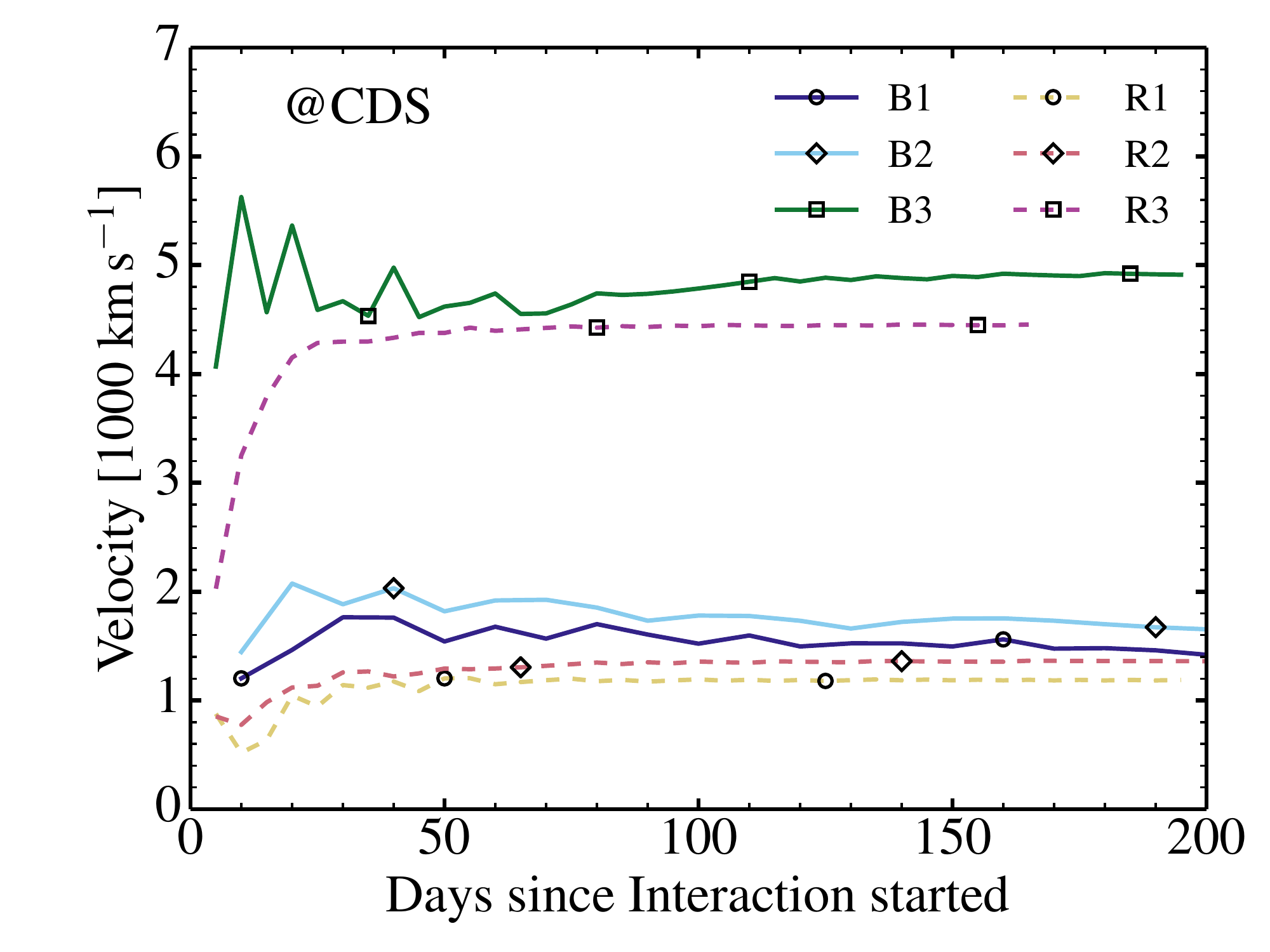, width=0.45\textwidth}
\caption{Left: Bolometric light curves for model B1, B2, B3, R1, R2, and R3 computed with \heracles.
Right: Evolution of the CDS velocity for the same set of simulations (see Section~\ref{sect_grid} for discussion).
\label{fig_lbol_bsg_rsg}
}
\end{figure*}

\section{Grid of models with different inner/outer shell mass and energy}
\label{sect_grid}

To determine the key inner/outer shell properties that influence the light curves
and spectra of interacting SNe, we perform a grid of simulations.
One important goal of these simulations is to identify configurations that can produce bright
displays even for low CDS velocities, since we anticipate this type of configuration
is warranted for SNe like 1994W \citep{dessart_etal_09}.

\subsection{Interaction configurations}

For the inner shell, we consider explosions taking place in BSG and RSG stars.
For the outer shell, we consider a dense CSM produced by wind mass loss.
Because expansion cooling has a stronger effect in more compact stars,
ejecta from BSG star explosions will be cooler than their RSG counterparts
when they reach a distance of $\sim$\,10$^{15}$\,cm (this is a rough estimate
of the interaction radius for SN\,1994W).
So, to start with a suitable ejecta temperature for the inner shell, we construct
BSG and RSG star models with \mesa\ and explode them with \v1d.

We use \mesa\ to evolve a star with a zero-age main-sequence mass of 20\,\msun.
The code is stopped when the surface radius increases to 100\,\rsun\ (BSG)
and again when it reaches 1000\,\rsun\ (RSG). One \mesa\ model is saved for each stage.
We then use \v1d\ to trigger explosions in these progenitors and evolve the resulting ejecta until they reach
a desired radius where we wish them to encounter some CSM. In all cases, the energy is deposited
within the H-rich envelope (1 to 10\,\msun\ below the surface) and leads to no explosive
nucleosynthesis --- the ejected material retains its original (H-rich) composition.

To generate a diversity of interactions we consider various combinations of
mass-loss rates, wind velocity, ejecta mass, and ejecta kinetic energy. For the wind
phase we adopt mass-loss rates of 0.1 and 1.0\,\msunyr\ that last for one year, and which
generate CSM masses of 0.1 and 1\,\msun. For RSG progenitors we use a wind velocity of 60\,\kms,
while for BSG progenitors we use a wind velocity of  600\,\kms. Due to the larger wind
velocity, and the fixed duration of the mass loss, the outer
shell thickness for the BSG case is 10 times that of the RSG shell.
For the inner shell, we simulate explosions with \v1d\ to eject approximately 1 or 10\,\msun\ with
an energy of  about 0.05, 0.1, or 1.0$\times$\,10$^{51}$\,erg. The corresponding models are
named B1, B2, B3 (BSG progenitors) and R1, R2, and R3 (RSG progenitors) --- see Table~\ref{tab_mod}
for a summary of properties.

\subsection{Dynamical properties}

The left panel of Fig.~\ref{fig_lbol_bsg_rsg} shows the bolometric luminosity computed with \heracles\ for
these 6 interaction models. Considerable diversity is exhibited by the light curves.

When the inner shell has a large energy (models R3 and B3), the high-brightness phase is very luminous.
Initially, the power stems from the interaction, but at later times, it can stem from the large stored energy in the inner shell.
For model R3, the explosion is from a RSG star and, as expected, we see a normal SN II-P luminosity of about 10$^9$\,\lsun\
after the initial interaction-powered peak (which lasts about 30\,d).
For model B3, the smaller progenitor star leads to a much smaller luminosity contribution
at late times (in these models of shell explosion and interaction, no \iso{56}Ni is produced so radioactive decay
is not a power source at any time),
so the late-time light curve drops faster and is powered exclusively by the interaction.

\begin{figure*}
\epsfig{file=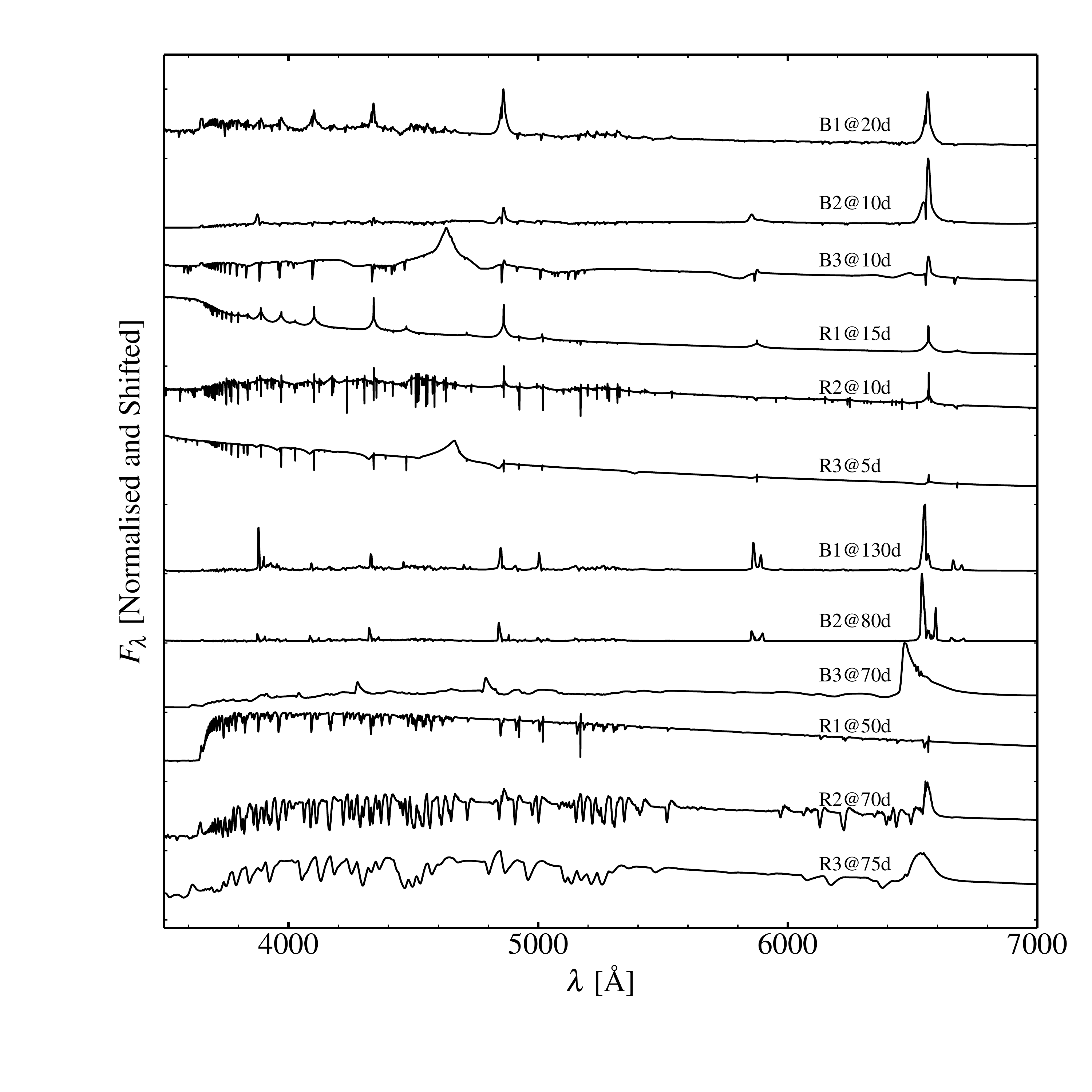, width=\textwidth}
\vspace{-2cm}
\caption{Early and late times spectra for models B1, B2, B3, R1, R2, and R3 computed with \cmfgen\ and based
on \heracles\ simulations.
\label{fig_bsg_rsg_spec}
}
\end{figure*}

Models B2 and R2 have similar ejecta masses to those of R3 and B3, but the ejecta energy is a factor of 10 lower.
As a consequence of the smaller kinetic energy the shock luminosity is much reduced and this leads to
a much lower peak luminosity.
The internal energy (i.e., left over by the original shock that produced the inner shell) is much lower
than in the R3/B3 model counterparts.
This leads to a smaller contribution to the light curve, although because of the lower expansion/evolution
of the whole system, the material stays optically thick longer and can be brighter at very late times (few 100 days)
than the R3/B3 model counterparts. Overall though, the lower energy models R2/B2 radiate a much lower time-integrated
bolometric luminosity than models R3/B3.

In weak explosions, it is nonetheless possible to power a bright light curve.
One way is to increase the CSM mass because this enhances the conversion efficiency from kinetic energy
of the inner shell into radiation.
Another way to boost the luminosity further is by reducing the mass of the inner shell (i.e., increasing the $E/M$
at fixed $E$). The reduced inertia of the inner shell facilitates its deceleration and thus enhances the conversion efficiency.
This is precisely what is achieved with models R1 and B1.

The inner shell of R1 has a kinetic energy a factor of 2 lower than R2, and a factor of 20 lower than R3.
However, model R1 reaches a peak bolometric luminosity of 2$\times$10$^9$\,\lsun, which is a factor
of about 10 larger than R2, and only a factor of 3 lower than R3. This occurs because in R1, 1\,\msun\ of
ejecta is ramming into 1\,\msun\ of CSM, and so the conversion of kinetic energy into radiative energy
is  very efficient. In models R2 and R3, a 9.54\,\msun\ ejecta rams into a CSM mass of only 0.1\,\msun.
Model B1 reaches a much fainter maximum than R1 because the CSM wind speed is 10 times larger (weaker shock) and the wind
density is 10 times smaller (this is because we adopt the same wind mass loss rate as model R1). However,
the interaction through the more extended CSM persists for longer and powers a luminosity $\gtrsim$\,10$^8$\,\lsun\
out to 250\,d (the luminosity from model R1 drops below 10$^7$\,\lsun\ at $\lesssim$\,100\,d).
Compared to model B2, model B1 has a peak luminosity a factor of 5 higher despite having a kinetic energy
almost a factor of 2 lower.

Thus it is straightforward to produce a very luminous interacting SN with a low-energy inner shell
simply by invoking a CSM mass that is comparable or larger than the inner shell mass. The conversion efficiency is
higher so that a larger fraction of the kinetic energy of the inner shell is extracted. The influence on the
luminosity is large because a typical
SN radiates a total of 10$^{49}$\,erg, so extracting 50\% of the 5$\times$10$^{49}$\,erg kinetic energy of
the inner shell can more than double the luminosity compared to a standard SN.

Another consequence of having an interaction between a low mass inner shell and a high mass outer shell
is that fast material will not survive the interaction --- the interaction will die out if the fast material
in the inner shell is entirely braked. As can be seen in the right panel of
Fig~\ref{fig_lbol_bsg_rsg}, models R3 and B3 show a CDS speed of 4000-5000\,\kms\ (comparable to the
3500\,\kms\ of our model A), while the CDS speed for models R1 and B1 is only $\sim$\,1000\,\kms.
Thus interactions between a low mass inner shell and a high mass outer shell, like those associated
with R1 and B1, may be able to explain events like SN\,1994W.

Finally, as can be seen in Fig.~\ref{fig_lbol_bsg_rsg}, the power from most of our interaction simulations
is of the order of a few 10$^7$\,\lsun\ at $\sim$\,200\,d (with no contribution from \iso{56}Ni), and thus can alone explain
the late-time luminosity of a SN IIn like 2009kn \citep{kankare_09kn}.

\subsection{Spectral properties}

We now discuss the spectral properties of these models at different epochs with
the goal of illustrating the diversity of spectral morphologies arising from different
configurations. To expedite the \cmfgen\ simulations for this section, we only include model atoms
for H\one, He\one--\two, and Fe\two--\five.

\begin{figure*}
\epsfig{file=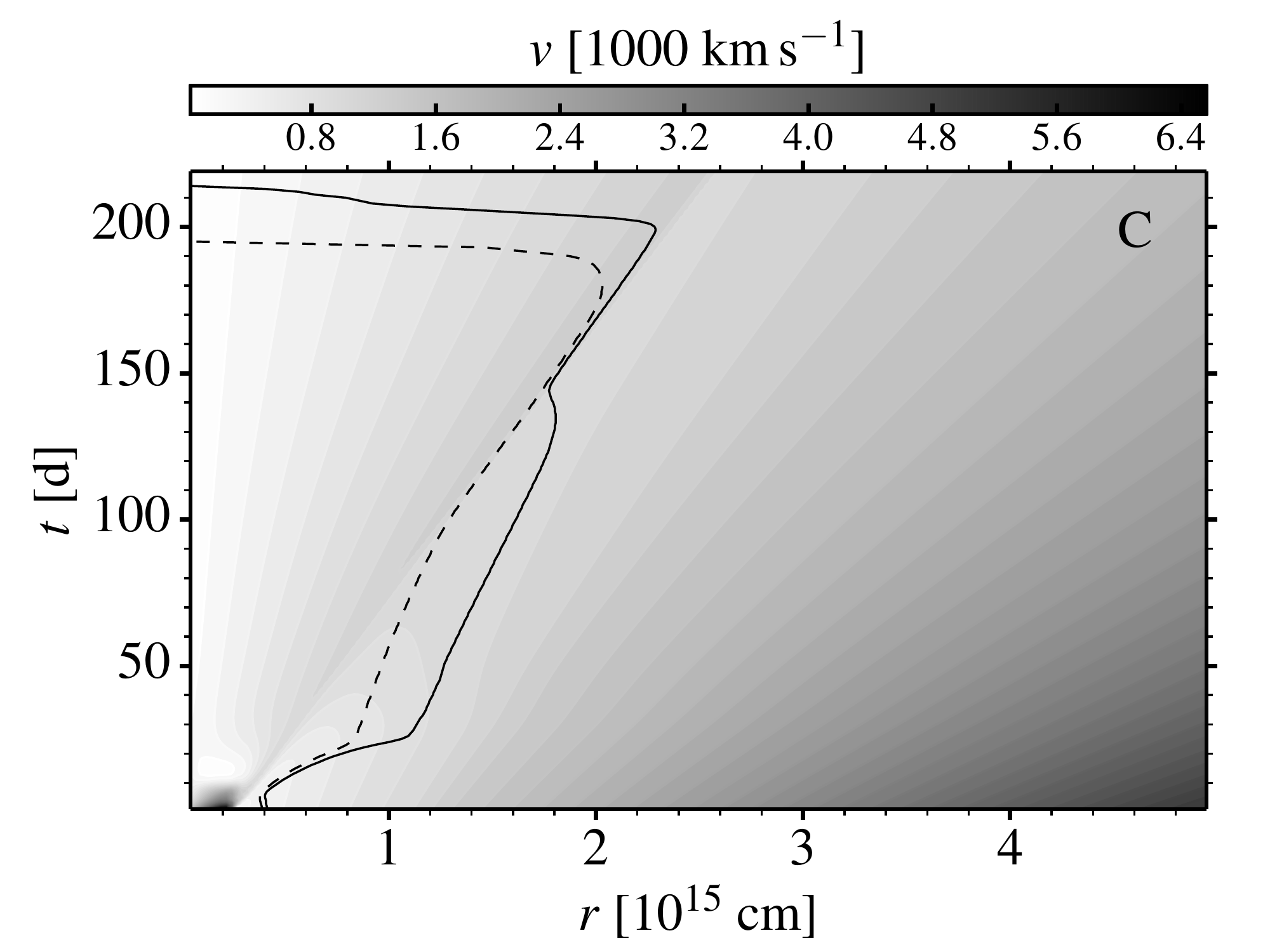, width=0.45\textwidth}
\epsfig{file=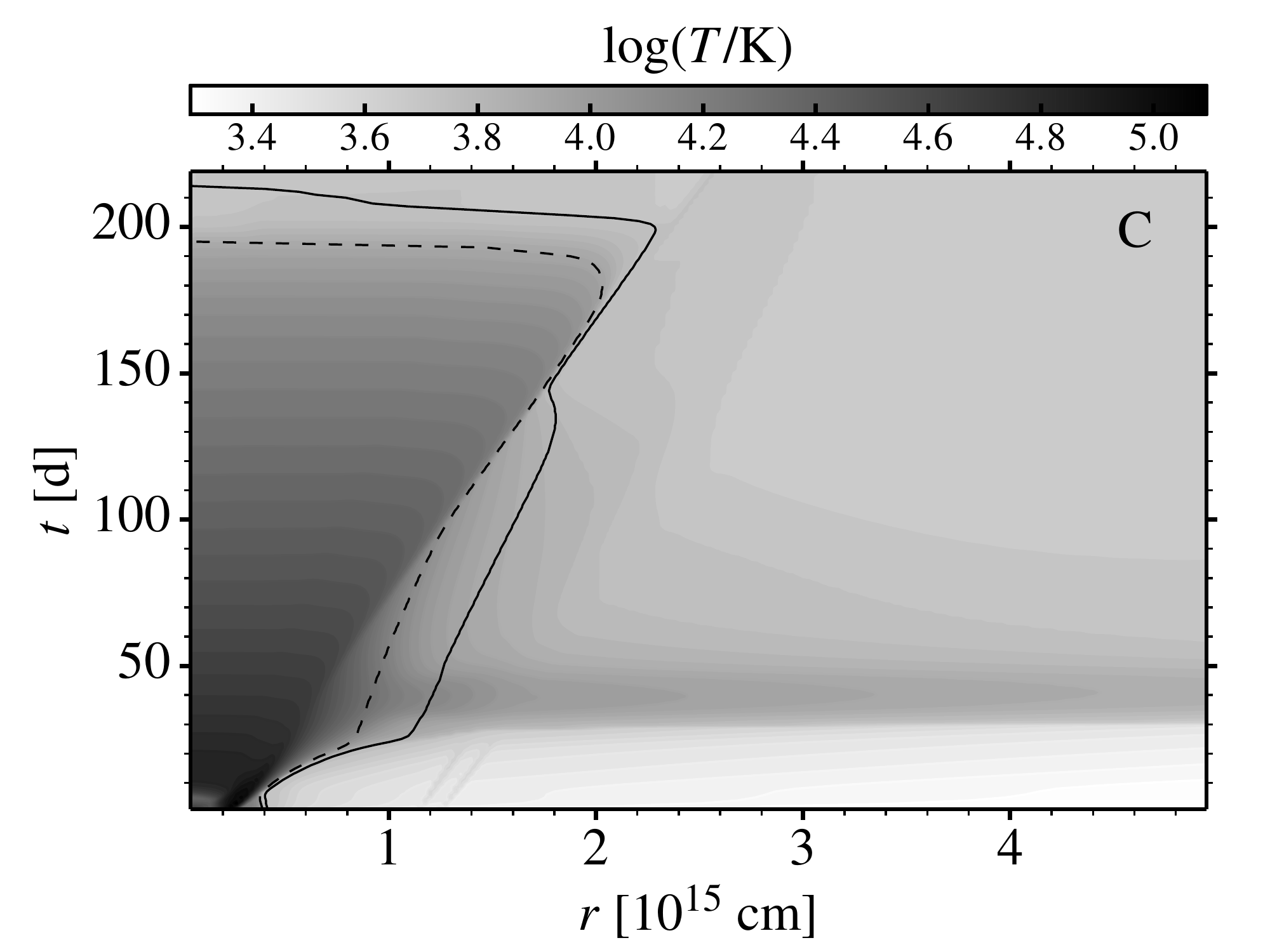, width=0.45\textwidth}
\epsfig{file=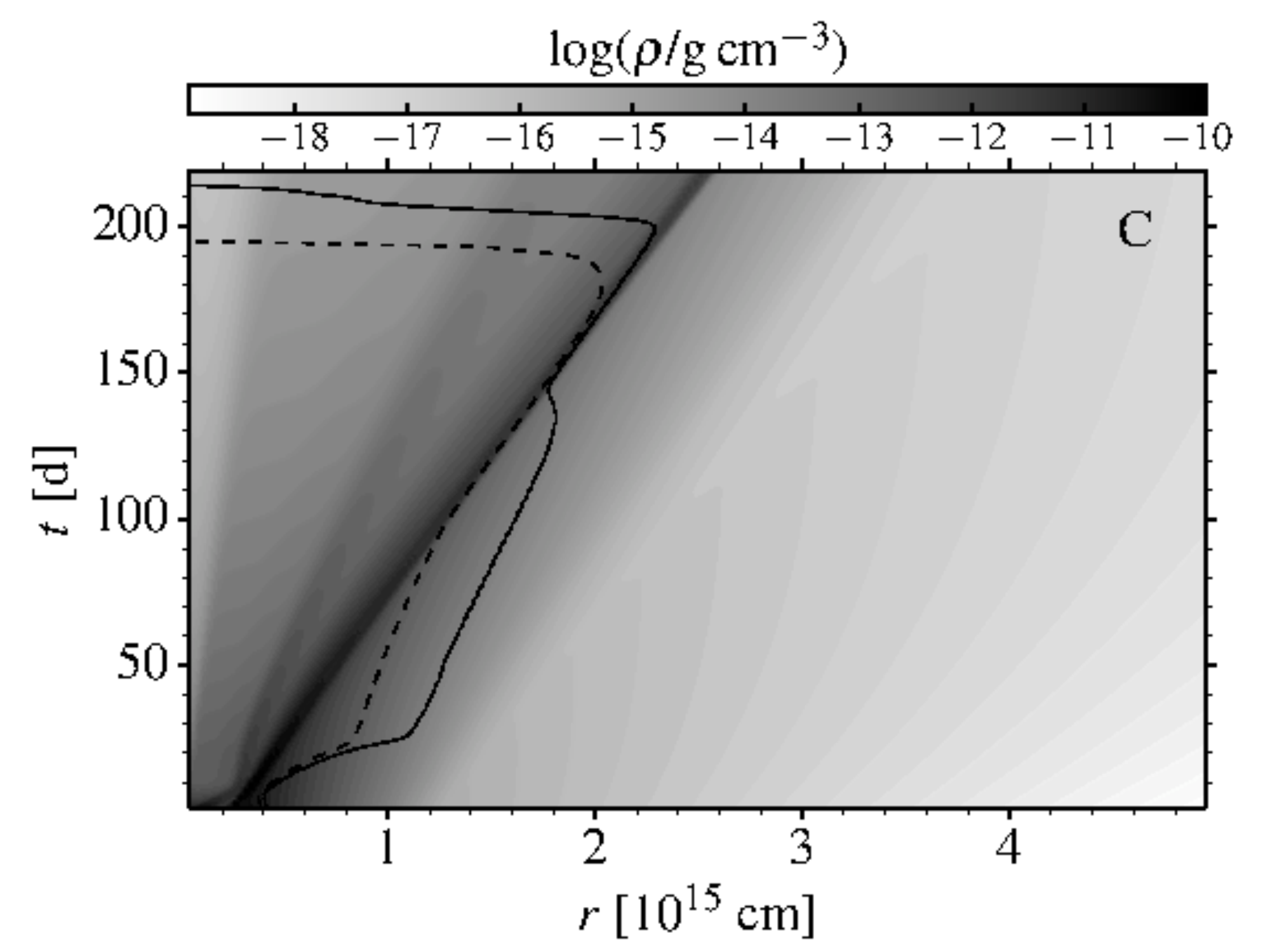, width=0.45\textwidth}
\epsfig{file=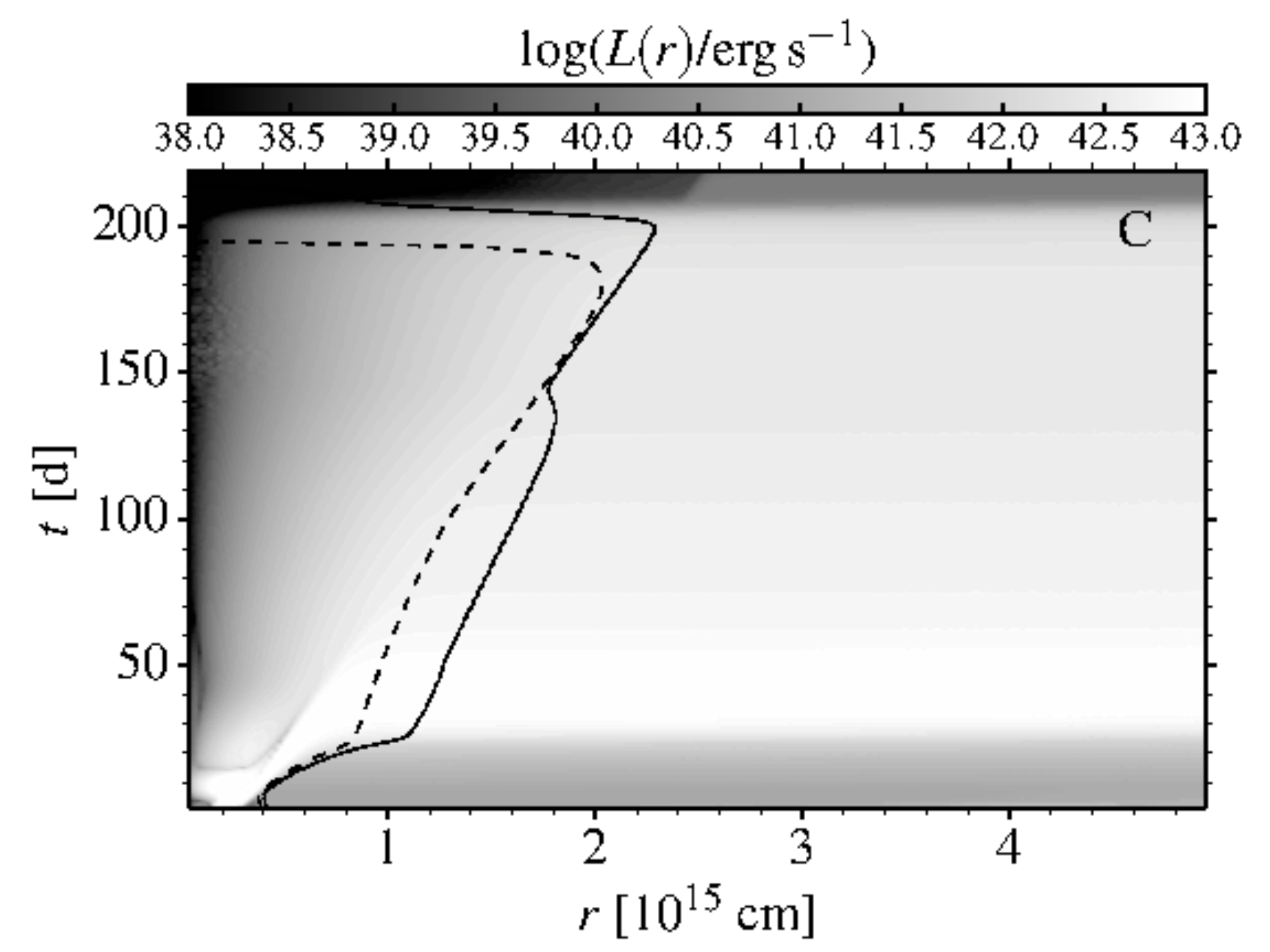, width=0.45\textwidth}
\caption{Greyscale images showing the evolution of the velocity, temperature, density, and local luminosity
for model C. The solid line traces the photosphere and the dashed line the location where the optical
depth is 10 (for both quantities, we use the opacity from electron scattering only).
\label{fig_map_modc}
}
\end{figure*}

Figure~\ref{fig_bsg_rsg_spec} shows a montage of spectra for models B1, B2, B3, R1, R2, and R3
computed with \cmfgen\ and based on \heracles\ simulations at early (top half) and late (bottom half) times
after the onset of interaction.
The spectra are quite disparate. At early times, all models except R3 and B3, show the distinctive IIn line profiles with a narrow
core and extended symmetric wings. In model R3 at 5\,d and B3 at 10\,d, the CSM optical depth is already low
(this can be indirectly inferred by the 10\,d rise time to bolometric maximum; Fig.~\ref{fig_lbol_bsg_rsg}).
Emission from the optically-thick CDS is thus weakly affected by the CSM and the spectra show the nearly unadulterated CDS radiation.
Some line profiles (e.g., H$\alpha$, H$\beta$, He\,\one\,5875\,\AA)
exhibit weak emission and blue-shifted absorption (the spectrum resembles that of model A at 20.0 and 34.7\,d),
while He\two\,4686\,\AA\ forms within the CDS and appears as broad and blueshifted emission.
This line was not predicted in our calculations for model A because the emitting region was too cool.
Its presence in model B3/R3 is not surprising, as it is observed in SNe like 1998S
\citep{fassia_98S_00, leonard_98S_00,shivvers_98S_15}.
Other models show the distinctive IIn properties.
Combined with the presence of lines from the hot CSM (broadened by electron scattering) or from the hot CDS
(broadened by Doppler effect), all 6 models show narrow lines arising from the cool outer CSM.

At late times models B3 show broad line profiles, in particular H$\alpha$, with a blue-shifted peak and emission skewed to the red.
Model R3 at 75\,d exhibits a spectrum more typical of a non-interacting SN II late in the plateau phase. The skewness of the H$\alpha$
profile is much reduced (i.e., the emission is more symmetric), although the peak emission is blue-shifted by about $-$1500\,\kms.
For model R3, the emission is primarily from the inner ejecta (unaffected by the interaction), as obtained for model A at corresponding epochs.

In model B2 at 80\,d, the emission is primarily from the CDS and gives rise to double-peaked line profiles (probably because
of an optical depth effect similar to limb darkening) for H\one\ and He\one\ lines, with peaks at about $\pm$1000\,\kms.
The blue component is stronger, which suggests a continuum optical depth effect acting on line photons.
Lower energy models show narrower lines. In model R1 at 50\,d, the spectrum forms in the optically thick slow-moving
CDS and only absorption lines are seen (this is similar to model A at 34.7\,d).
In model B1, only narrow emission lines are seen.

This diversity of early/late time line profiles conveys a wealth of information on the interaction configuration.
We defer to a subsequent paper the study of all these profile properties. What we take from this study is that model
R1 can produce a bright display while showing broad lines neither at early times nor at late times. In the next
section, we will elaborate a more suitable model that shares these properties and reflects those
observed for SN\,1994W.

\begin{figure}
\epsfig{file=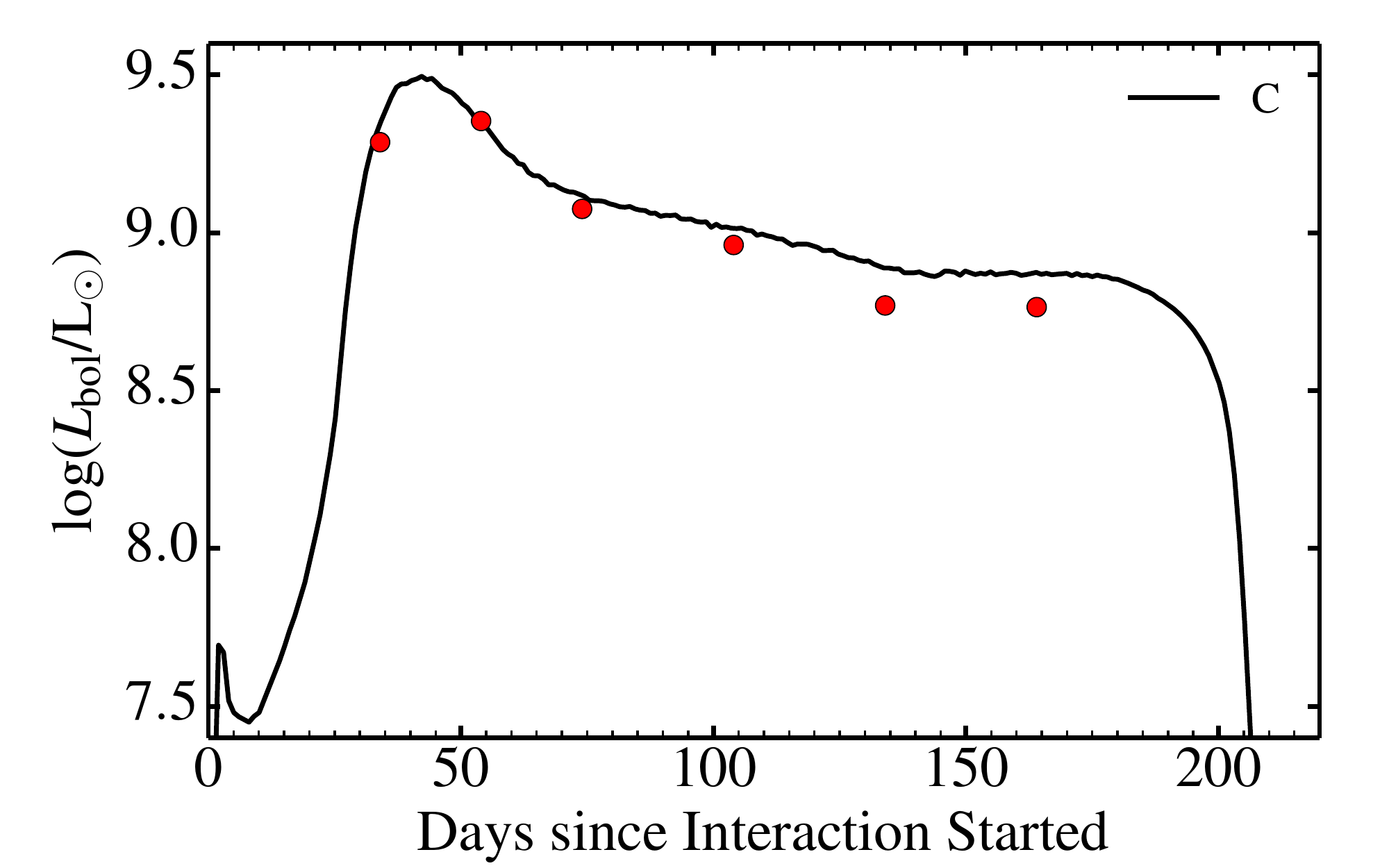, width=0.47\textwidth}
\caption{Bolometric light curve for model C (the dots
correspond to the luminosity computed at corresponding epochs with \cmfgen).
\label{fig_lbol_modc}
}
\end{figure}

\section{SN\,1994W: a low-mass energetic inner shell ramming into a slow massive outer shell}
\label{sect_modc}

\subsection{Preamble}

In this final modelling section, we setup a more physical model to explain events like
SN\,1994W. Building upon the results from the preceding sections, the configuration
we are after is a massive slow outer shell and a low/moderate mass inner shell
with a modest kinetic energy (by SN standards).

Stars in the range 8-12\,\msun\  sit at the junction between two completely different fates.
The lower mass ones may ultimately produce a bare
white dwarf, while the more massive ones will form a degenerate core (made of ONeMg
or Fe-group elements) that collapses to a neutron star \citep{poelarends_sAGB_08}.
For stars just above that mass cut, O, Ne, or Si may ignite off-center, under degenerate conditions.
If/when this occurs, burning is dynamical and leads to a nuclear flash \citep{weaver_woosley_79, WH15}.
Because the phenomenon is only encountered in the late stages of evolution, when these low-mass massive stars
are in a RSG phase, even a very small energy release from nuclear burning is sufficient to eject the
loosely bound envelope \citep{dlw10a}.
If the delay until core collapse is weeks to months, the subsequent explosion will lead to an interaction between
two detached shells.
If the delay is short (say, of the order of days), only one explosion may be seen because the SN shock
will overtake the first shock before it reaches the stellar surface \citep{WH15}.

Such nuclear flashes have been proposed as a potential mechanism leading to interacting SNe such
as SN\,1994W \citep{chugai_etal_04, dessart_etal_09}. Our simulations of Sections~\ref{sect_moda}--\ref{sect_grid}
show that interactions in which the inner shell is more energetic than, but less or as massive as,
the outer shell can lead to a luminous SN IIn event.
Here, we compute light curves and spectra for such configurations, and confront them with the observations of SN\,1994W.

\subsection{Interaction configurations}

The possible parameter space is large since we do not know the final star mass, the energy of each explosion,
nor the delay between them --- significant variations are to be expected for these quantities.
So, our model is one realisation out of many possibilities, but it
is chosen to match roughly what was inferred for SN\,1994W. In particular the interaction must
take place at a large distance from the progenitor star \citep{chugai_etal_04, dessart_etal_09}.
We need the inner shell to be less massive than the outer
shell and to have a larger kinetic energy (Sections~\ref{sect_moda}--\ref{sect_grid}).

Because of off-center ignition (in particular of Ne), we could not evolve any \mesa\ model for an $<$\,12\,\msun\ star
until core collapse. So, we use instead a higher mass model of 12\,\msun\ and enhance the mass loss during the RSG phase
to compensate for the higher initial mass.
Our 12\,\msun\ \mesa\ model has then a final mass of 9.87\,\msun\ and a final radius of 520\,\rsun\
(our conclusions would hold if these values were changed by few 10\%). Using
\v1d\ and starting from this model, we first trigger a 10$^{49}$\,erg explosion at the base of the H-rich
envelope (outer edge of the He core), at a mass cut of 3.5\,\msun\ and let this ejecta evolve to late times.
We then trigger a 10$^{50}$\,erg explosion in this H-deficient remnant. We use a mass cut of 3.1\,\msun\
(and not 1.4-1.5\,\msun) because this second explosion is meant to mimic core collapse of a lower-mass star,
which would have a lower He-core mass than the 12\,\msun\ model we use. This second ejecta is then evolved to late times.

We set up the interaction using both ejecta calculated with \v1d.
To test the influence of the initial interaction radius  $R_{\rm t}$,
which is connected to the timing of the two explosions, we set
up a model C with $R_{\rm t}= 2 \times 10^{14}$\,cm,
and a model D with $R_{\rm t} = 1.8 \times 10^{15}$\,cm.

For the outer shell, we use the structure of the first exploded shell
when its base reaches $R_{\rm t}$. For the inner shell, we use the structure of the second
exploded shell when its outer part reaches $R_{\rm t}$. We join the two shells at that radius $R_{\rm t}$.
Because each shell was exploded and evolved
with the radiation-hydrodynamics code \v1d, the fluid variables (radius, velocity, density, temperature) are
physically consistent.

\begin{figure*}
\epsfig{file=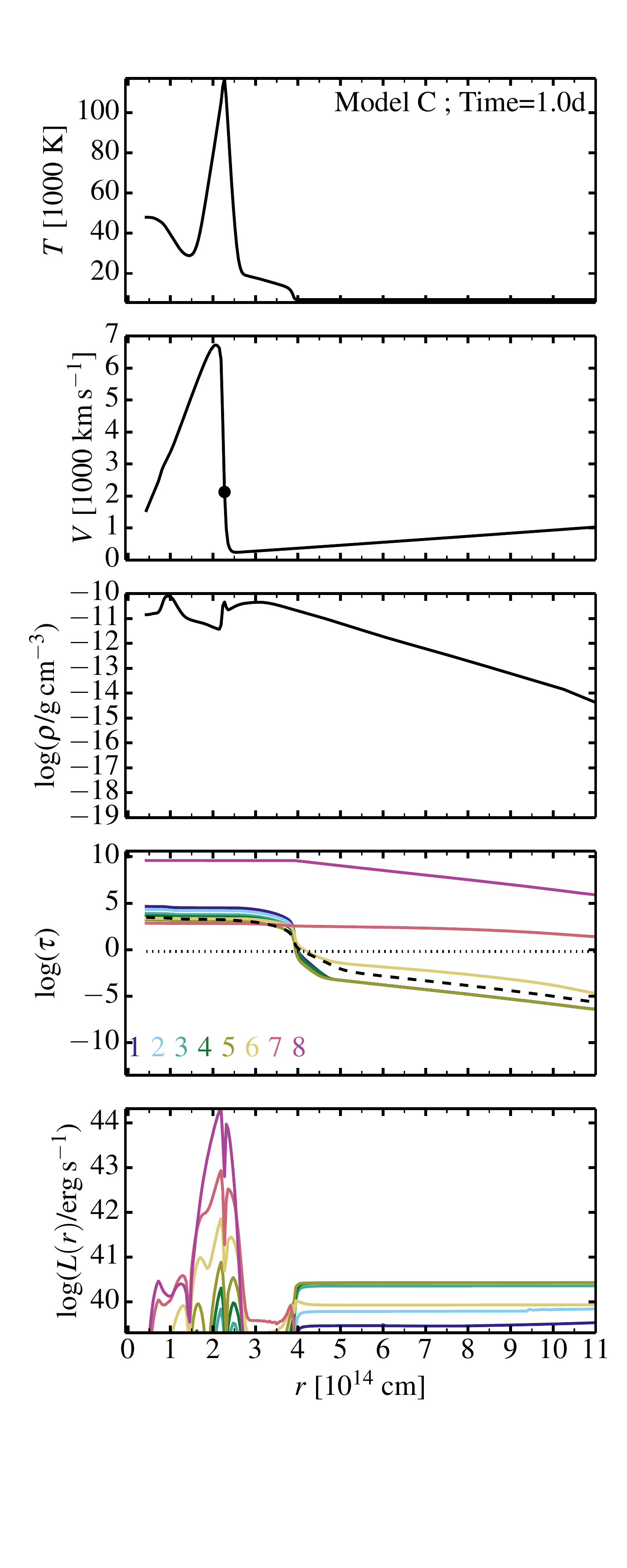, width=0.47\textwidth}
\epsfig{file=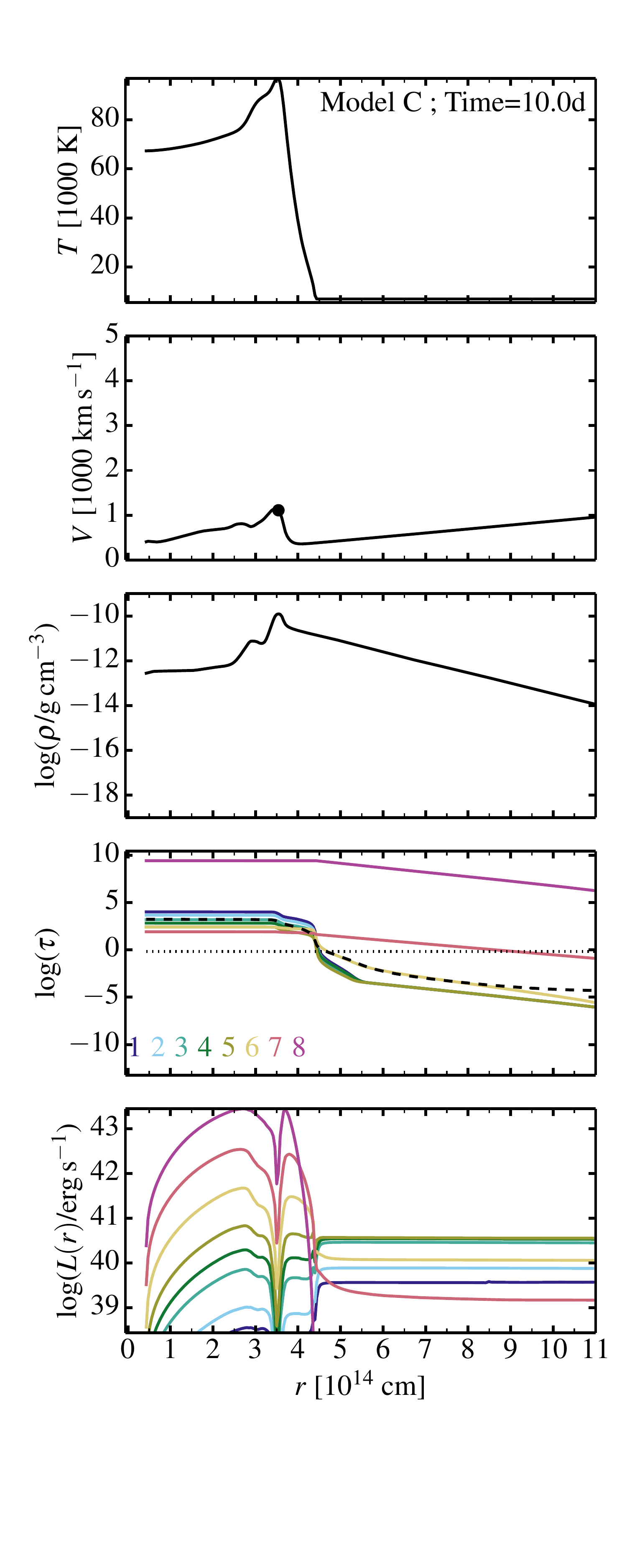, width=0.47\textwidth}
\vspace{-2cm}
\caption{Left:
Properties of the interaction Model C computed with \heracles\ at 1.0\,d after the start of the interaction.
We show the gas temperature, the velocity (the dot corresponds to the velocity of the CDS),
the mass density, the optical depth, and the radiative luminosity versus radius.
For the optical-depth panel, we show that quantity for each energy group (coloured line;
the group energy increases with the group number; see Section~\ref{sect_setup} for details)
and for electron scattering (dashed line) --- the dotted line corresponds to an optical depth of 2/3.
At this time, the outer shell is optically thick out to 4$\times$10$^{14}$\,cm.
The radiation emerging through the photosphere is very small compared to the luminosity
injected at the shock.
Right: Same as left, but now at 10.0\,d after the onset of the interaction.
\label{fig_modc_1}
}
\end{figure*}

\begin{figure*}
\epsfig{file=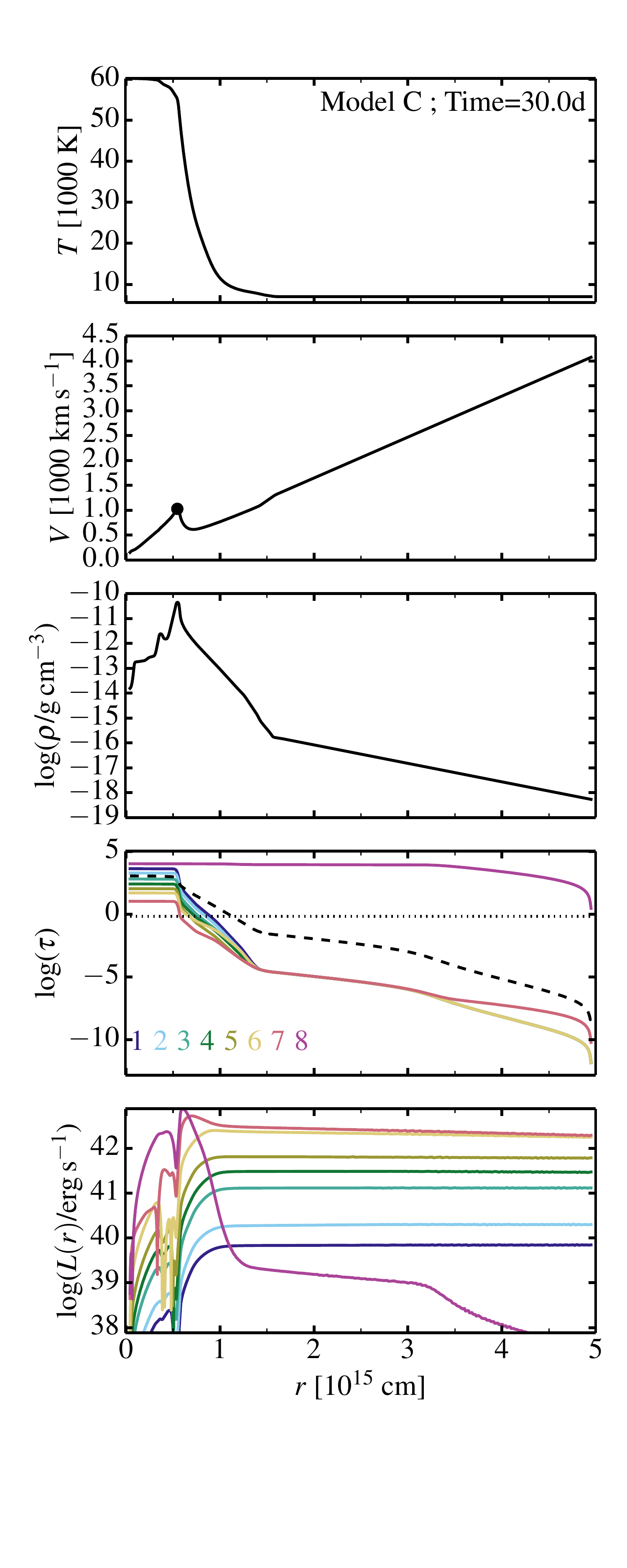, width=0.47\textwidth}
\epsfig{file=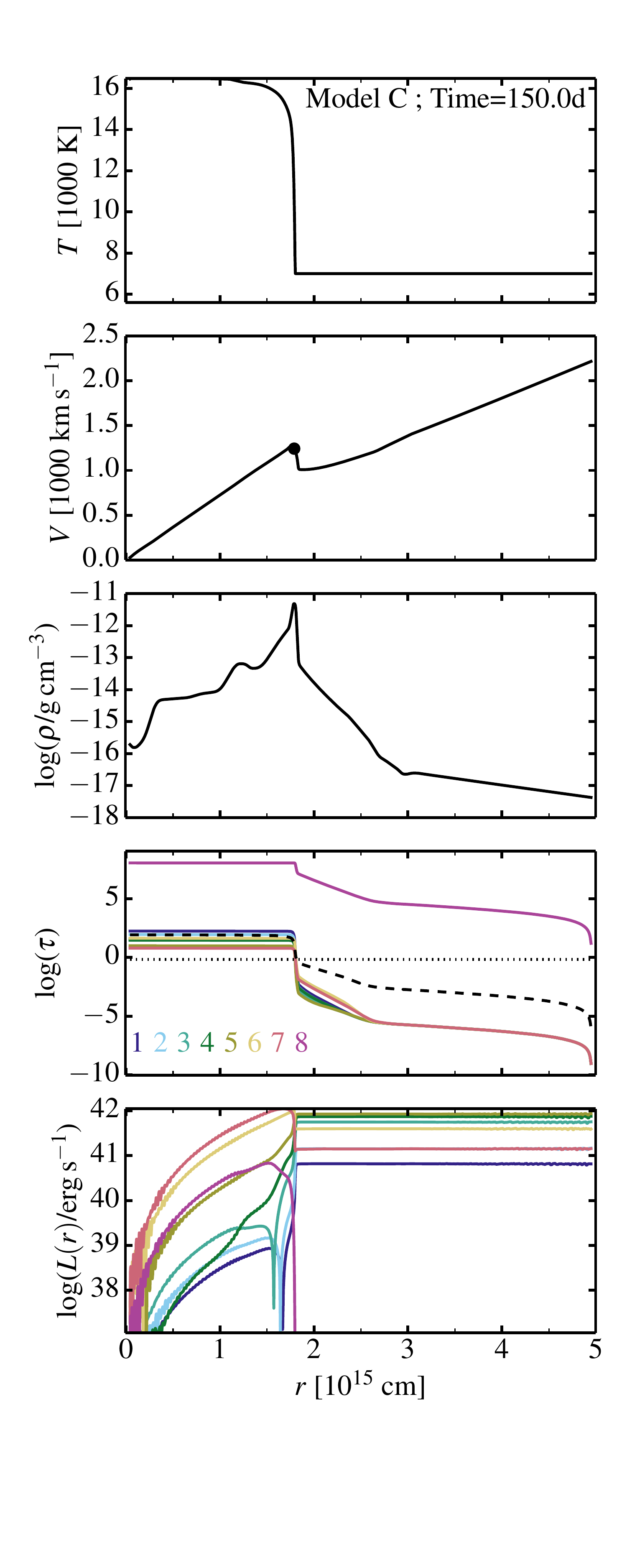, width=0.47\textwidth}
\vspace{-2cm}
\caption{Same as for Fig.~\ref{fig_modc_1}, but now at 30.0\,d (left) and 150.0\,d (right) after the onset of the interaction.
\label{fig_modc_2}
}
\end{figure*}

The properties of the initial configuration for models C and D are given in Table~\ref{tab_mod}.
The initial velocity, density and temperature structures are illustrated in Figs.~\ref{fig_modc_init}--\ref{fig_modd_init}
in the appendix.
When setting up the interaction configuration, some grid zones are trimmed
in the outer parts of the inner shell and the inner parts of the outer shell in order to avoid creating a gap.
This explains the slight differences in mass/energy for each shell of models C and D.
The inner shell is $\approx$\,0.3\,\msun\ and expands very fast, with a mean mass-weighted velocity of
$\approx$\,4750\,\kms.
In model C, it is only 2.3\,d old at the start of the interaction so the inner shell is initially hot and mostly optically thick.
In model D, the inner shell is 23.4\,d old, cold and optically thin.
The outer shell of models C/D is $\approx$\,6.3\,\msun\ (20 times more massive than the inner shell), with a mean mass-weighted velocity
of  $\approx$\,380\,\kms.
Its age at the start of the interaction is 127.9\,d in model C, and 941.7\,d in model D.
Because of its slow expansion, the inner parts of that shell are still optically thick and hot (about 20000\,K) in model C.
In model D, the outer shell is much older, and therefore cold and optically thin.
In the absence of interaction, the outer shell alone would exhibit a very long and faint plateau (with a bolometric luminosity
of the order of 10$^7$\,\lsun; see the low-energy explosions of RSG stars discussed in \citealt{dlw10a}).

In model A and the grid of models presented in Section~\ref{sect_grid}, the outer shell was
formed through a phase of intense mass loss. So, the CSM in that case was pre-SN wind material.
Here, the outer shell is a very different type of CSM (described by an homologous expansion, a
complex density structure etc). It is explosively formed by a nuclear flash
and subsequently affected, within days to months, by the terminal collapse of the remaining star
(which in this case can only produce a very low mass ejecta made of the fraction of the He core that
does not collapse into the remnant, so $\lesssim$\,1\,\msun\ in such progenitors).

\begin{figure*}
\epsfig{file=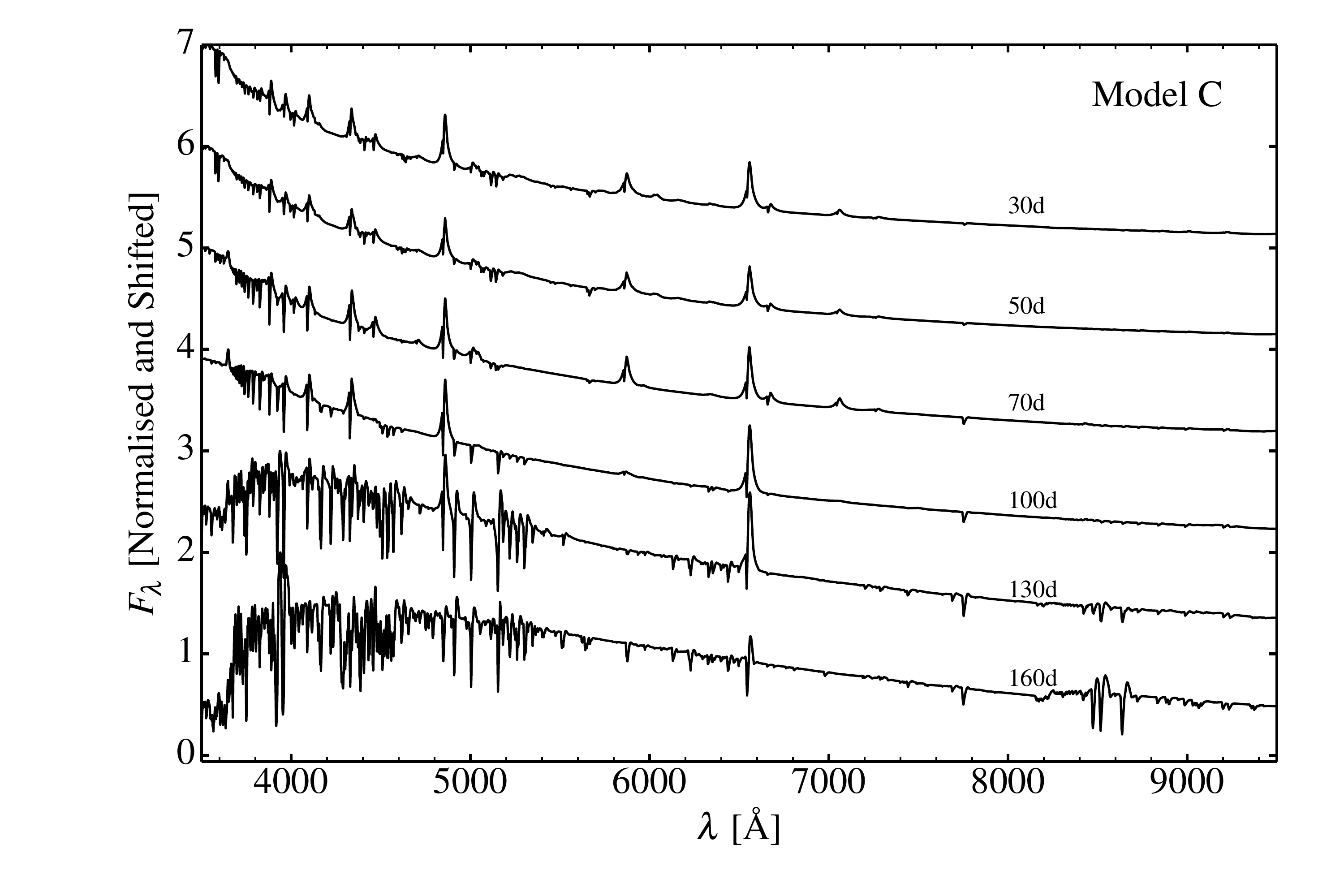, width=\textwidth}
\epsfig{file=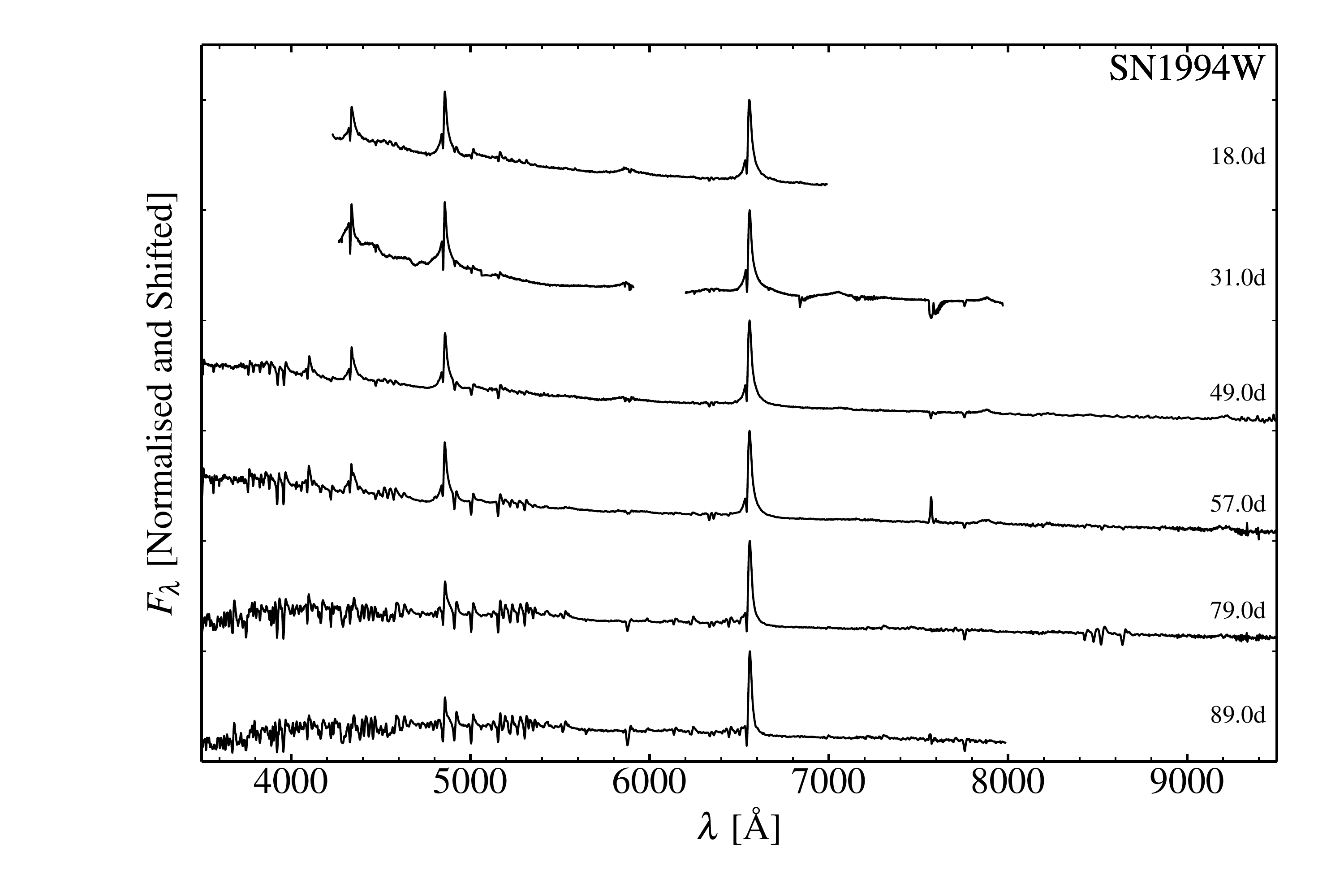, width=\textwidth}
\vspace{-1.cm}
\caption{
Top: Spectral evolution of the double explosion explosion model (C).
Bottom : Multi-epoch spectra of SN\,1994W, corrected for redshift but not for extinction.
The spectrum shows narrow lines with broad wings early on, but, unlike SN\,1998S, never
develops broad lines at late times.
\label{fig_spec_modc_94w}
}
\end{figure*}

\begin{figure*}
\epsfig{file=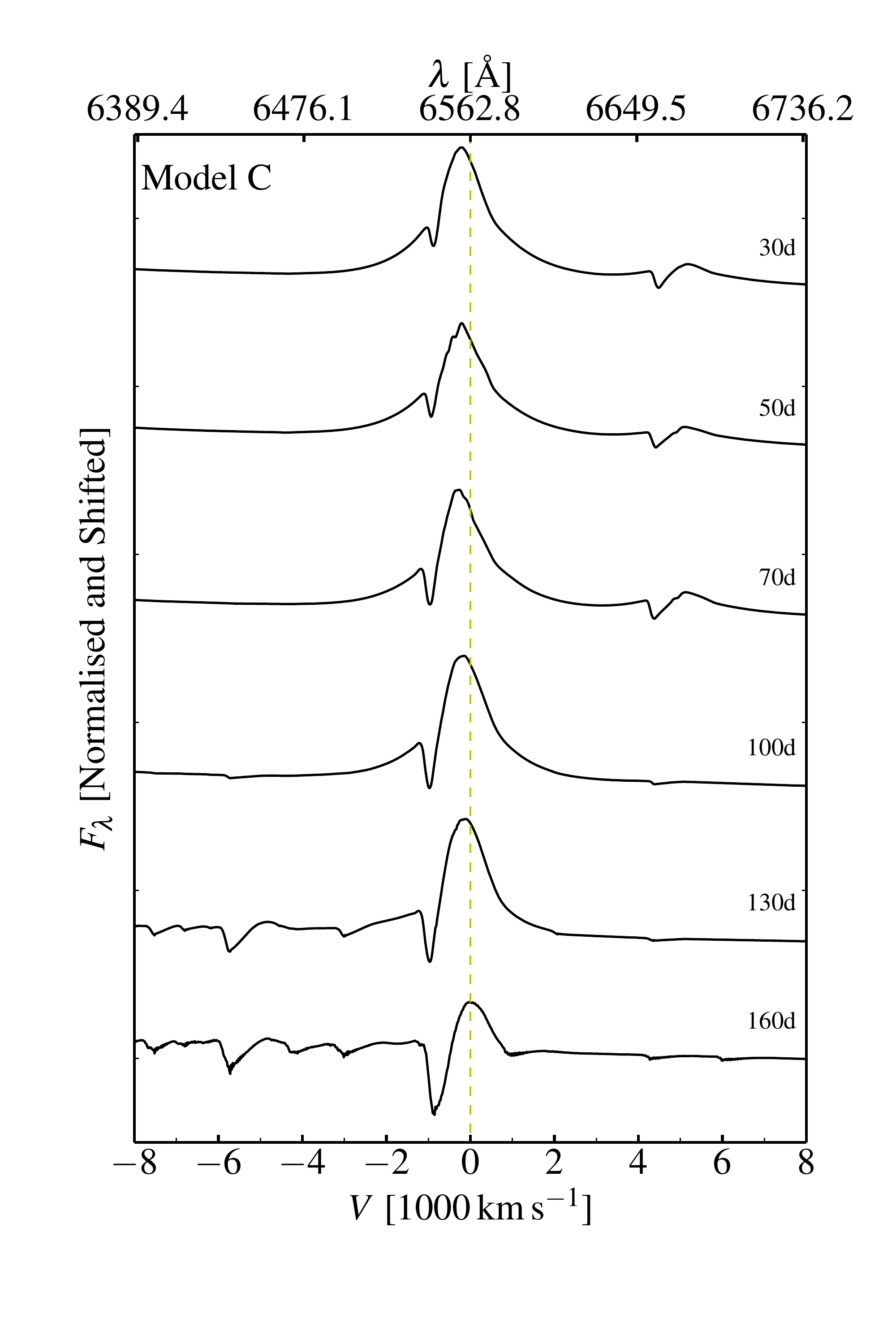, width=0.45\textwidth}
\epsfig{file=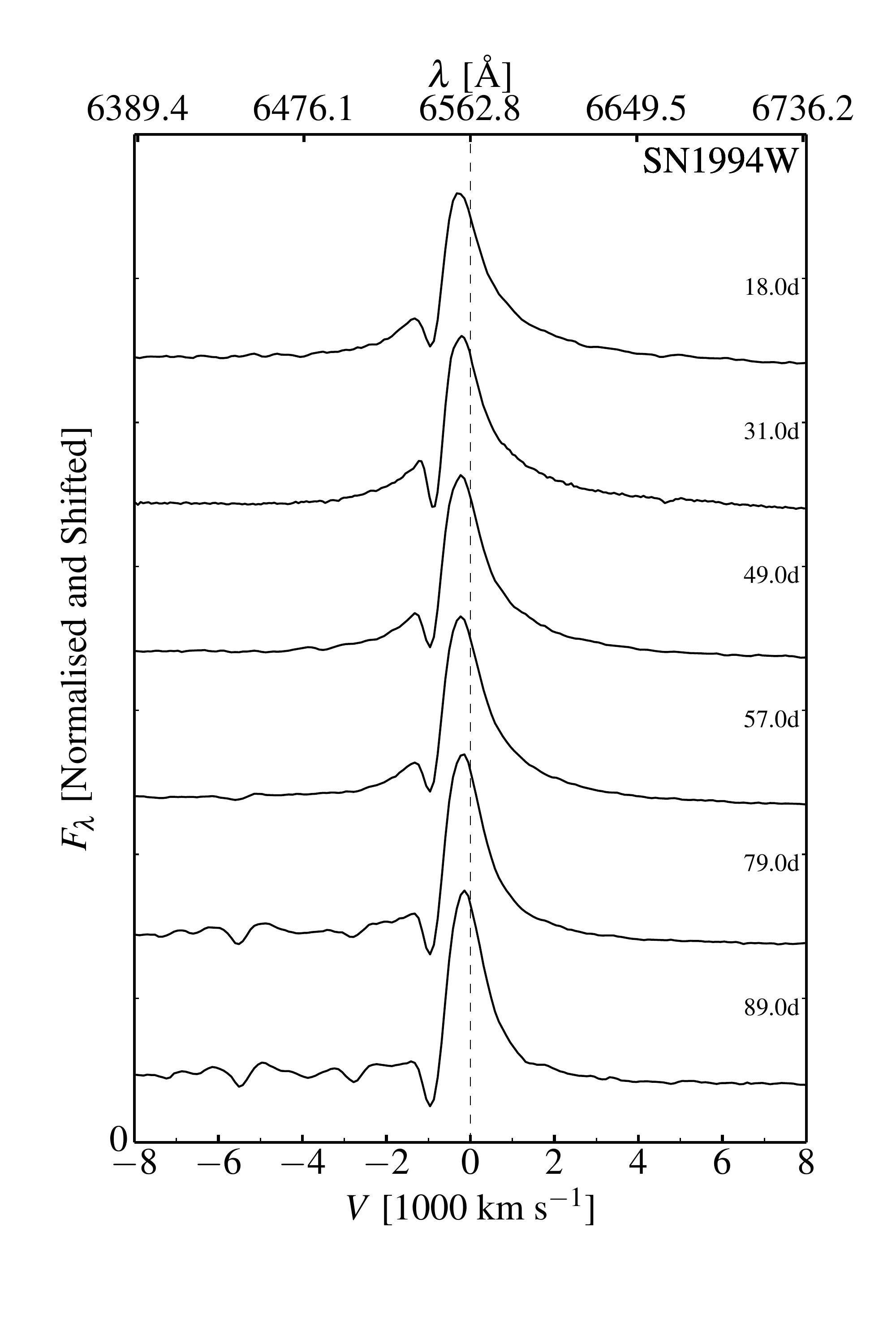, width=0.45\textwidth}
\vspace{-1.cm}
\caption{
Left: Spectral evolution of the double explosion explosion model C.
Right: Multi-epoch spectra of SN\,1994W. The spectrum shows
narrow lines with broad wings early on, but, unlike SN\,1998S, never
develops broad lines at late times.
\label{fig_spec_modc_94w_ha}
}
\end{figure*}

\subsection{Results for model C}
\subsubsection{Dynamical properties}

The dynamical evolution for model C is very different from standard interacting SNe --
in particular the strong interaction generating the shock has ceased before the peak brightness is reached.
We illustrate these differences in a series of greyscale images and snapshots.
Figure~\ref{fig_map_modc} shows greyscale images of the velocity, temperature, density, and radiative luminosity
versus radius and time computed with \heracles\ for our model C.
Figure~\ref{fig_lbol_modc} shows the bolometric light curve of model C computed with \heracles.
Figures~\ref{fig_modc_1}--\ref{fig_modc_2} shows this evolution at selected snapshots,
from 1.0 to 150.0\,d after the onset of the interaction.

The interaction is very strong initially because the density at the base of the outer shell is $\sim$\,10$^{-11}$\,g\,cm$^{-3}$
and the velocity of the fast material in the inner shell $\sim$\,10000\,\kms.
The outer shell being optical thick, the radiation generated by the shock cannot escape. All the energy
is trapped. The temperature at the base of the outer shell rises from 20\,000\,K to $\gtrsim$\,100\,000\,K.
As time progresses, the shock continues to progress through the outer shell, sweeping up more material
and building a hot optically-thick dense shell, but the deceleration is severe. By 10\,d, no material in the inner shell
moves faster than 1000\,\kms\ -- the shock is essentially gone by then (Fig.~\ref{fig_modc_1}).
This implies that the rest of the outer
shell will not be shocked, and also that the inner parts of the inner shell will not be decelerated further.
The effect of the shock has been to pile up a lot of material around 1000\,\kms, leaving slower and faster material
(dynamically) unaffected.

Within 10\,d, the interaction as a power source for the SN has vanished (the shock is already weak).
The bulk of the kinetic energy
in the inner shell has been transformed into internal energy, which is primarily radiation.
Most of this radiation energy is trapped within the dense shell, so the subsequent radiation
is the slow release of radiation energy from this optically thick ejecta.
At 10.0\,d, the flux emerging from
the dense shell is huge, much in excess of the flux radiated at the photosphere: the two will become
comparable after a diffusion time of about 30\,d (see, e.g., the bottom right panel of Fig. ~\ref{fig_map_modc}
and Fig.~\ref{fig_modc_2}).

Before the interaction starts, the photosphere in the outer shell is at $\lesssim$4$\times$10$^{14}$\,cm.
Once the interaction starts, the energy
deposited by the shock heats up the ejecta, shifting the spectral energy distribution to the blue,
facilitating the re-ionisation of the outer shell material, and pushing the photosphere out to larger radii.
At 10.0\,d, the photosphere has moved out to
$\sim$\,4.2$\times$10$^{14}$\,cm, and following the (slow) ejecta expansion, it keeps moving
out to reach $\sim$\,1.1$\times$10$^{15}$\,cm at 30\,d, and $\sim$\,1.6$\times$10$^{15}$\,cm at 100\,d.
The thickness of the ionised material in the outer shell (or, equivalently, the size of the region between the photosphere
and the CDS) is fairly constant between 30.0 and 150.0\,d (Fig~\ref{fig_map_modc}) --- this region is co-moving with the
ejecta. So, this situation is analogous to
a low energy faint SN II-P, but here energy is freshly supplied once the ejecta has expanded to a large
radius, in this case beyond 2$\times$10$^{14}$\,cm. The fresh supply of energy boosts the SN luminosity,
transforming this originally faint transient into a very luminous type II SN. The ejecta retains
its low velocity and thus, unlike a typical SN II-P, the event is very luminous but has a low expansion rate.

\begin{figure}
\epsfig{file=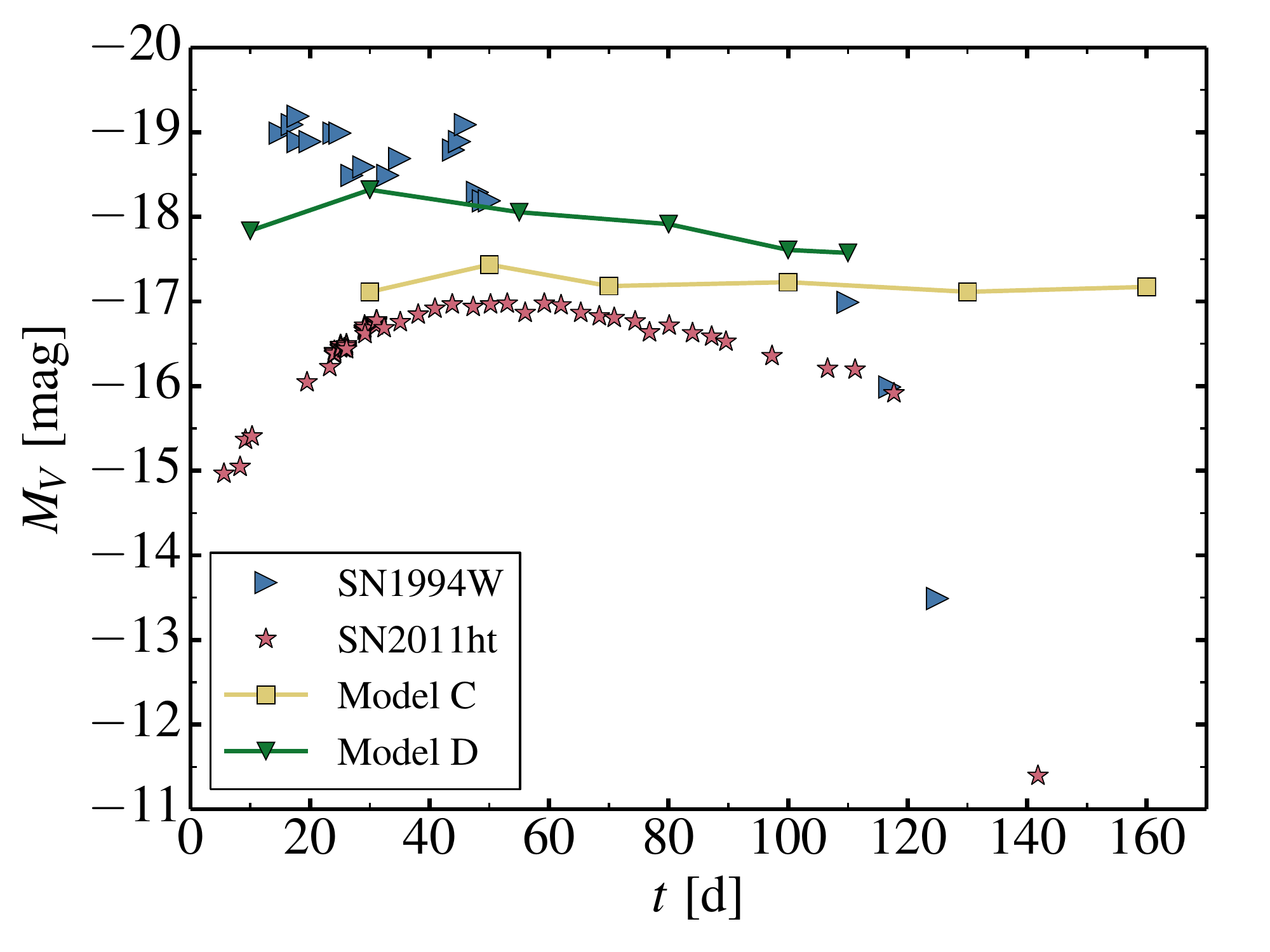, width=0.45\textwidth}
\caption{Absolute $V$-band light curve for SN\,1994W and SN\,2011ht compared to our models C and D.
See Section~\ref{sect_obs} for details on the source of observations, distances, reddening, and inferred
time of explosion.
For the models, the time is with respect to the onset of interaction.
\label{fig_lc_mod_c_d_obs}
}
\end{figure}

The very slow and dense material at the base of the outer shell has been swept up into a dense shell, undergoing
an acceleration from $\sim$\,200\,\kms\ to $\sim$\,1000\,\kms.
This corresponds to about half the total kinetic energy of the inner shell initially.
The other half, $\sim$\,5$\times$10$^{49}$\,erg, is radiated away, producing a very luminous event,
typically with 5 times the total integrated luminosity of a standard SN.

\begin{figure*}
\epsfig{file=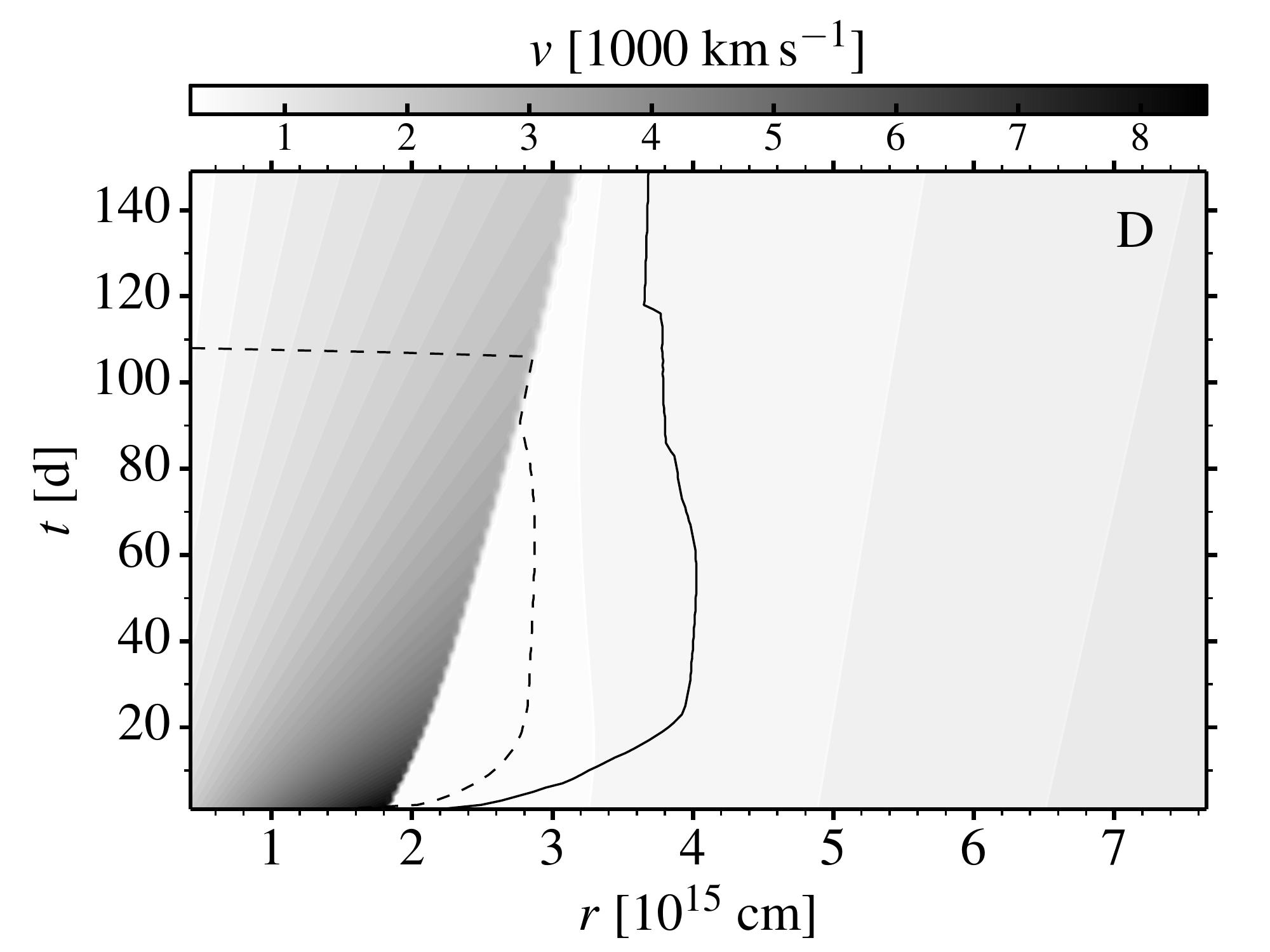, width=0.45\textwidth}
\epsfig{file=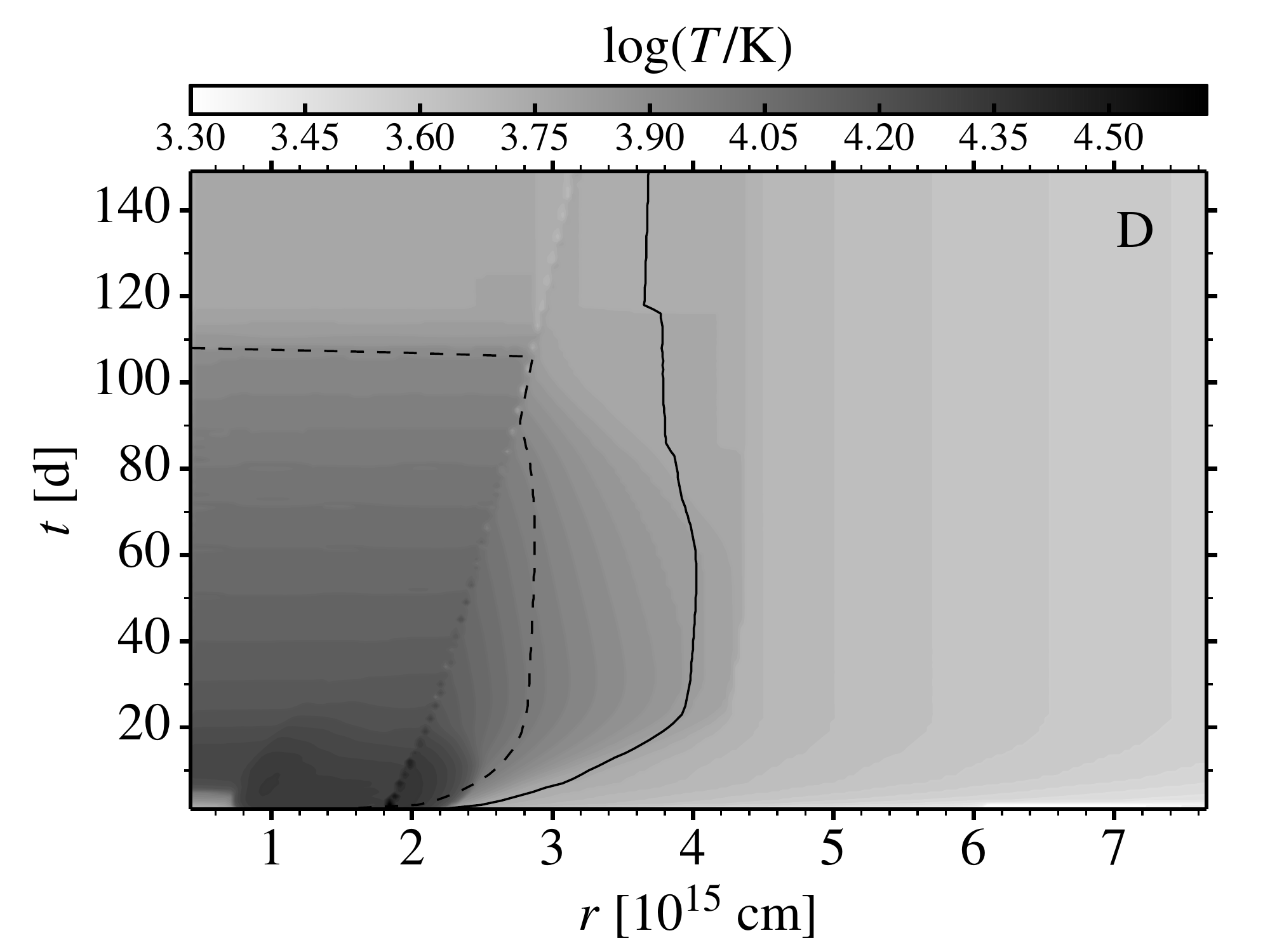, width=0.45\textwidth}
\epsfig{file=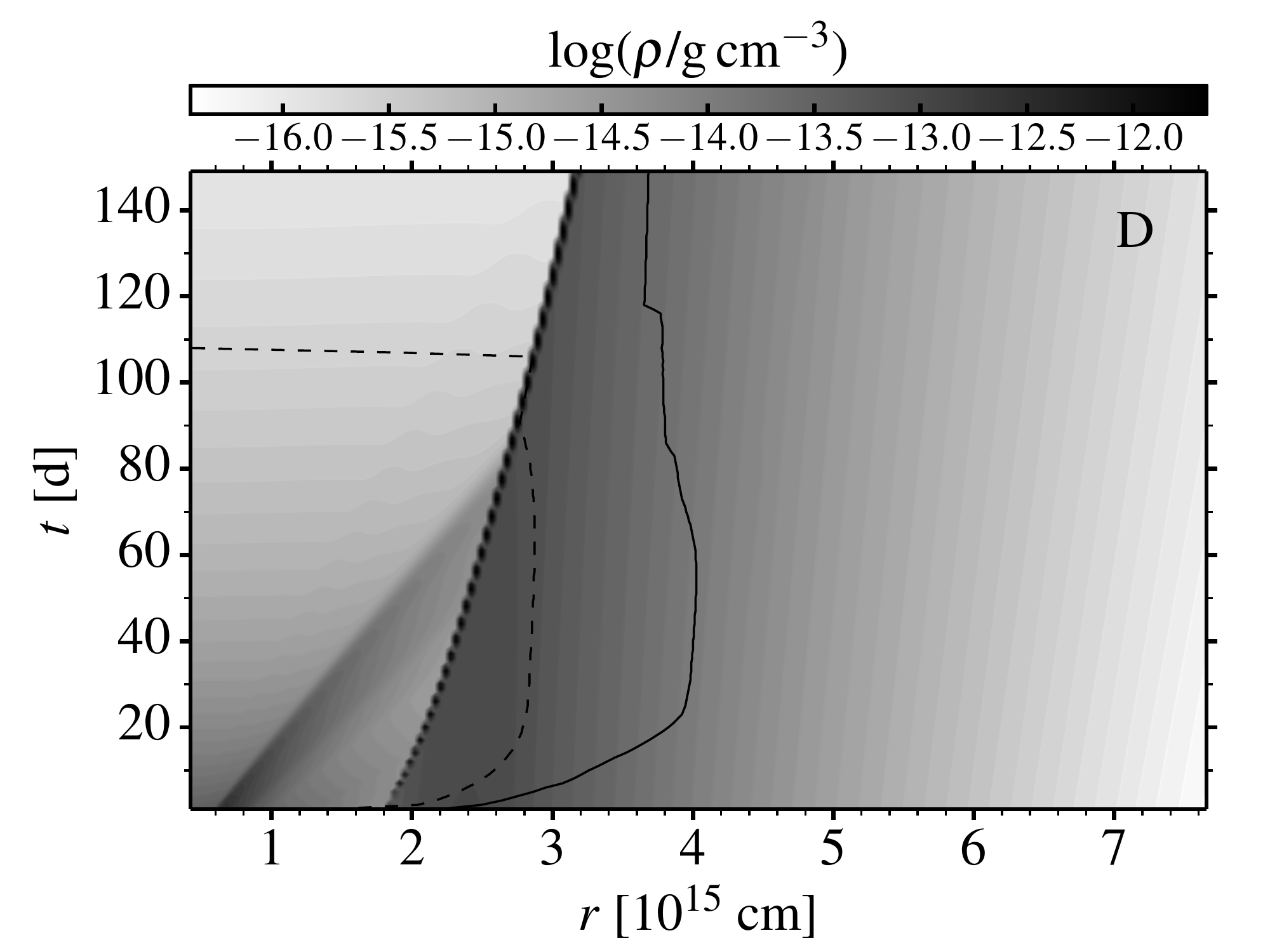, width=0.45\textwidth}
\epsfig{file=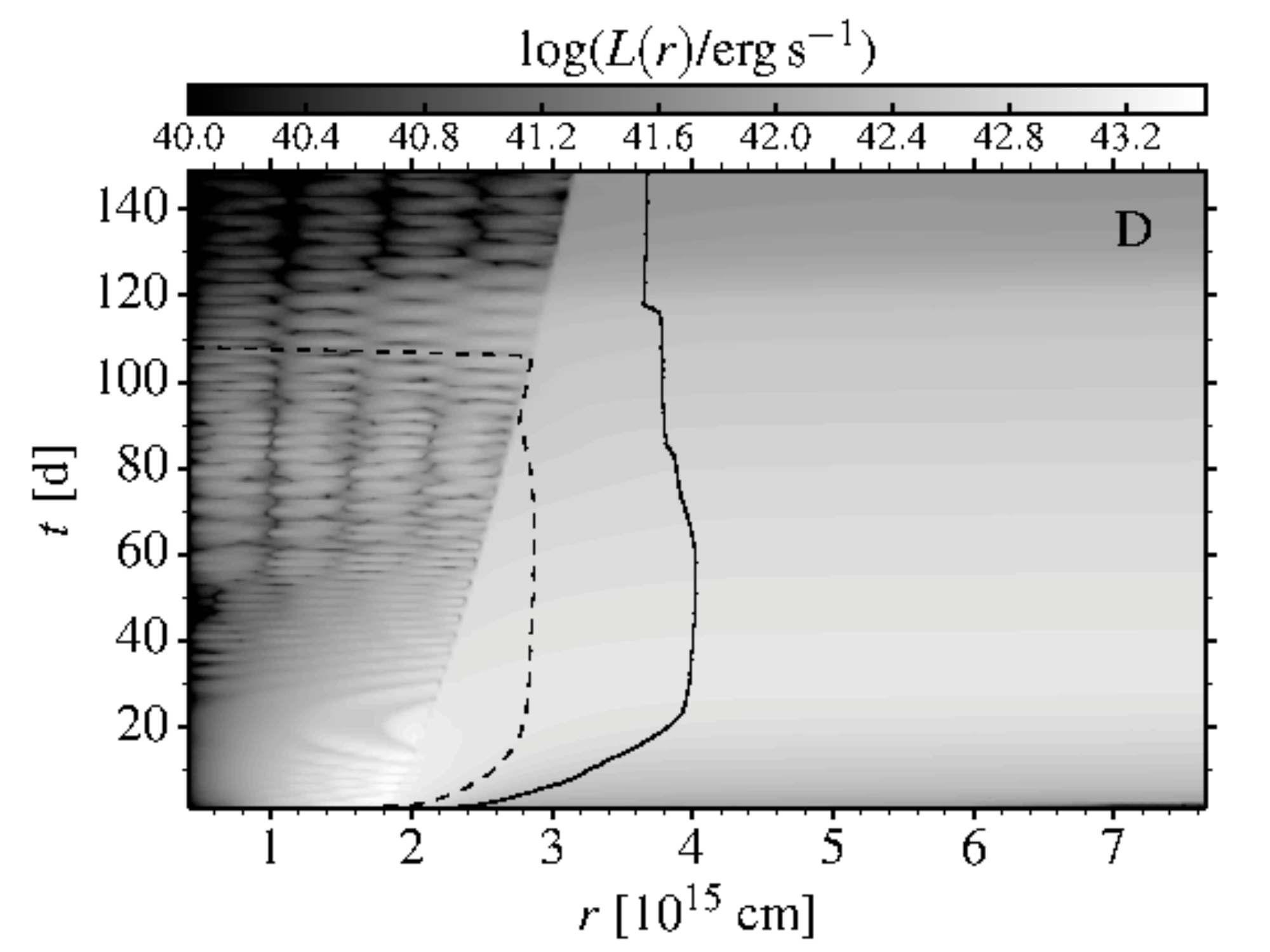, width=0.45\textwidth}
\caption{Greyscale images showing the evolution of the velocity, temperature, density, and local luminosity
for model D. The solid line traces the photosphere and the dashed line the location where the optical
depth is 10 (for both quantities, we use the opacity from electron scattering only).
\label{fig_map_modd}
}
\end{figure*}

\subsubsection{Spectral simulations for model C and comparison to SN\,1994W}

Using the same procedure as in Section~\ref{sect_moda}, we have post-processed the \heracles\
simulations of model C with \cmfgen\ to compute the emergent spectra at 30, 50, 70, 100, 130, and 160\,d
after the onset of interaction.
For each epoch, we obtain a good correspondence between the bolometric luminosity computed by
\heracles\ (line) and \cmfgen\ (Fig.~\ref{fig_lbol_modc}).

Figure~\ref{fig_spec_modc_94w} shows the \cmfgen\ spectral sequence for our model C (top panel) as
well as the multi-epoch spectral observations of SN\,1994W (bottom panel).
The model C shows the distinctive IIn line profile signatures at 30, 50, 70, and 100\,d on H\one\ and He\one\ lines,
as well as the presence of very narrow lines associated with Fe\two.
The spectrum forms within the outer-shell, in a region not affected dynamically by the interaction.
This outer shell shows a steep declining density (comparable to a power-law density profile with an
exponent $-$8),
is partially-ionised, and moves with a velocity of 500-1000\,\kms\ (Fig.~\ref{fig_map_modc}).
It is the scattering with free electrons in these photospheric layers that cause the presence at early times
of the broad symmetric wings on H\one\ and He\one\ lines. This configuration, which is based
on a radiation-hydrodynamics model, is physically equivalent to the heuristic setup
of \citet{dessart_etal_09} for SN\,1994W.

As time progresses, the line profiles lose their extended wings and become narrower (see, e.g.,
H$\alpha$; Fig.~\ref{fig_spec_modc_94w_ha}).
This arises because the spectrum formation region is more confined to the recombination front, with hot/ionised
material below and cool/neutral material above (Fig.~\ref{fig_modc_2}).
The spectral energy distribution becomes redder, and
lines from Ca\two, Na\one, Sc\two, Ti\two, and Fe\two\ strengthen. Broad lines are absent at all times
because the bulk of the radiation arises from slow moving material (shocked material from the interaction
but also unshocked material in the outer shell) --- the representative expansion rate is $\sim$\,1000\,\kms.
This is also the origin of the absorption in all line profiles at a Doppler velocity $\sim-$\,1000\,\kms\
(see, e.g., H$\alpha$; Fig.~\ref{fig_spec_modc_94w_ha}).
Initially, the absorption is partially filled-in (the residual flux is above the continuum level) because of
photon frequency redistribution by scattering with thermal electrons.
At later times (spectra for days 130 and 160), this absorption appears as a standard P-Cygni
absorption (with a minimum flux below the continuum level).
All these properties match the spectral evolution of SN\,1994W (bottom panel), and are in stark contrast
with model A (and SN\,1998S).

\begin{figure}
\epsfig{file=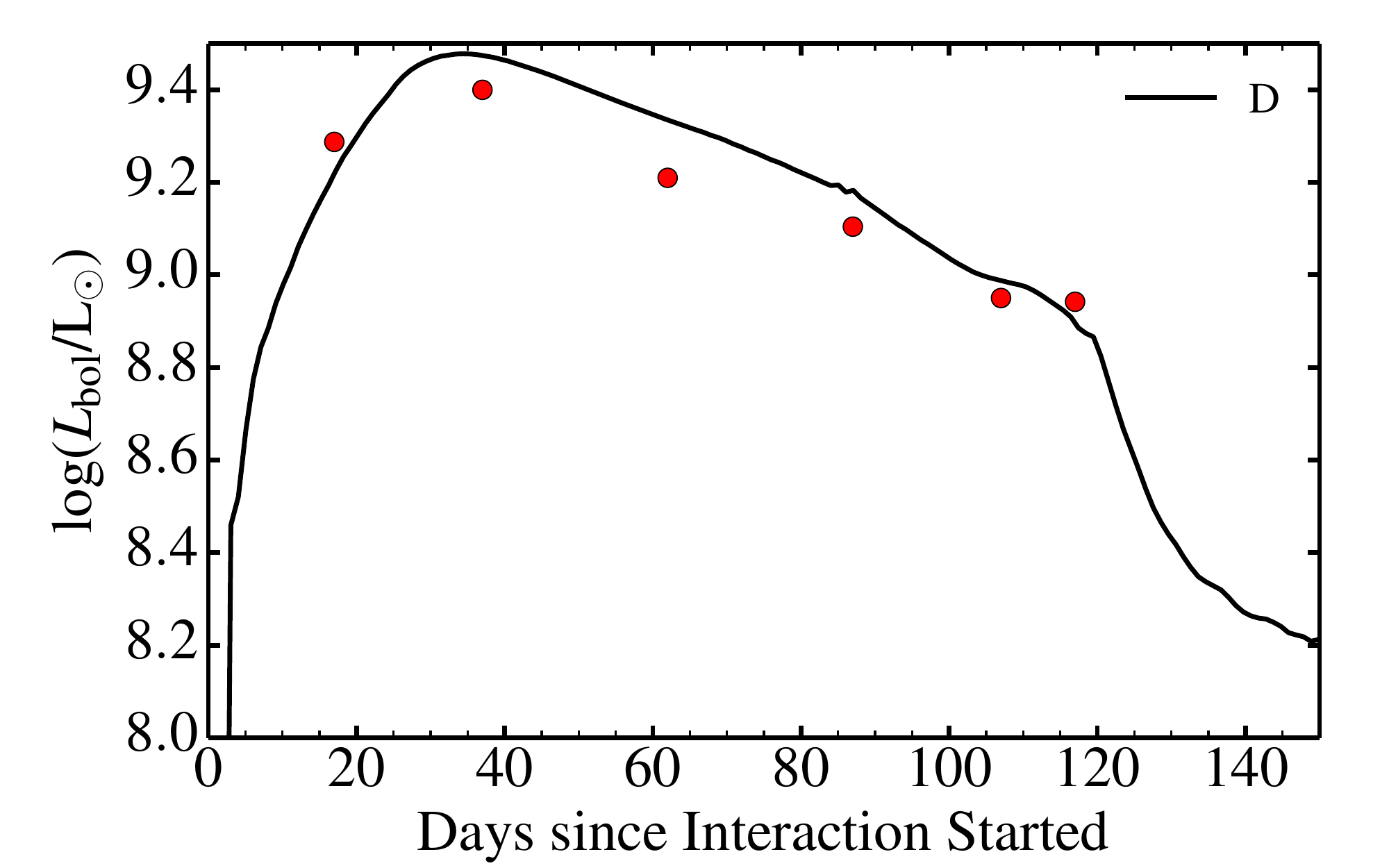, width=0.47\textwidth}
\caption{Bolometric light curve for model D (the dots
correspond to the luminosity computed at corresponding epochs with \cmfgen).
\label{fig_lbol_modd}
}
\end{figure}

Despite the good agreement between the synthetic spectra of model C and the observations
of SN\,1994W, the model light curve is somewhat discrepant. While Model C predicts a sustained
bolometric luminosity for up to 200\,d, the high brightness phase of SN\,1994W lasts for about
100\,d (Fig.~\ref{fig_lc_mod_c_d_obs}).
As mentioned earlier, the parameter space covered by interacting SNe is likely very large.
Nuclear flashes in 8-12\,\msun\ stars are an attractive scenario for interacting SNe because they
offer a natural explanation for the short delay between the ejection of the H-rich envelope and the
terminal explosion following core collapse. So, rather than considering alternate scenarios (different
progenitor mass, shell eruptions etc), we have
slightly modulated the initial configuration of model C to test how we could reconcile the
light curve of model C with the observations of SN\,1994W, while retaining the same spectral evolution
properties. We find that the discrepancy is reduced if we shift the interaction to larger radii. In the next
section, we illustrate the results for one such model, named model D.

\begin{figure*}
\epsfig{file=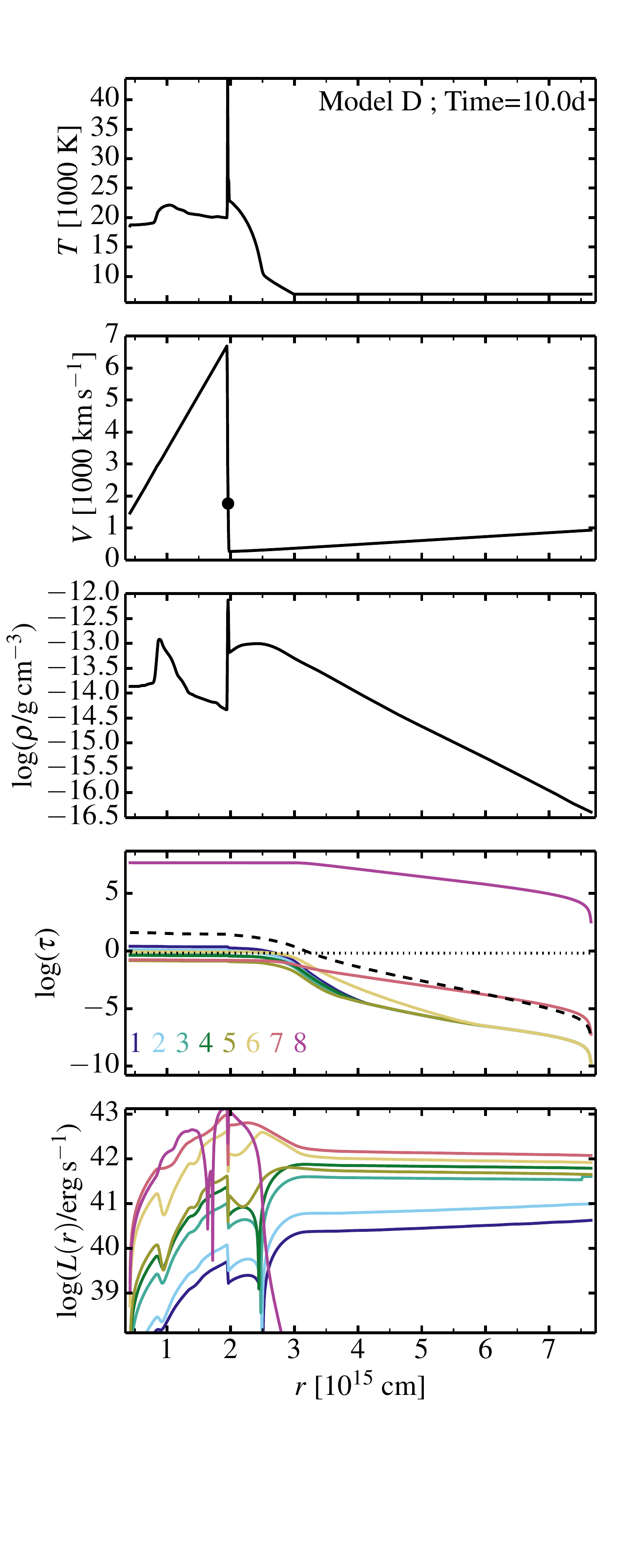, width=0.47\textwidth}
\epsfig{file=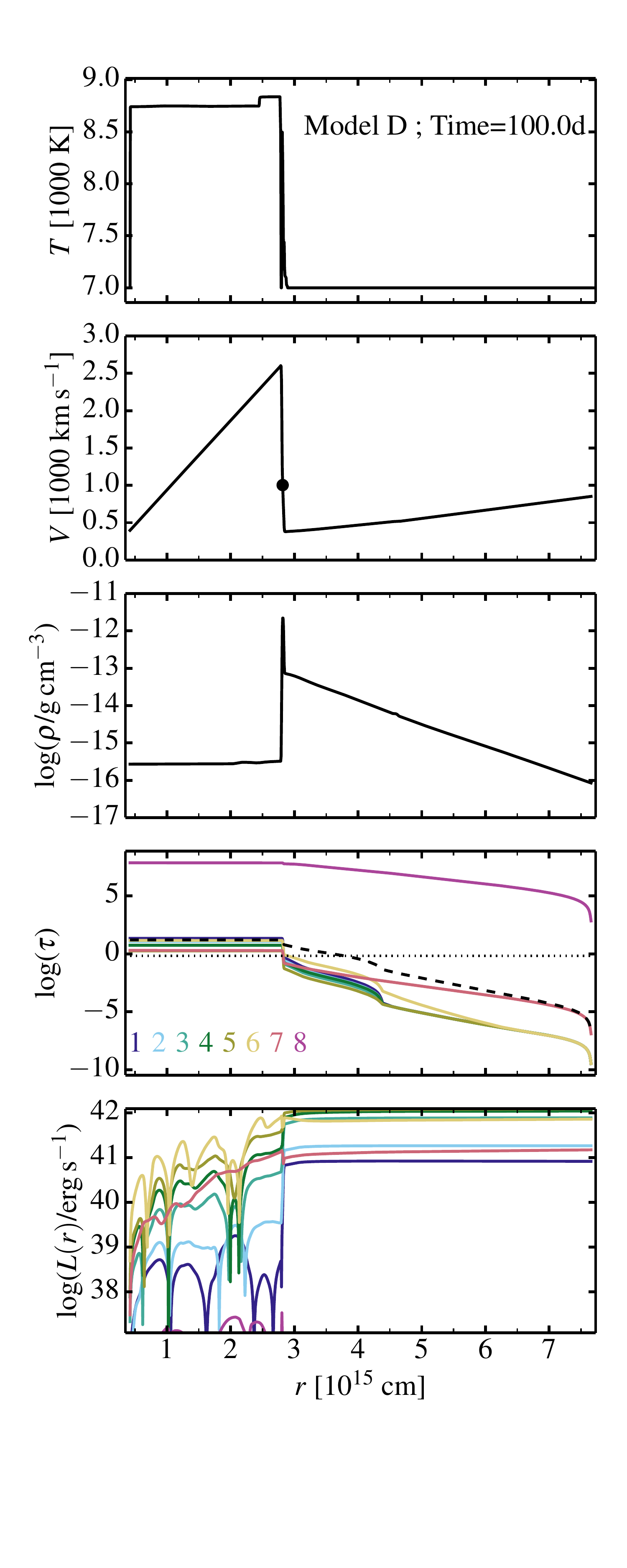, width=0.47\textwidth}
\vspace{-2cm}
\caption{Same as for Fig.~\ref{fig_modc_1}, but now for model D at 10.0\,d (left) and 100.0\,d (right)
after the onset of the interaction.
At late times, the unshocked inner shell mass is $\lesssim$\,0.01\,\msun\ while the CDS mass is $\approx$\,3\,\msun.
The dot in the velocity panel (second row from top) corresponds to the velocity of the CDS.
\label{fig_modd_her}
}
\end{figure*}

\subsection{Results for model D}

  The initial configuration of model D is very similar to that of model C except for the initial interaction radius $R_{\rm t}$,
  which we increase from $2 \times 10^{14}$\,cm to $1.8 \times 10^{15}$\,cm.
  Compared to model D, the interaction region in model C is initially much more compact so the inner/outer shell densities
  and temperatures are higher.
  Interaction starts when the outer shell is still optically thick. The length scales are smaller so the deceleration of the
  inner shell is very rapid --- the shock in model C dies within 10\,d of the onset of interaction.
  In model D, the interaction region is much more extended. The outer shell age is $\gtrsim$\,7 times greater than in model C,
  so its density is about 400 times lower (each shell is explosively produced and in homologous expansion at the onset of
  interaction). Were it not cold and recombined, the outer shell optical depth would be  $\gtrsim$\,50 times
  lower than in model C. The characteristic time for the CDS to sweep up a given mass of CSM is also $\gtrsim$\,7 longer.
  So, even though the shell masses and energies are comparable in models C and D, the dynamics and radiative properties
  of the interaction are distinct.

  We show in Fig.~\ref{fig_map_modd} a set of greyscale images for the velocity, temperature, density, and luminosity
  versus radius and time for model D.
  Figure~\ref{fig_lbol_modd} shows the bolometric light curve of model D computed with \heracles.
  We also show snapshots of these interaction properties at 10.0 and 100.0\,d
  (Fig.~\ref{fig_modd_her}).

   Compared to model C, the shock is still present at the end of the simulation at 150\,d. But just like in model C,
   the inner shell material is strongly decelerated. Because of the contrast in inner/outer shell mass, the fast material
   in the inner shell at late times has a very low density.\footnote{Recall here that the \heracles\ code is Eulerian
   so we actually inject low density
   material at the ``inflow" inner boundary. The innermost point of the inner shell is initially at 1500\,\kms. At 150.0\,d,
   the only part of the original inner shell that remains is between 1500 and 2000\,\kms.}
   At 150.0\,d, the CDS mass is about 3\,\msun\ and the residual fast inner shell material contains $\lesssim$0.01\,\msun.

\begin{figure*}
\epsfig{file=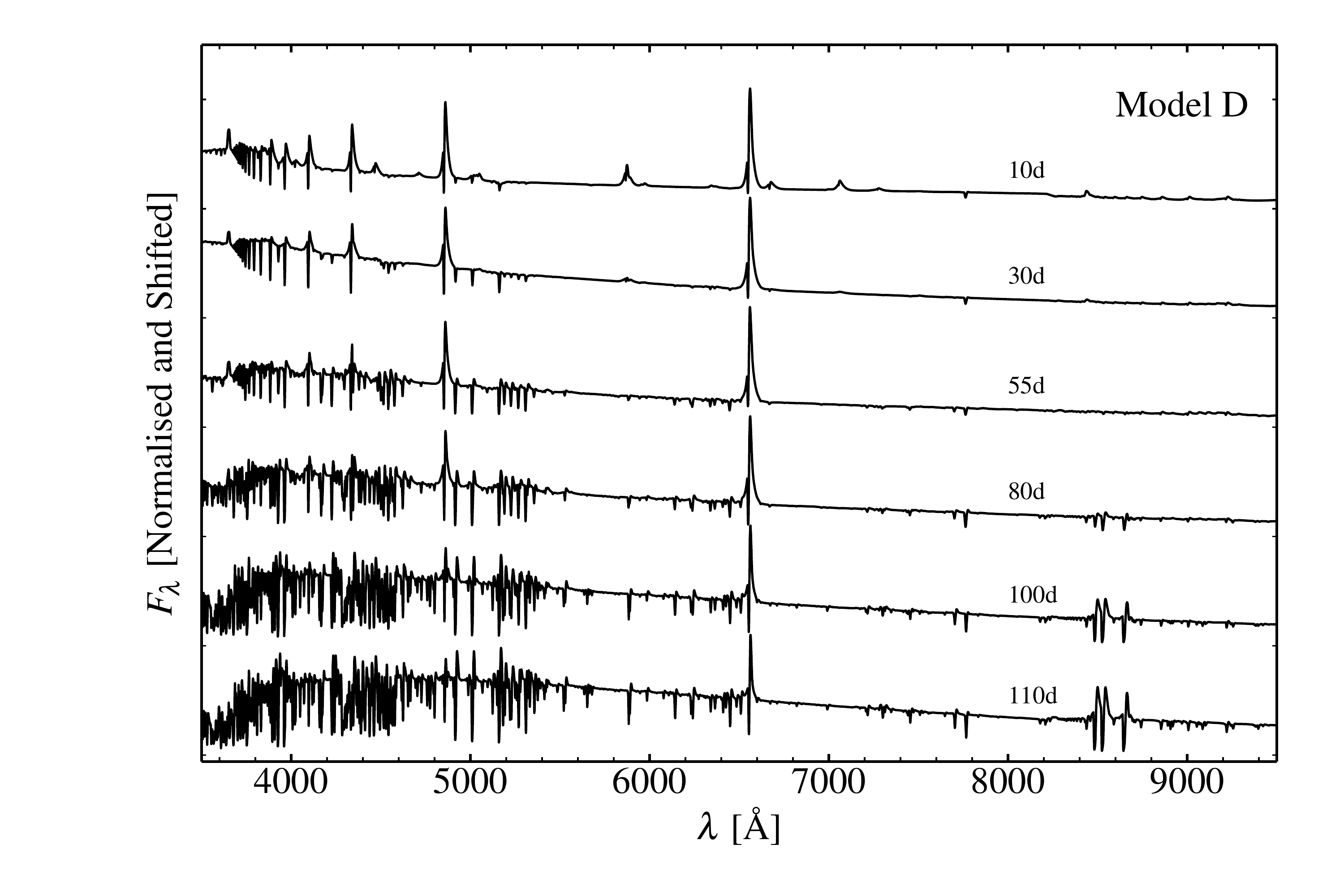, width=\textwidth}
\vspace{-1.cm}
\caption{Spectral evolution of the double explosion model D. For each spectrum, the label indicates
the time since the onset of interaction.
\label{fig_spec_modd_94w}
}
\end{figure*}

   Because the shock persists for the entire duration of the simulation, the shock luminosity is a continuous power source
   for the light curve. Initially, about half of the outer shell mass is quickly ionised,
   not directly by Lyman photons because they suffer a large optical depth,
   but indirectly through absorption of lower energy photons. Within 20\,d, the photosphere has migrated
   out to $4 \times 10^{15}$\,cm in the outer shell and will remain at that location until the CDS overtakes that radius at $\gtrsim$150\,d.
   Relative to model C, the volume of ionised material is larger and the energy from
   the shock is released for longer so the temperature in the interaction region
   is lower (it peaks at $\sim$\,30\,000\,K, rather than $\gtrsim$\,100\,000K).

   Because of the slow expansion of the outer shell (in both model C and D), the densities are large enough to make the material
   optically thick to electron scattering. The limiting factor is the ionisation. In model D, the balance between heating by the
   shock and cooling from radiation and expansion permits the ionisation of a sizeable fraction of the outer shell.
   This ionised layer is present during the entire high-brightness phase of model D. So, we expect to see
   persistent signatures of electron scattering. Furthermore, by 60\,d after the onset of the interaction, the CDS velocity is
   down to $\approx$1000\,\kms\ while its optical depth is about 25. So, just like in model C, this configuration should
   not give rise to broad spectral lines at any time.

   The lower optical depths achieved in model D produce a faster rising light curve, a brighter peak, and a shorter
   high brightness phase than in model C. The correspondence with the observed light curve of SN\,1994W is
   now satisfactory (Fig.~\ref{fig_lc_mod_c_d_obs}). Besides kinetic energy, the transition radius seems an important
   tuning parameter controlling the light curve of this type of interactions and it may be in part
   responsible for the observed diversity of SNe IIn (the light curves of SN\,1994W and SN\,2011ht differ
   sizeably, yet they exhibit a very similar spectral evolution; \citealt{humphreys_etal_11ht}).

\begin{figure}
\epsfig{file=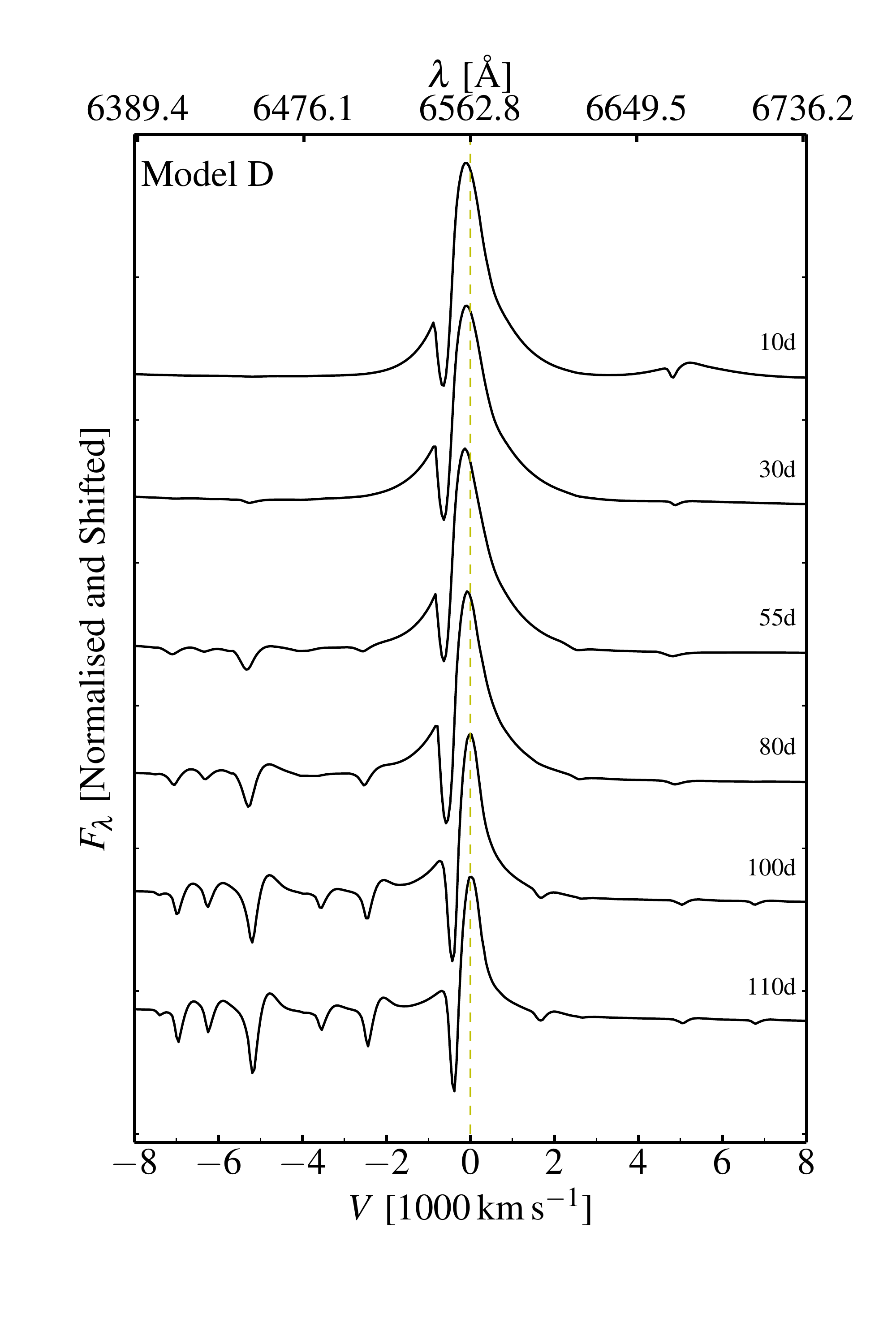, width=0.45\textwidth}
\vspace{-1.cm}
\caption{
Evolution of the H$\alpha$ region for the double  explosion model D.
For each spectrum, the label indicates the time since the onset of interaction.
\label{fig_spec_modd_94w_ha}
}
\end{figure}

\begin{figure}
\epsfig{file=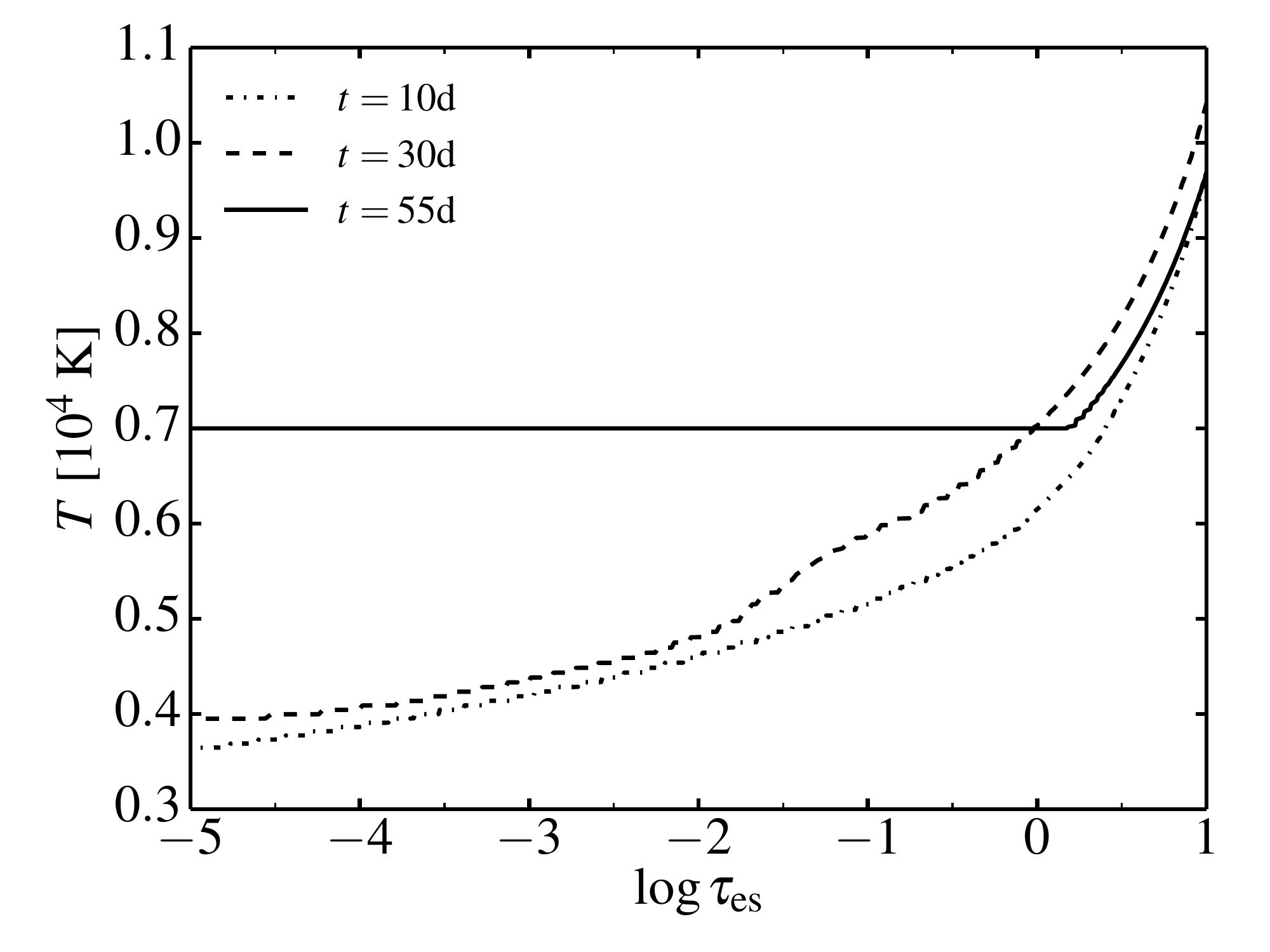, width=0.45\textwidth}
\caption{Temperature profile versus electron-scattering optical depth in model D at
10, 30, and 55\,d after the onset of interaction. Notice the higher gas temperature in
the spectrum formation region (at optical depth between 1 and 10) at the time of maximum ($t\sim$\,30\,d),
which is also when the spectrum is bluer.
\label{fig_f22}
}
\end{figure}

   Figure~\ref{fig_spec_modd_94w} shows the spectral evolution for model D, which is very analogous
   to that of model C. Because of the larger radius for interaction, the temperature of the ionised outer shell and of the CDS
   is lower. This causes the spectral energy distribution to be redder than for model C early on.
   Importantly, broad lines are never seen, as typified by H$\alpha$ (Fig.~\ref{fig_spec_modd_94w_ha}).
   Before light curve maximum, the rate at which radiation escapes the CSM is lower than the rate at which
   it is released at the shock because of optical-depth effects. Consequently, the temperature rises in the optically-thick
   CSM until a maximum around bolometric maximum. Afterwards, as long as the optical depth remains sizeable,
   the rate at which radiation escapes the CSM is greater than the rate at which it is released at the shock and the
   temperature decreases (Fig.~\ref{fig_f22}). This effect, characteristic of the diffusion process here, is seen in model D
   (and also in model X of D15) and is also observed in SN\,1994W \citep{sollerman_etal_98,chugai_etal_04}.
   Indeed, the spectra progress towards higher colour temperatures until maximum and towards
   lower colour temperatures afterwards (they increasingly redden with time), showing signs of recombination and increasing
   line blanketing. All these properties reproduce satisfactorily the observations of SN\,1994W.

\section{Conclusions}

In this paper, we have presented a series of radiation-hydrodynamics and radiative-transfer simulations
in order to explore the origin of the spectral diversity of interacting SNe.
The study is not exhaustive but it captures the salient features seen in events like SN\,1998S on the
one hand, and SN\,1994W on the other hand.

The main impetus for this work was the perplexing absence of broad lines at all times in the spectra
of SN\,1994W. In \citet{dessart_etal_09}, we obtained a very good match to the spectral observations of SN\,1994W
by invoking a hydrogen-rich slow optically-thick shell, in which the bulk of the radiation is emitted.
We also proposed that the electron-scattering wings were formed internally to this thick shell, rather
than externally into some CSM.
Our proposed scenario was not easily adaptable to the results of \citet{chugai_etal_04}, who argued
for an interaction between a 7-12\,\msun\ $\sim$\,10$^{51}$\,erg inner shell with a 0.4\,\msun\ dense wind CSM.
The difficulty with that model is to explain the lack of broad lines at late times.
In a recent study on super-luminous SNe like 2010jl \citep{d15_10jl}, we found that the emission
from the dense shell eventually appears, even with a massive and dense CSM, and the associated line profiles
are Doppler broadened, with a velocity representative of the velocity of the dense shell.
For SN\,1994W, the Chugai et al. model predicts a dense shell moving at $\sim$\,4000\,\kms, which
is in tension with the $\sim$\,1000\,\kms\ width of spectral lines observed at late times.

Here, we resolve this issue. Revisiting the model of Chugai et al. we find that it is more
suitable to explain events like SN\,1998S.
At early times, when the CSM optical depth is large, the CDS and the SN ejecta are all obscured.
Some spectral lines appear with a distinct IIn morphology, with a narrow core and extended wings,
all arising from photons re-processed within the slow unshocked CSM.
Most lines show a second
and more narrow component, associated with the more distant and cool CSM.
However, as time progresses and the CSM optical depth becomes only a few, a large fraction of
the photons emitted by the CDS is no longer re-processed by the CSM. For a CDS velocity
of a few 1000\,\kms, the Doppler effect dominates line broadening.
In our model A (and by extension the corresponding model of Chugai), the SN spectrum evolves
from a IIn morphology to a pure absorption spectrum, where the flux forms within the CDS.
The lack of an emission component stems from the very steep density profile at the (outer) edge of the CDS.
The SN enters a third phase when the CDS  becomes optically thin.
The SN radiation then comes primarily from the inner slower unshocked ejecta, while the CDS located further out
contributes primarily through broad emission in H$\alpha$ and the Ca\two\ triplet at 8500\,\AA.

These three phases  predicted by our model A are observed in SN\,1998S (with slight temporal offsets).
Our analysis demonstrates that the CDS is not obscured at late times. More generally, the
CDS is likely seen in all interacting SNe at late times.
But the implication is that SNe like 1994W, which never show broad lines, cannot be explained
by an interaction involving a massive energetic ejecta ramming into a less massive CSM since it would produce
a fast moving CDS. As our grid of explosions has shown, configurations involving an inner shell
of moderate mass and energy interacting with a massive slow outer shell  (e.g., models R1 and B1)
produce a CDS with a low velocity, compatible with the observations of SN\,1994W.

Such a configuration can probably arise in many different ways in nature.
Here, we invoke the context of nuclear flashes in low-mass massive stars \citep{weaver_woosley_79, WH15}.
We produce two subsequent explosions in a low-mass
RSG star of  12\,\msun\ on the main sequence. The first explosion, which mimics the effect of a nuclear flash,
ejects the loosely-bound H-rich envelope, producing a $\approx$\,6\,\msun\ ejecta with a mean mass-weighted velocity
of $\approx$\,400\,\kms.
The second explosion corresponds to the terminal gravitational collapse of the star left behind,
producing a 0.3\,\msun\ ejecta with a kinetic energy of the order of 7-8$\times$10$^{49}$\,erg
ejecta with a mean mass-weighted velocity of 4740\,\kms.
We perform two simulations (models C and D) with an initial radius of interaction at 0.2 and $1.8 \times 10^{15}$\,cm.

This mass/energy configuration leads to a very different type of interaction compared to model A,
while the differing interaction radius between models C and D introduce more subtle differences.
In model C, the low-mass inner shell is completely
decelerated within 10\,d of the onset of the interaction. About 50\% of its kinetic energy is converted
to internal energy, trapped in the interaction region.
Without this interaction, the first exploded shell would appear as a very faint ($\sim$\,10$^7$\,\lsun)
narrow-line SN II-P (such events likely exist and are missed by current surveys).
With the interaction the internal energy of the outer shell is boosted, causing its temperature to increase.
A sizeable part of the outer shell material that had recombined
is then re-ionised, pushing the photosphere out in both mass and radius.
The combination of low velocity, large radii, and partial ionisation above the photosphere lead
to a IIn spectral morphology at early times. Importantly, as the ejecta cools and recombines,
the spectrum reddens, the electron scattering wings weaken, the photosphere recedes, and
the line profiles narrow. All these properties are observed in SN\,1994W and shown by model C.
Model C is, however, somewhat discrepant because it underestimates the bolometric luminosity
of SN\,1994W, and then overestimates it at late times: model C remains luminous for 200\,d after the
onset of interaction. This discrepancy is cured with model D by invoking a larger initial interaction radius.

In model D, the length scales are much larger so the deceleration of the inner shell takes longer.
The interaction region has a lower optical depth, producing a light curve that now matches satisfactorily
the observations of SN\,1994W. Although the shock survives throughout the high brightness phase,
the CDS velocity is never large, and quickly drops to $\sim$\,1000\,\kms, as in model C.
The spectral evolution of models C and D are therefore comparable, and compatible to that of SN\,1994W
(because of a lower gas temperature in the CSM, the spectral evolution of Model D reproduces
better the observations of SN\,1994W than model C, for example with weaker and
shorter-lived He\one\ lines, like He\one\,6678\,\AA).
The spectral simulations of \citet{dessart_etal_09}, which were not based on a radiation-hydrodynamical model,
correspond to a configuration similar to model D.

While we were finalising this manuscript and preparing for submission,
a study of SN\,2011ht by \citet{chugai_15} came out. Chugai's work has a lot of overlap
with the present work but it uses an independent and very different approach. Still, the conclusions
by and large agree and give further support to the notion that SNe like 1994W or 2011ht result from
the interaction of a low mass ejecta with an extended, slow, and massive outer shell.
One distinction is that Chugai argues for fragmentation to explain the $\sim$\,100\,d high-brigthness
phase of SN\,2011ht while we argue that the duration of the high-brightness phase can be
modulated simply by invoking a different interaction radius (model D versus model C).

Models that invoke an interaction of two shells can produce large luminosities.
This is because SNe generally have a huge kinetic energy reservoir, which exceeds by a factor
of about 100 the time-integrated luminosity of a standard SN.
Interaction with CSM is an efficient process to tap into this energy reservoir.
The norm in an interacting SN is therefore to be super-luminous.
In the model of Chugai et al., the CSM is much less massive than the inner shell, so the
deceleration is weak and the high energy available in the inner shell is hardly tapped. But the consequence
is the formation of a fast moving CDS, which we demonstrated is incompatible with the
observations of SN\,1994W.
Both SN\,2011ht and SN\,2011A \citep{dejaeger_11A_15}
show a similar spectral evolution to SN\,1994W
but are characterised by even lower peak visual magnitudes (of the order of $-$16\,mag).
Our models C and D, and possible incarnations of that scenario that retain the principle of a low/moderate-mass
energetic inner shell exploding within a massive slow moving outer shell, seem much more
suited to explain these events than models in which a massive inner shell
rams into a CSM of much lower mass.
One plausible circumstance for this type of interaction is a nuclear flash
shortly before core collapse in a low mass massive star \citep{WH15}.

The designation of SN\,1994W-like events as SN IIn-P \citep{mauerhan_11ht_13}
aims to distinguish them from the broader diversity of SNe IIn. But the designation is somewhat controversial,
in part because light curves tend to be degenerate --- distinct models can produce identical light curves.
The nebular flux of SN\,1994W may be explained
by invoking a low amount of \iso{56}Ni, but it can also be explained by invoking
an interaction with a lower density CSM and no \iso{56}Ni at all.
The origin of the nebular flux is ambiguous, especially in an object that showed strong signs of interaction early on,
and therefore it cannot be decisive.
The ``P'' might suggest a connection to SNe II-P, but these result from a point explosion
in a RSG star and are thus distinct.
In our SN IIn simulations (and in observations), the post-maximum brightness can show a range of decline rates,
often incompatible with a plateau designation.
Interestingly, the light curves of SN\,1998S and SN\,1994W are somewhat similar. What is strikingly
different between the two types of events is the spectral evolution. SN\,1994W-like
events differ from all other SNe by the presence of narrow lines {\it at all times} --- SN\,1998S shows narrow lines
at early times but broad lines (reminiscent of a SN II-P/II-L) at late times.
This feature is missed in the IIn classification since many SNe IIn exhibit narrow lines only for a short time.
So, perhaps a better designation for SN\,1994W-like events would be as type IInn, to emphasize the
persistence of narrow lines at all times.

What distinguishes narrow line SNe II-P from SNe IIn like 1994W are their low luminosity:
a fresh supply of energy to a low-energy SN II-P boosts its luminosity but hardly affects its
expansion rate. So SNe IIn that would form through the scenario of model C should be characterised
by a large brightness for their inferred expansion rate, which would go against the correlation
between optical brightness and expansion velocity seen in SNe II-P \citep{HP02}.
LSQ13fn is a SN that exhibits such peculiar properties, including evidence for ejecta/CSM interaction
at early times \citep{polshaw_lsq13fn_15}.

\section*{Acknowledgments}

We thank Jesper Sollerman and Robert Cumming for providing the photometric data for SN\,1994W.
We also thank Roberta Humphreys for providing the spectra for SN\,2011ht.
Part of this work was done during a one-month
visit to ESO-Santiago as part of the ESO Scientific Visitor Programme.
LD acknowledges financial support from ``Agence Nationale de la Recherche"
grant ANR-2011-Blanc-SIMI-5-6-007-01.
DJH acknowledges support from STScI theory grant HST-AR-12640.01, and NASA theory
grant NNX10AC80G.
This work was also supported in part by the National Science Foundation under
Grant No. PHYS-1066293.
This work utilised computing resources of the mesocentre SIGAMM,
hosted by the Observatoire de la C\^ote d'Azur, Nice, France.

\appendix

\section{Initial interaction configurations for Models A and C}
\label{sect_init_mod}

This section contains figures describing the initial conditions for the
\heracles\ simulations for models A, C, and D.

\begin{figure*}
\epsfig{file=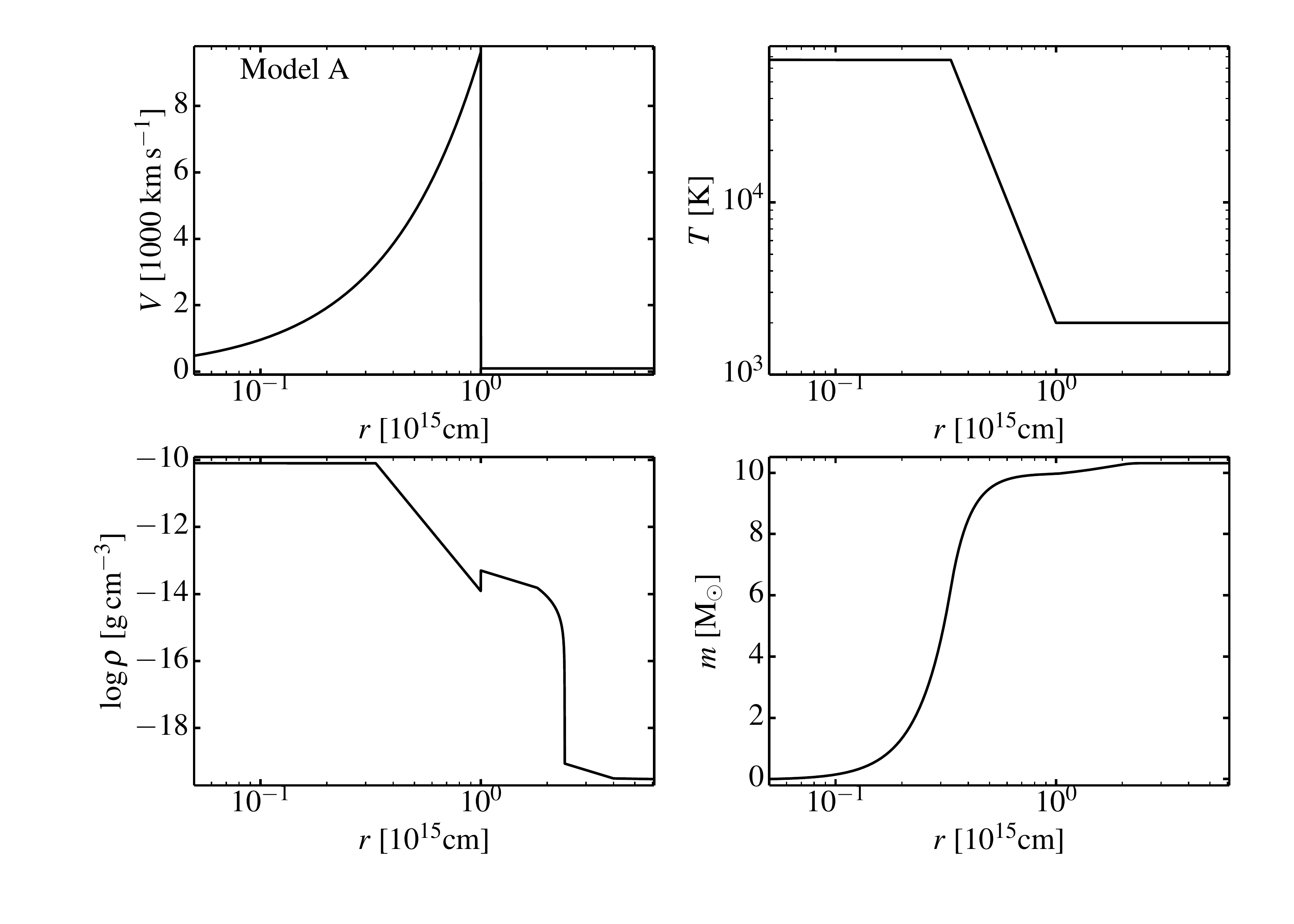,width=0.8\textwidth}
\vspace{-0.7cm}
\caption{Interaction configuration of model A used as initial conditions for the \heracles\ simulations.
\label{appendix_fig_moda_init}
}
\end{figure*}

\begin{figure*}
\epsfig{file=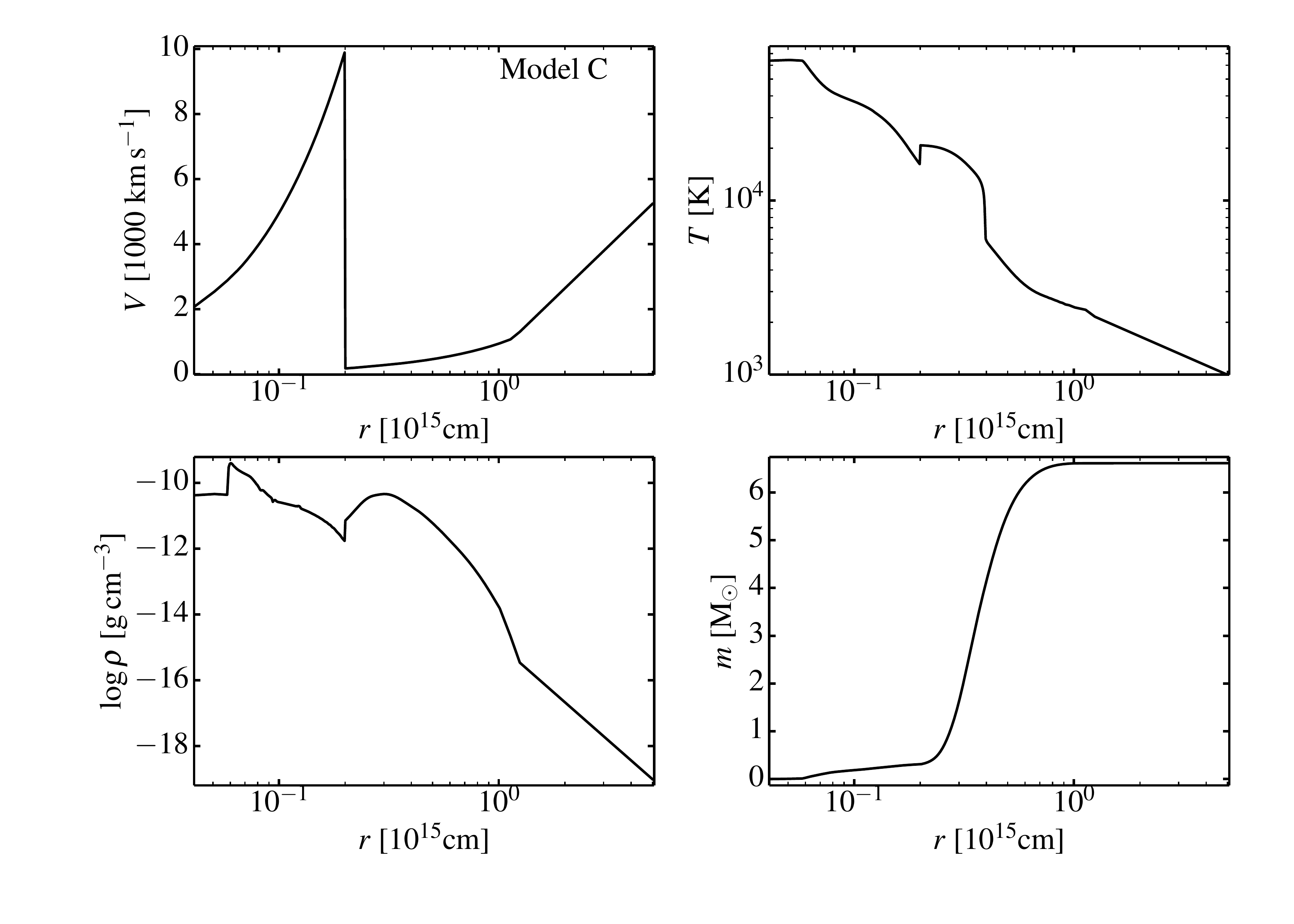, width=0.8\textwidth}
\vspace{-0.7cm}
\caption{Interaction configuration of model C used as initial conditions for the \heracles\ simulations.
The outer shell and the inner shell were simulated separately with \v1d\ to mimic a nuclear flash
at the base of the H-rich envelope and the final core collapse, respectively.
\label{fig_modc_init}
}
\end{figure*}

\begin{figure*}
\epsfig{file=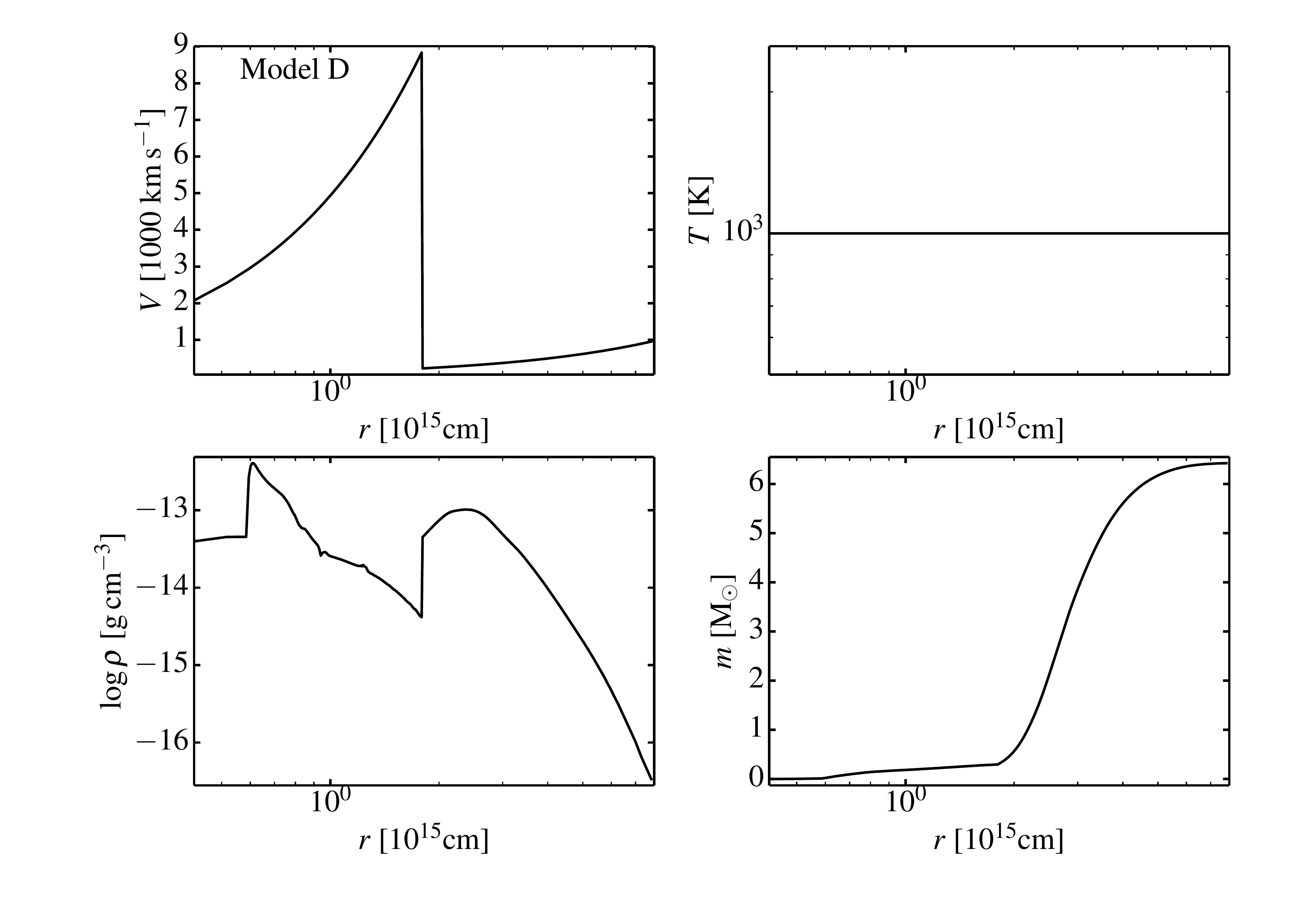, width=0.8\textwidth}
\vspace{-0.7cm}
\caption{Interaction configuration of model D used as initial conditions for the \heracles\ simulations.
The outer shell and the inner shell were simulated separately with \v1d\ to mimic a nuclear flash
at the base of the H-rich envelope and the final core collapse, respectively.
This model is similar to model C, but the radius where the two shells start interacting is located
further out, at about 1.7$\times$\,10$^{15}$\,cm. Both shells are optically thin
and cold initially --- for convenience we reset the initial temperature to a floor value of 1000\,K.
\label{fig_modd_init}
}
\end{figure*}

\section{dfr plots for models A}
\label{sect_dfr_moda}

This section provides additional information on the spectrum formation computed by \cmfgen\
for the interaction configuration of model A simulated with \heracles. Details on how to interpret these
figures can be found in \citet{d15_10jl}.

\begin{figure*}
\epsfig{file=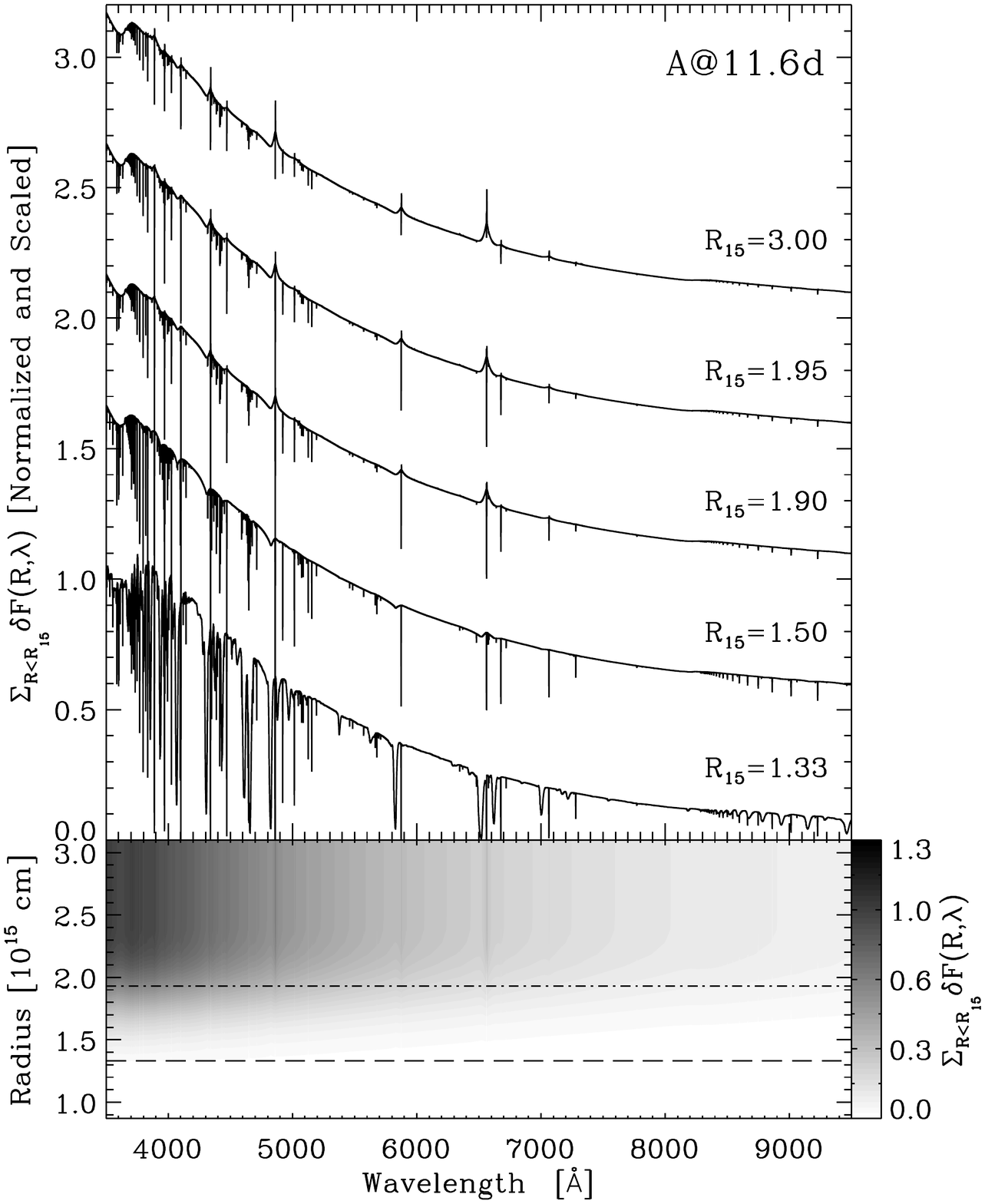,width=0.4\textwidth}
\epsfig{file=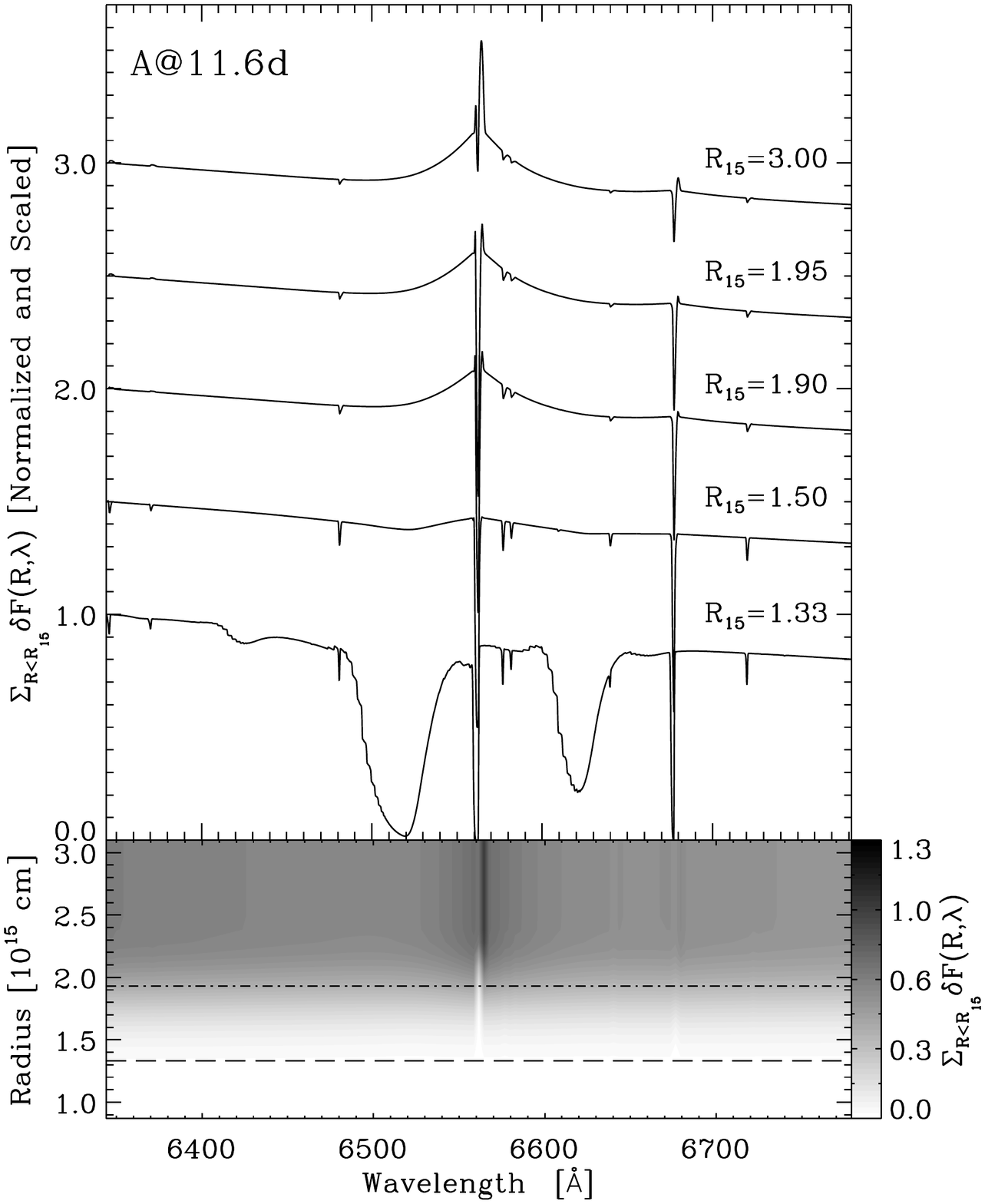, width=0.4\textwidth}
\vspace{-0.7cm}
\caption{
Illustration of the wavelength ($\lambda$) and depth ($R$)
dependence of the quantity $\sum_{R<R_{15}} \delta F(R,\lambda)$ for model A
at 11.6\,d after the onset of interaction.
Here, $\delta F(R,\lambda)$ represents the contribution to the observer's flux at wavelength
$\lambda$ originating in a shell of width $\Delta R$ at radius R.
It is defined by
$\delta F(R,\lambda) = (2\pi/D^2) \int_{\Delta R} \, \Delta z  \, \eta(p, z,\lambda) \, e^{-\tau(p, z,\lambda)} p dp$;
$R_{15}$ is $R$ in units of 10$^{15}$\,cm; $D$ is the distance; $\Delta R$ and $\Delta z$ are the shell thickness
in the radial direction and along the ray with
impact parameter $p$, respectively; $\eta(p, z,\lambda)$  and $\tau(p, z,\lambda)$ are the emissivity and the
ray optical depth at location $(p,z)$ and wavelength $\lambda$.
The grey scale in the bottom panel shows how $\sum_{R<R_{15}} \delta F(R,\lambda)$ varies as we progress outwards
from the inner boundary, indicating the relative flux contributions of different regions.
If we choose $R_{15}$ as the maximum radius on the \cmfgen\ grid, we recover the total flux.
The dash-dotted line corresponds to the radius of the electron-scattering photosphere,
and the dashed line  corresponds to the CDS radius.
The top panel shows selected cuts (see right label) of the quantity $\sum_{R<R_{15}} \delta F(R,\lambda)$.
From top to bottom ($R_{15}=$\,3.0, 1.95, 1.90, 1.50, and 1.33), $\sum_{R<R_{15}} \delta F(R,\lambda)$ represents
1, 0.83, 0.73, 0.19, and 0.003 of the total emergent flux at 6800\,\AA.
{\it Right:} Same as left, but now zooming in on the H$\alpha$ region.
At this epoch, the bulk of the observed radiation originates in the CSM.
\label{fig_dfr_w0s_11p6}
}
\end{figure*}

\begin{figure*}
\epsfig{file=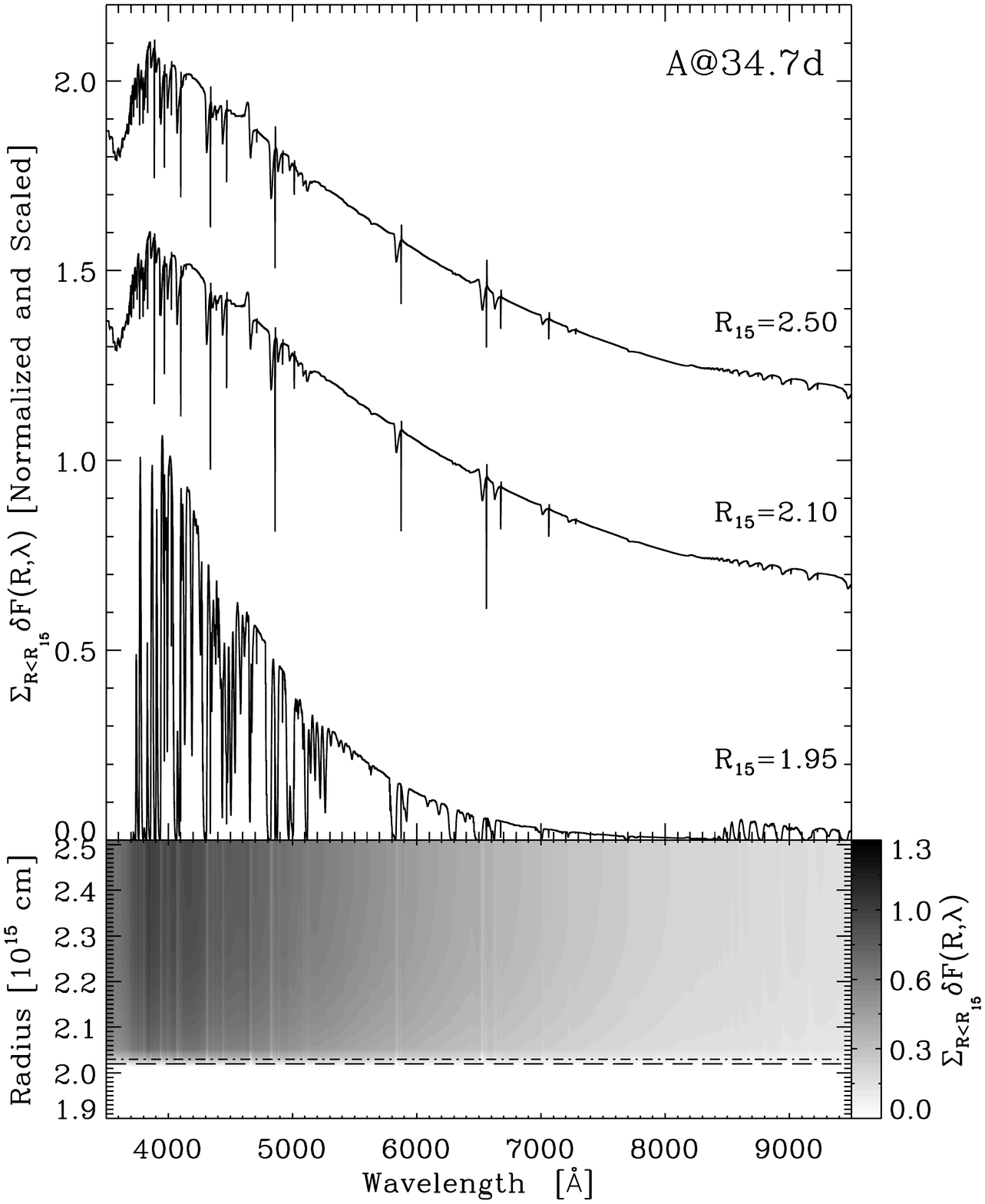, width=0.4\textwidth}
\epsfig{file=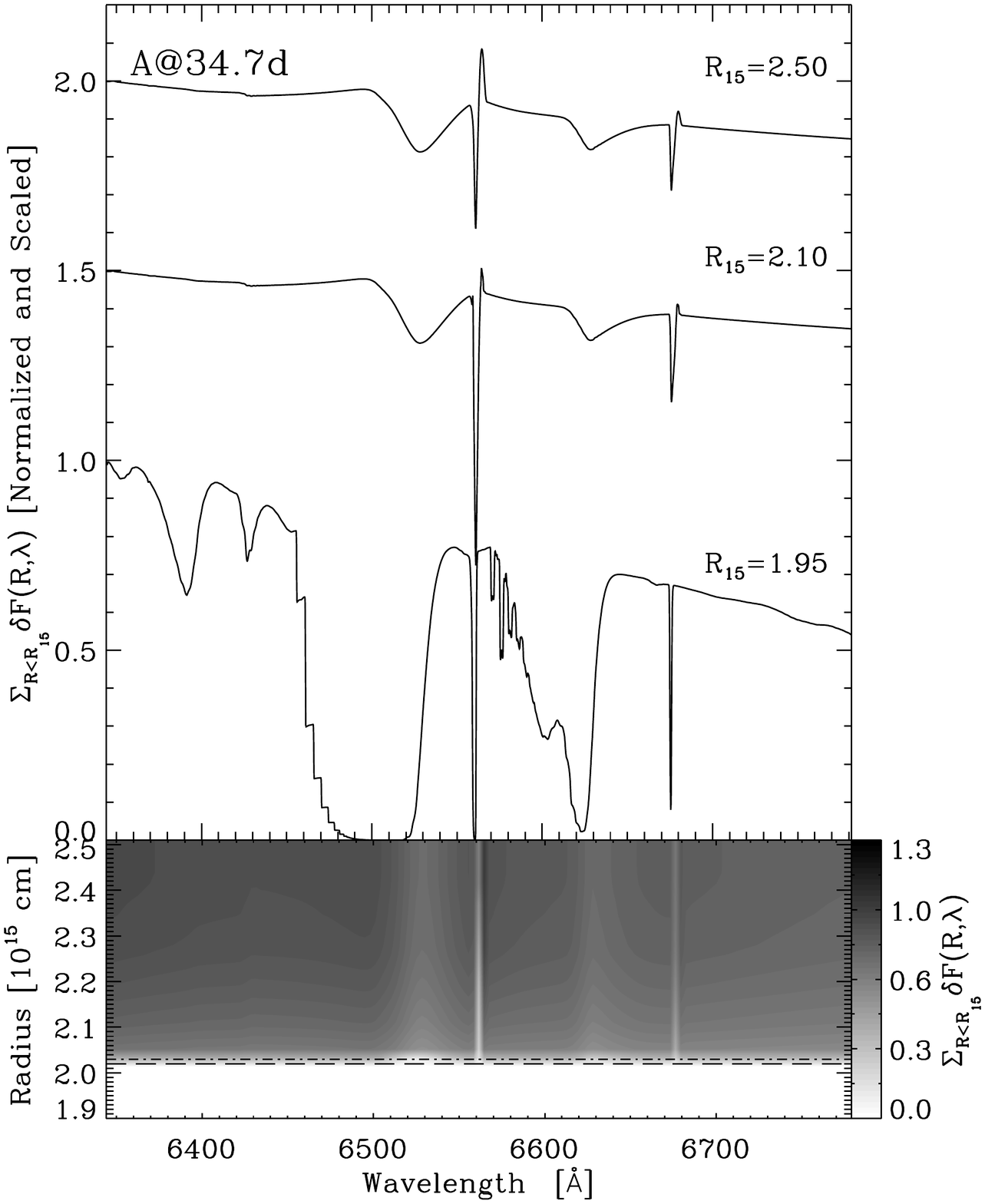, width=0.4\textwidth}
\vspace{-0.7cm}
\caption{Same as Fig.~\ref{fig_dfr_w0s_11p6}, but now at 34.7\,d after the onset of interaction.
In the bottom panel and from top to bottom ($R_{15}=$\,2.5, 2.1, and 1.95), $\sum_{R<R_{15}} \delta F(R,\lambda)$
represents 1, 0.95, and 0.0001 of the total emergent flux at 6800\,\AA.
The CDS is optically thick, and most of the observed radiation
originates in the CDS. The electron scattering photosphere (i.e., where $\tau_{\rm es}=2/3$)
now lies within the CDS.
\label{fig_dfr_w0s_34p7}
}
\end{figure*}

\begin{figure*}
\epsfig{file=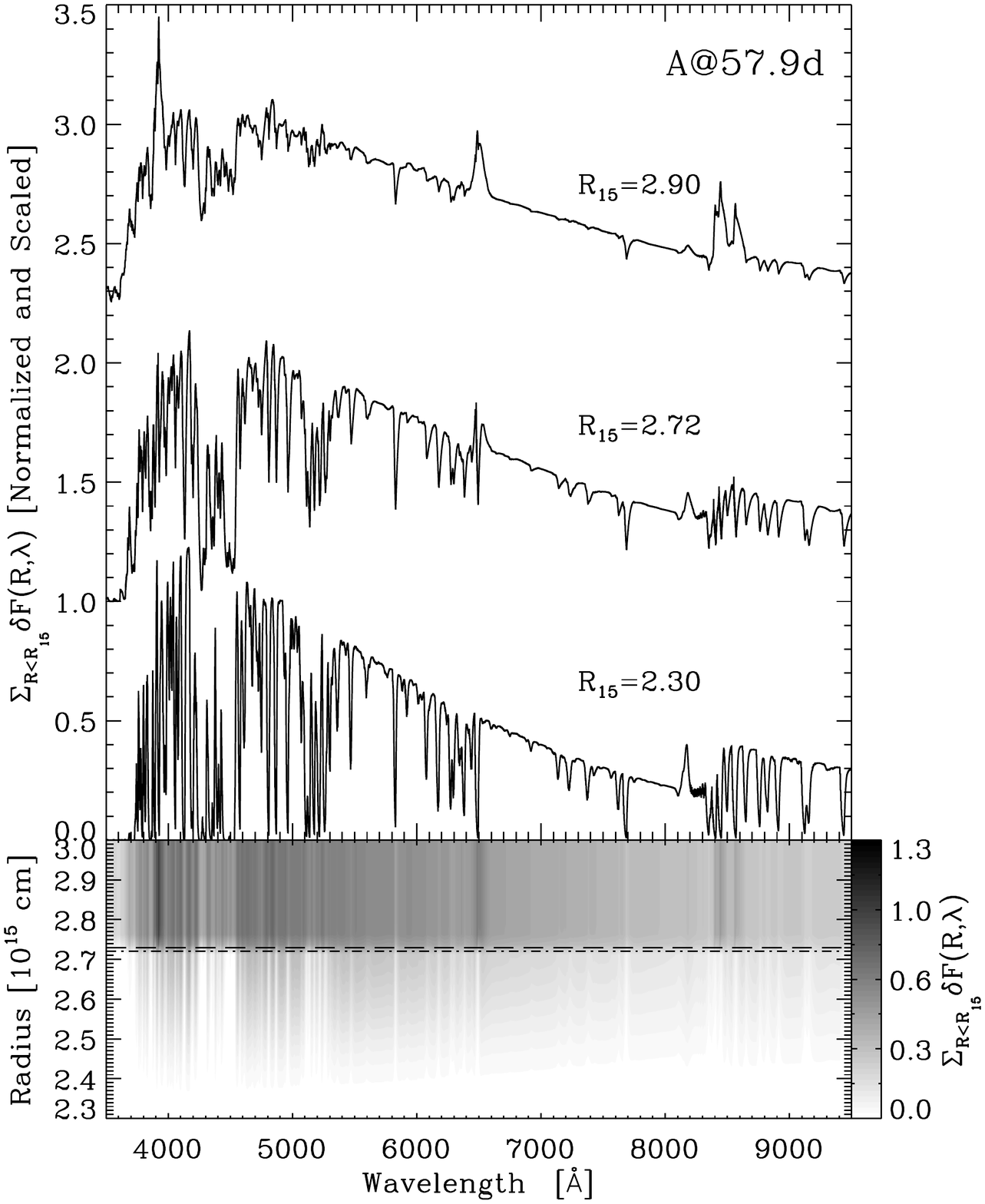, width=0.4\textwidth}
\epsfig{file=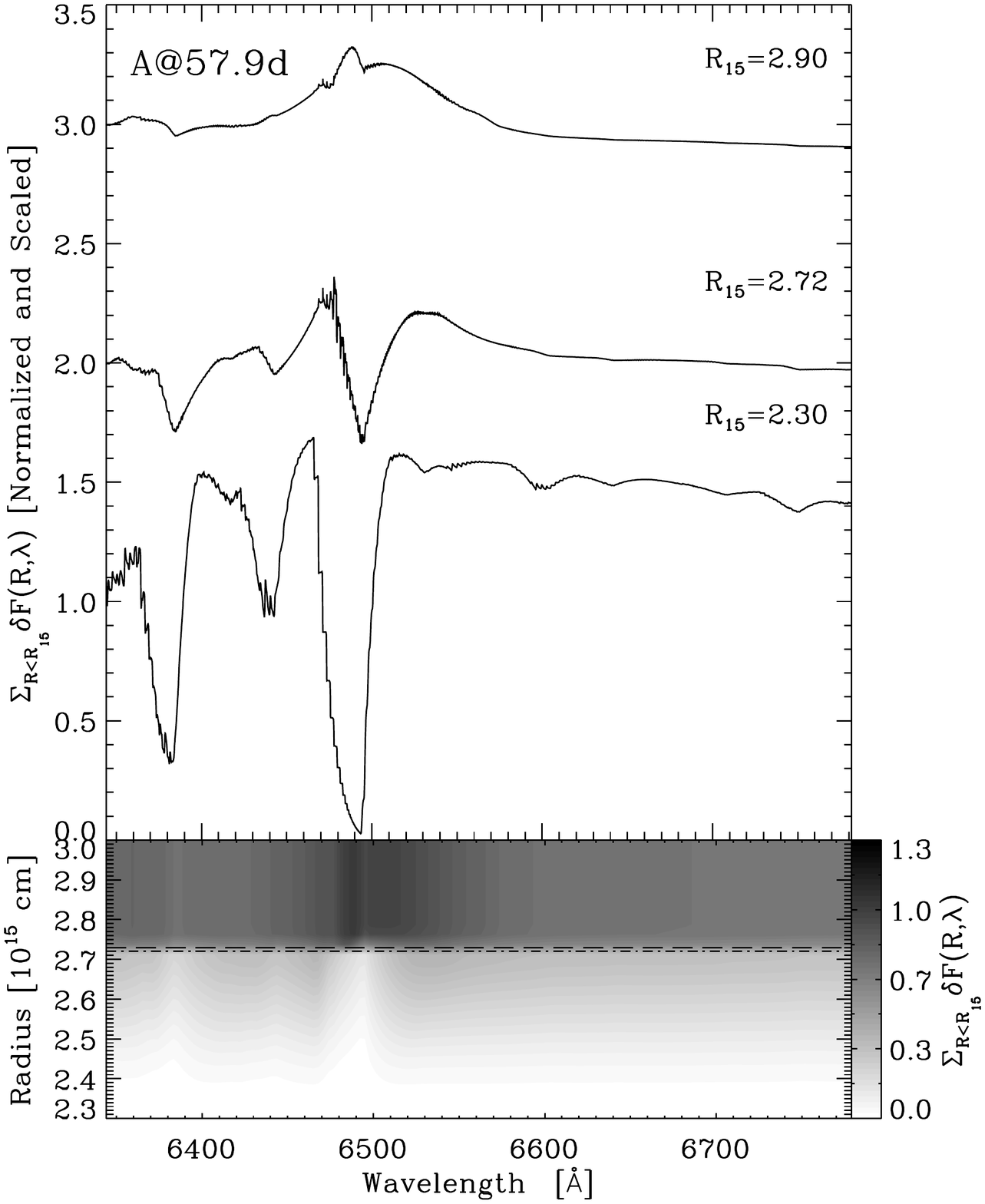, width=0.4\textwidth}
\vspace{-0.7cm}
\caption{
Same as Fig.~\ref{fig_dfr_w0s_11p6}, but now at 57.9\,d after the onset of interaction.
In the bottom panel and from top to bottom ($R_{15}=$\,2.9,2.725, and 2.3),
$\sum_{R<R_{15}} \delta F(R,\lambda)$ represents
1, 0.52, and 0.0007 of the total emergent flux at 4010\,\AA.
The CDS is becoming transparent, and emission from the inner
shell is starting to directly escape to the observer. However, at this epoch, the observed
emission is still dominated by the CDS.
\label{fig_dfr_w0s_57p9}
}
\end{figure*}

\begin{figure*}
\epsfig{file=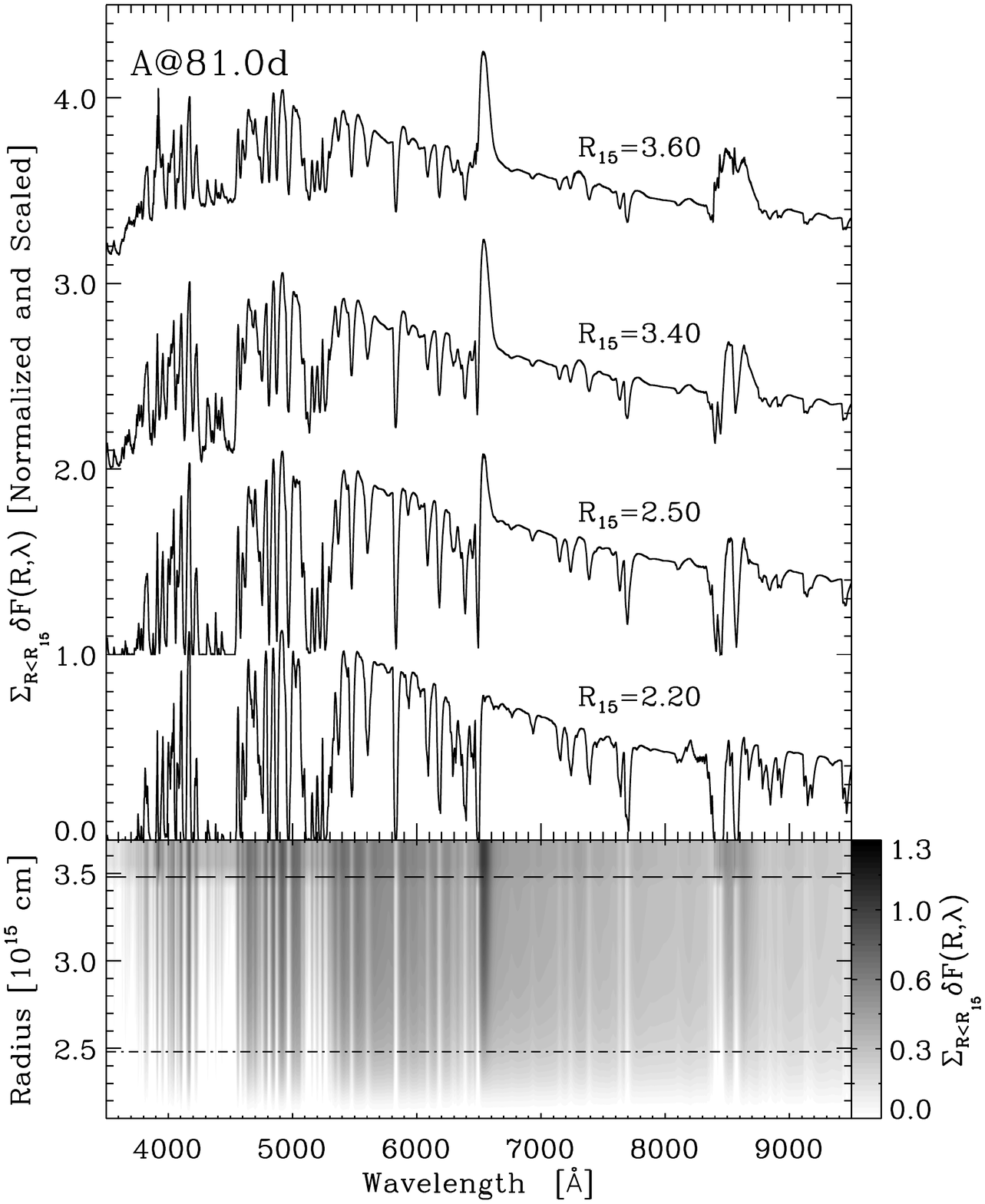, width=0.4\textwidth}
\epsfig{file=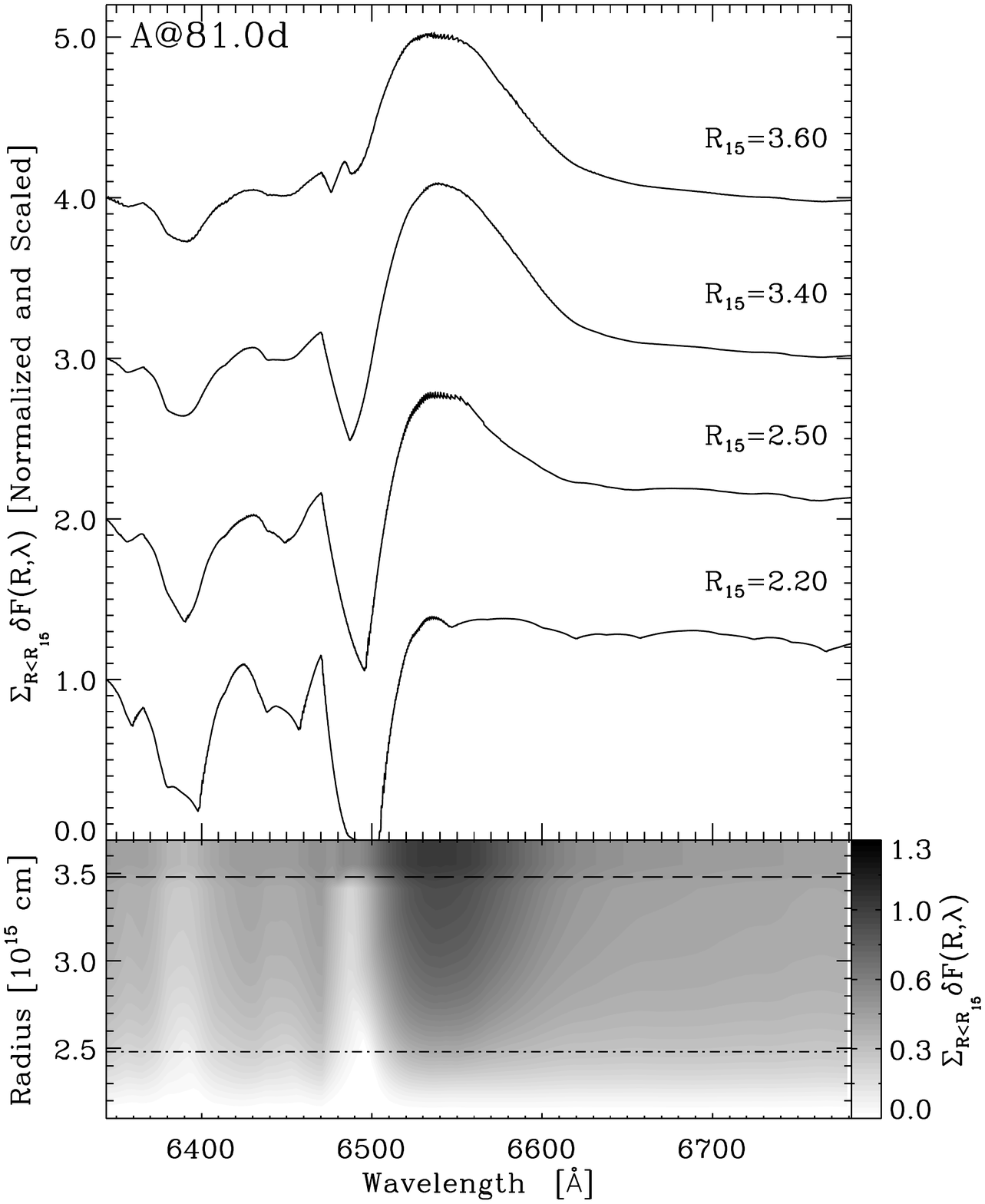, width=0.4\textwidth}
\vspace{-0.7cm}
\caption{
Same as Fig.~\ref{fig_dfr_w0s_11p6}, but now at 81.0\,d after the onset of interaction.
In the bottom panel and from top to bottom ($R_{15}=$\,3.6, 3.4, 2.5, and 2.2),
$\sum_{R<R_{15}} \delta F(R,\lambda)$
represents  1.0, 0.93, 0.58, and 0.16 of the total emergent flux at 4930\,\AA.
At this epoch, the CDS is almost fully transparent and most of the emission arises from
the inner ejecta (shell). Emission form the CDS fills in many of the absorption features
normally associated with a Type II SN.
\label{fig_dfr_w0s_81p0}
}
\end{figure*}

\section{dfr plots for models C}

This section provides additional information on the spectrum formation computed by \cmfgen\
for the interaction configuration of model C simulated with \heracles. Details on how to interpret these
figures can be found in \citet{d15_10jl}.

\begin{figure*}
\epsfig{file=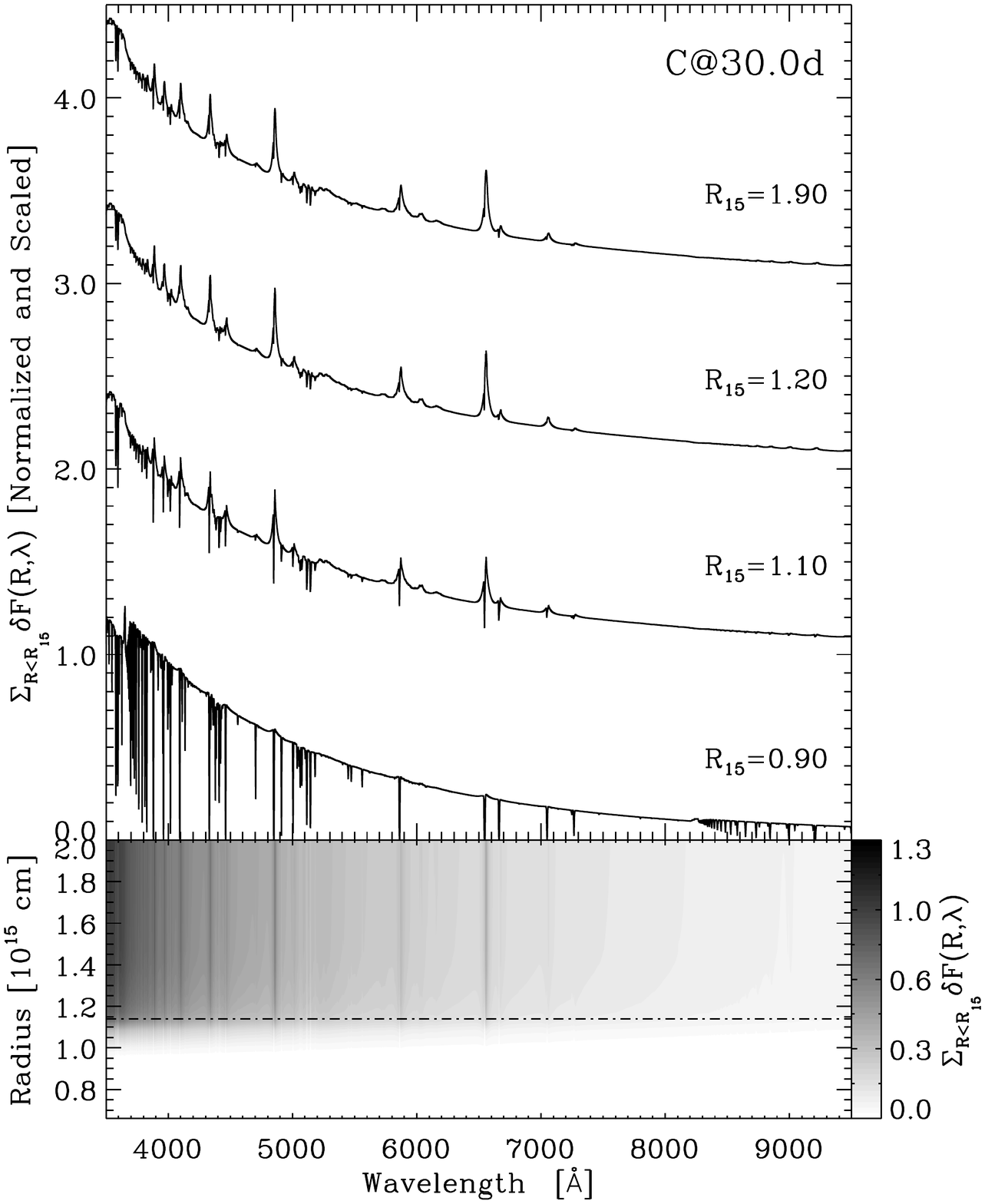, width=0.4\textwidth}
\epsfig{file=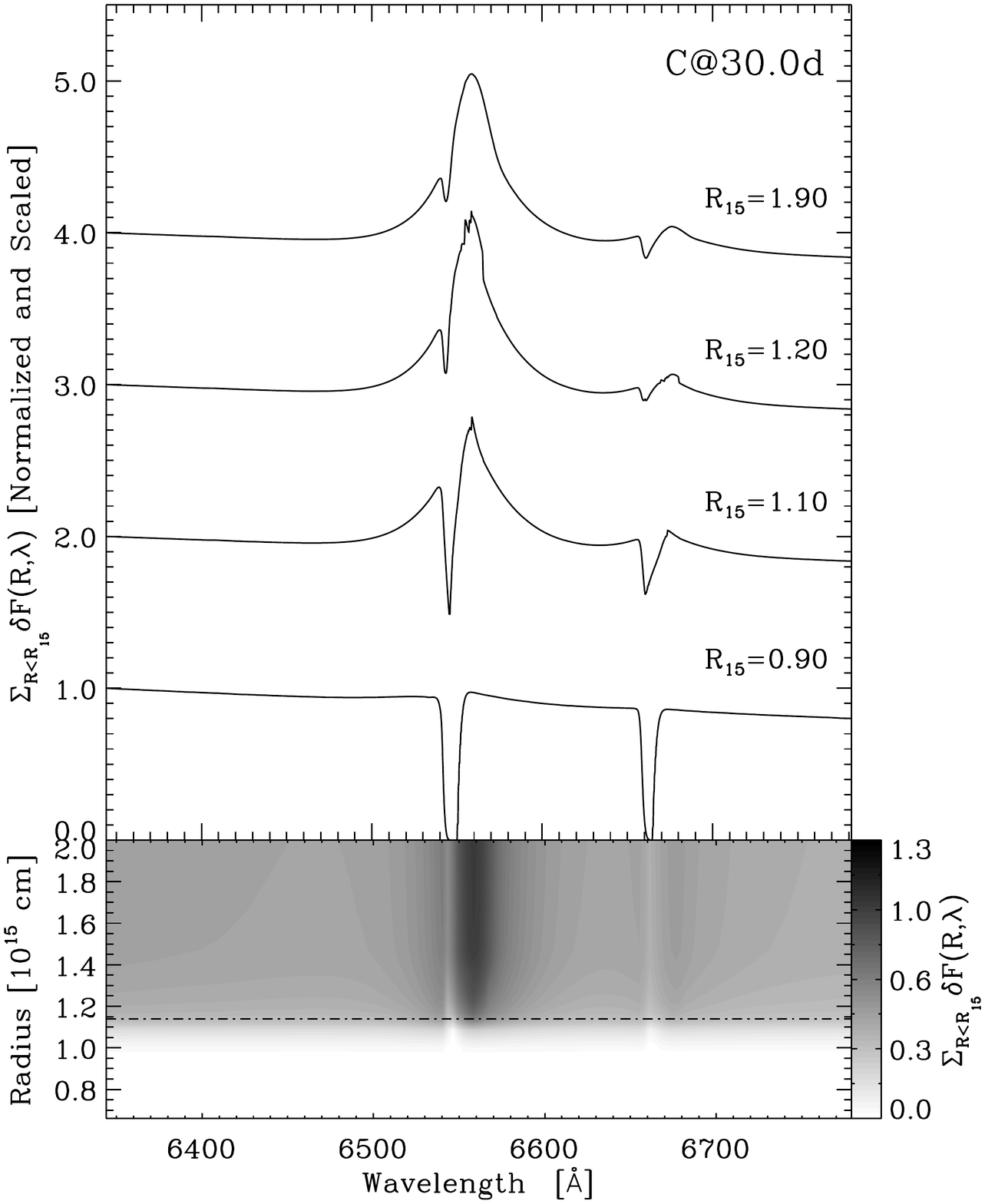, width=0.4\textwidth}
\vspace{-0.7cm}
\caption{
Same as Fig.~\ref{fig_dfr_w0s_11p6}, but now for model C at 30.0\,d after the onset of interaction.
In the bottom panel and from top to bottom ($R_{15}=$\,1.9, 1.2, 1.1, and 0.9),
$\sum_{R<R_{15}} \delta F(R,\lambda)$
represents  1.0, 1.07, 0.72, and 0.002 of the total emergent flux at 3950\,\AA.
At this epoch, the CDS is located deep within the otically thick CSM.
\label{fig_dfr_modc_30p0}
}
\end{figure*}

\begin{figure*}
\epsfig{file=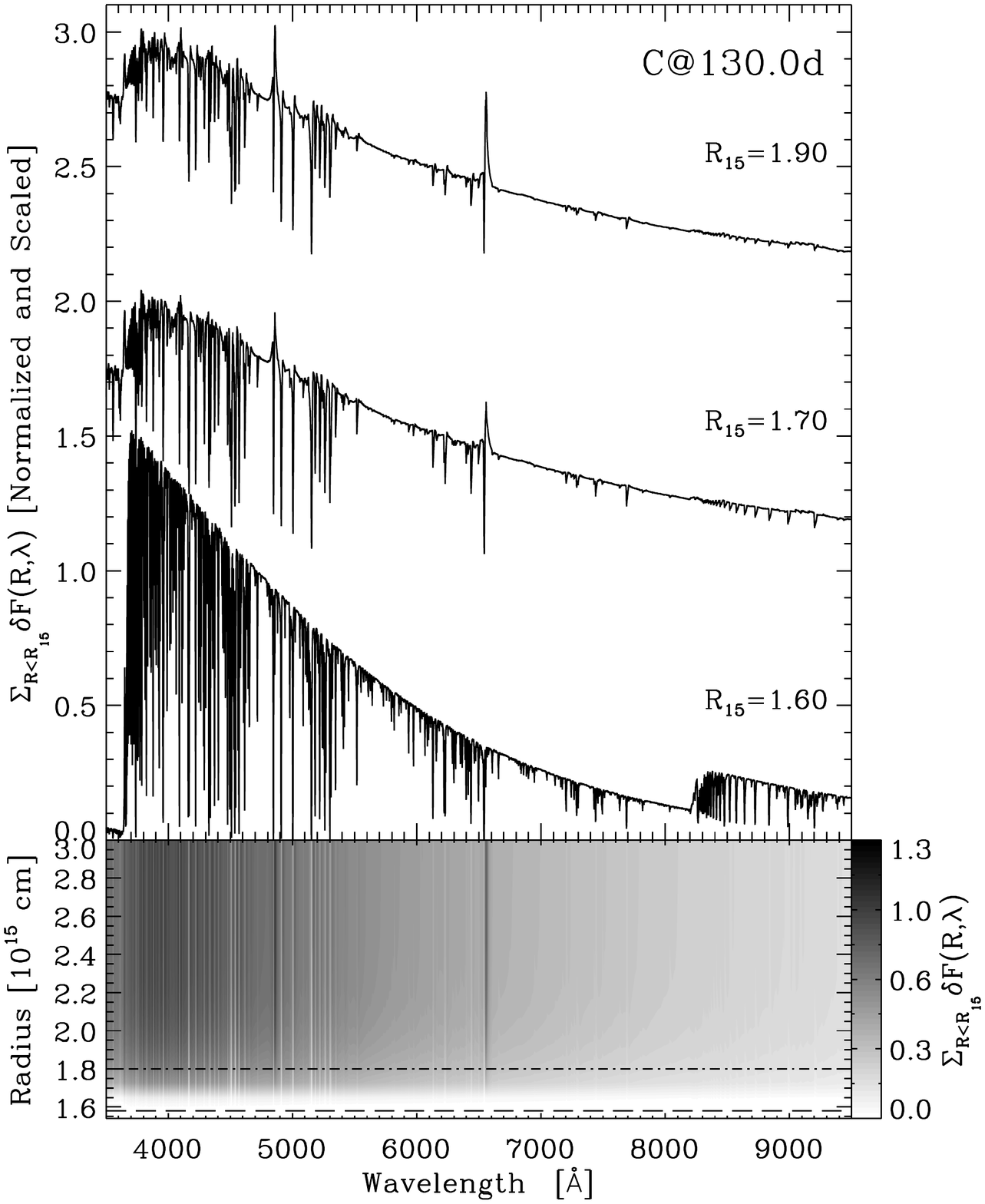, width=0.4\textwidth}
\epsfig{file=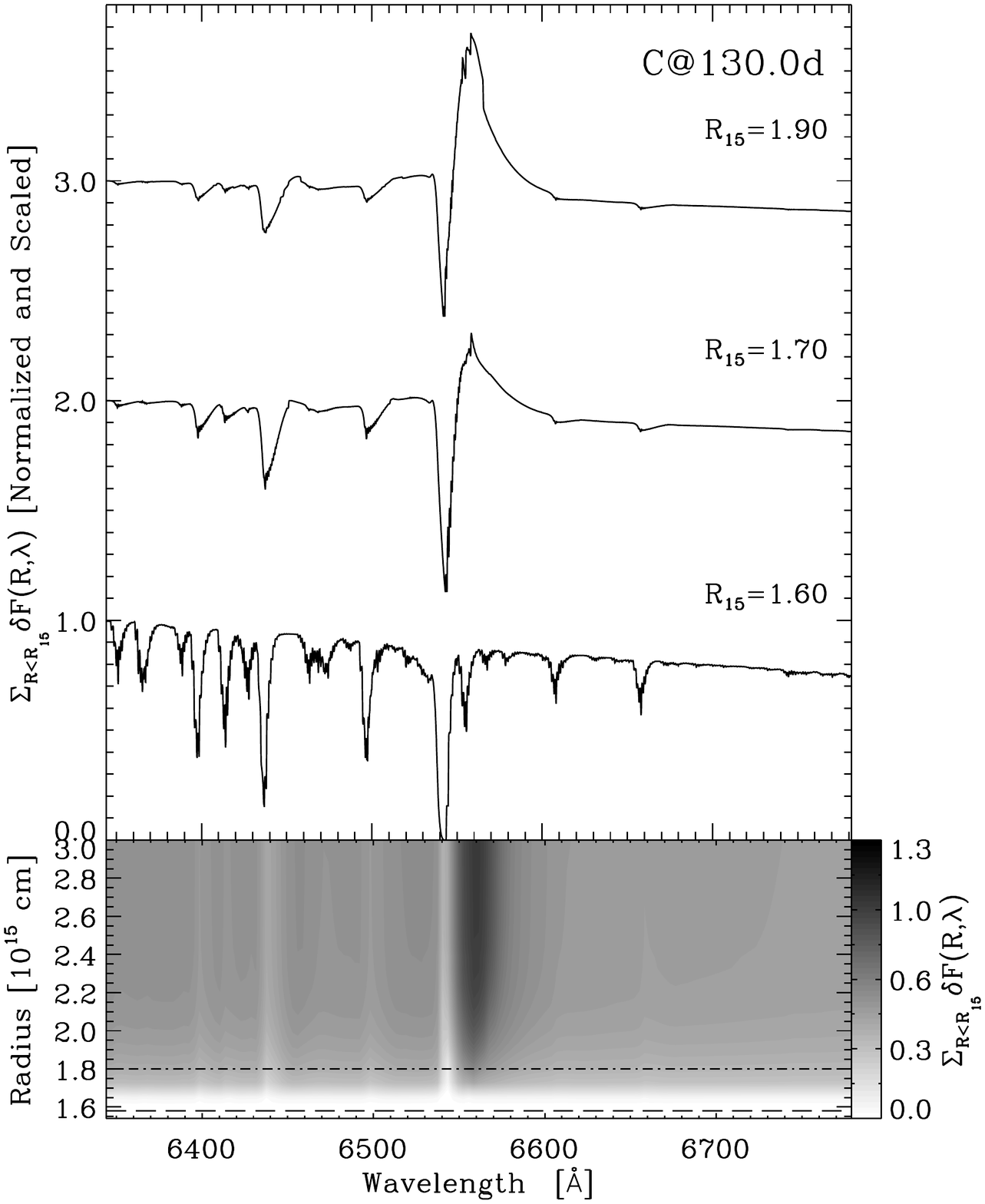, width=0.4\textwidth}
\vspace{-0.7cm}
\caption{
Same as Fig.~\ref{fig_dfr_modc_30p0}, but now at 130.0\,d after the onset of interaction.
In the bottom panel and from top to bottom ($R_{15}=$\,1.9, 1.7 and 1.6),
$\sum_{R<R_{15}} \delta F(R,\lambda)$
represents 0.99, 0.81, and 0.006 of the total emergent flux at 4100\,\AA.
At this epoch, the CDS is located deep within the otically thick CSM.
\label{fig_dfr_modc_130p0}
}
\end{figure*}

\label{lastpage}

\end{document}